\documentstyle[11pt,epsfig]{article}
\def\PR#1#2#3{{\rm Phys.~Rep.} {\bf#1} (19#2) #3}

\newcommand{\gs}      {\mbox{$ G_{\mathrm s}                          $}}
\newcommand{\gl}      {\mbox{$ G_{\mathrm L}                          $}}
\newcommand{\gr}      {\mbox{$ G_{\mathrm R}                          $}}
\newcommand{\gn}      {\mbox{$ G_{\mathrm \nu}                        $}}
\newcommand{\gmk}     {\mbox{$ G_{\mathrm k}                          $}}
\newcommand{\gml}     {\mbox{$ G_{\mathrm l}                          $}}
\newcommand{\gsg}      {\mbox{$ \Gamma_{\mathrm s}                        $}}
\newcommand{\glg}      {\mbox{$ \Gamma_{\mathrm L}                        $}}
\newcommand{\grg}      {\mbox{$ \Gamma_{\mathrm R}                        $}}
\newcommand{\gng}      {\mbox{$ \Gamma_{\mathrm \nu}                      $}}
\newcommand{\gmkg}     {\mbox{$ \Gamma_{\mathrm k}                        $}}
\newcommand{\gmlg}     {\mbox{$ \Gamma_{\mathrm l}                        $}}

\newcommand{\sqs}     {\mbox{$ \sqrt{s}                                    $}}
\newcommand{\ee}      {\mbox{$ {\mathrm e}^+ {\mathrm e}^-                 $}}

\newcommand{\ctw}     {\mbox{$ \cos\theta_{\mathrm W}                    $}}
\newcommand{\cwd}     {\mbox{$ \cos^2\theta_{\mathrm W}                  $}}
\newcommand{\cwq}     {\mbox{$ \cos^4\theta_{\mathrm W}                  $}}
\newcommand{\thw}     {\mbox{$ \theta_{\mathrm W}                          $}}

\newcommand{\sel}     {\mbox{$ \tilde{e}                                $}}
\newcommand{\sell}     {\mbox{$ \tilde{e_L}                                $}}
\newcommand{\selr}     {\mbox{$ \tilde{e_R}                                $}}
\newcommand{\smu}     {\mbox{$ \tilde{\mu}                              $}}

\newcommand{\stau}    {\mbox{$ \tilde{\tau}                             $}}

\newcommand{\snu}      {\mbox{$ \tilde{\nu}                                $}}
\newcommand{\ssup}     {\mbox{$ \tilde{u}                                  $}}
\newcommand{\sdn}     {\mbox{$ \tilde{d}                                   $}}

\newcommand{\stp}     {\mbox{$ \tilde{t}                                   $}}
\newcommand{\sbt}     {\mbox{$ \tilde{b}                                   $}}
\newcommand{\squ}     {\mbox{$ \tilde{q}                                   $}}

\newcommand{\snue}     {\mbox{$ \tilde{\nu_e}                              $}}

\newcommand{\schi}    {\mbox{$ \tilde{\chi}                               $}}

\newcommand{\sfe}       {\mbox{$ \tilde{f}                              $}}
\newcommand{\mxo}       {\mbox{$ m_{\chi_o}                              $}}

\newcommand {\rp } {${R}_{p}$\ }
\newcommand {\rpv} {\( \not \!\! {R}_{p} \) }
\newcommand {\emiss} {\( \not \!\! {E} \) }
\newcommand{\cpmtwo}    {\mbox{$ {\chi}^{\pm}_{2}                    $}}
\newcommand{\cpmone}    {\mbox{$ {\chi}^{\pm}_{1}                    $}}

\newcommand{\newc}{\newcommand}
\newc{\beq}{\begin{equation}}
\newc{\eeq}{\end{equation}}
\newc{\barr}{\begin{eqnarray}}
\newc{\earr}{\end{eqnarray}}
\newc{\ra}{\rightarrow}
\newc{\chio}{{\tilde\chi_0}}
\newc{\lam}{\lambda}
\newc{\dbar}{{\bar d}}
\newc{\ubar}{{\bar u}}

\newc{\dpr}[2]{({#1}\cdot{#2})}
\newc{\rpvm}{{\not \! R_p}}
\newc{\Hbar}{{\bar H}}
\newc{\Ubar}{{\bar U}}
\newc{\Dbar}{{\bar D}}
\newc{\Ebar}{{\bar E}}
\newc{\eg}{{\it e.g.}\ }

\def\st{\widetilde{t}}

\def\sth{\st_{2}}
\def\beqd            {\begin{displaymath}}
\def\eeqd            {\end{displaymath}}
\def\baa             {\begin{array}}
\def\eaa             {\end{array}}
\def\beqaa           {\begin{eqnarray}}
\def\eeqaa           {\end{eqnarray}}
\def\beqaad          {\begin{eqnarray*}}
\def\eeqaad          {\end{eqnarray*}}
\def\btabu           {\begin{tabular}}
\def\etabu           {\end{tabular}}
\def\bfig            {\begin{figure}}
\def\efig            {\end{figure}}
\def\bce             {\begin{center}}
\def\ece             {\end{center}}

\def\noi             {\noindent}

\newcommand{\eq}[1]  {\mbox{Eq.~(\ref{#1})}}

\def\ti              {\tilde}

\def\b               {\beta}

\def\l               {\lambda}
\def\t               {\theta}

\def\x               {\chi}

\def\st              {\ti t}

\def\sb              {\ti b}

\def\stau            {\ti \tau}

\def\sf              {\ti f}

\def\ch              {\ti \x^\pm}

\def\nt              {\ti \x^0}

\newcommand{\msf}[1]   {m_{\sf_{#1}}}
\newcommand{\mch}[1]   {m_{\ti \x^\pm_{#1}}}
\newcommand{\mnt}[1]   {m_{\ti \x^0_{#1}}}

\def\tW              {\t_W}
\def\sth             {\sin\t}
\def\cth             {\cos\t}

\def\onehf               {\small{\frac{1}{2} }} 
\def\oneth               {\small{\frac{1}{3} }}    
\def\twoth               {\small{\frac{2}{3} }}

\def\mnue{m_{\tilde{\nu}_{e}}^2}

\def\Gamn{\Gamma_{\tilde{\nu}}}
\def\mwi{m_{\tilde{\chi}^{\pm}}^2}

\def\mneu{m_{\tilde{\chi}^{0}}^2}

\begin{document}
\pagenumbering{roman}

\title{{\bf {\tt SUSYGEN}~2.2} \\
 A Monte Carlo Event Generator for MSSM Sparticle Production at $e^+ e^-$ Colliders}
\author{Stavros Katsanevas$^{a}$ and Peter Morawitz$^{b,c}$}
\date{{\small \it 
$^a$  Universit\'e Claude Bernard Lyon I, 
         47 11 Novembre 1917,          
         VILEURBANNE 69222, FRANCE \\
$^b$ Imperial College, HEP Group, London SW7 2AZ, England  \\
$^c$ Now at IFAE, Universitat Aut\`{o}noma de Barcelona, 08193 Bellaterra
(Barcelona), Spain  \\
e-mail: Stavros.Katsanevas@cern.ch, Peter.Morawitz@cern.ch}}
\maketitle

\vspace{-9.5cm}
\rightline{IC/HEP/97-5}
\rightline{IFAE-UAB/97-01}
\rightline{LYCEN 9744}
% \rightline{GDR-SUSY 97-01}
\vspace{+8cm}

\begin{center}
\bigskip
\noindent
Program classification: 11.1 \\
\bigskip

\abstract{{\tt SUSYGEN} is a  Monte Carlo program  designed for computing
 distributions and generating events for MSSM sparticle production in $e^+ e^-
 $ collisions. The Supersymmetric (SUSY) mass spectrum may either be supplied by
 the user, or can  alternatively be calculated in two different models of SUSY
 Breaking: gravity
 mediated supersymmetry breaking (SUGRA), and  gauge mediated supersymmetry
 breaking (GMSB).  The program incorporates the most important production processes 
 and decay modes, including  the full set of R-parity violating  decays, and the decays
 to the gravitino in GMSB models. Initial state  radiation corrections take into
 account  $p_T/p_L$ effects in the Structure Function formalism, and  an 
 optimised hadronisation interface to {\tt JETSET} 7.4 including 
 final state radiation is also provided.}
\end{center}

\vspace{1cm}
\begin{center}
To be submitted to Computer Physics Communications. \\
\end{center}

\newpage

\leftline{\Large{\bf NEW VERSION SUMMARY}}
\vskip 15pt

\leftline{{\it Title of new version:} {\tt SUSYGEN}~2.2}
\vskip 8pt

\leftline{{\it Catalogue number:} }
\vskip 8pt

\noindent
{\it Program obtainable from:} CPC Program Library, Queen's
University of Belfast, N. Ireland (see application form in this issue)
\vskip 8pt

\noindent
{\it Reference to original program:} {\tt SUSYGEN~1.0}; S.~Katsanevas and
S.~Melanchrinos, {``\tt SUSYGEN''},  in the proceedings of the Workshop {\em Physics at LEP~2}, 
eds. G.~Altarelli, T.~Sj\"ostrand, and  F.~Zwirner, CERN~96-01, page~328,
Geneva, 19 February 1996.
\vskip 8pt

\noindent
{\it Authors of original program:} Stavros Katsanevas, Stavros Melachroinos
\vskip 8pt

\noindent
{\it The new version supersedes the original program}
\vskip 8pt

\noindent
\leftline{{\it Licensing provisions:} none}
\vskip 8pt

\noindent
\leftline{\it Computer for which the new version is designed:}
DEC ALPHA, HP 9000/700, IBM AIX series; 
\vskip 8pt

\noindent
\leftline{{\it WWW site:} http://lyohp5.in2p3.fr/delphi/katsan/susygen.html}
\vskip 8pt

\noindent
\leftline{{\it Operating system under which the new version has been tested:}
UNIX, VMS}
\vskip 8pt

\leftline{{\it Programming language used in the new version:} FORTRAN 77}
\vskip 8pt

\noindent
\leftline{{\it Memory required to execute with typical data:} 
$\approx$~250 -- 450Kb }
\vskip 8pt

\leftline{{\it No. of bits in a word: } 32}
\vskip 8pt

\leftline{{\it No. of processors used: } 1}
\vskip 8pt

\leftline{\it The code has not been vectorised}
\vskip 8pt

\leftline{{\it Subprograms used:}
{\tt CERNLIB}~[i],  {\tt JETSET}~[ii], {\tt RANLUX}~[iii], {\tt PHOTOS~[iv]}}
\vskip 8pt

\leftline{{\it No. of lines in distributed program, including test data,
etc.:} $\approx$~18000}

\vskip 8pt
\leftline{\it Correspondence to: Stavros.Katsanevas@cern.ch, Peter.Morawitz@cern.ch}
\vskip 8pt


\vskip 8pt

\leftline{{\it Nature of physical problem} }
\noindent
The complexity and the interdependence of the MSSM signals demand a fast and  
precise  simulation program which treats the production and
 decay of all possible sparticles in different models  of supersymmetry
 breaking, and in models in which R-parity is either conserved or violated.
\vskip 8pt

\leftline{{\it Method of solution} }
\noindent
SUSYGEN 2.2 is a Monte Carlo generator for the tree-level production
and decay of all MSSM sparticles at $\ee$ colliders.  It includes 
pair-production of gauginos, sfermions and squarks, the 
production of a gravitino plus a neutralino in Gauge-mediated SUSY breaking
models,  the single production of gauginos through s-channel resonance in 
R-parity violating models, and the production of Higgses.
The program implements all important decay modes of sparticles relevant to LEP
energies, including cascade decays to lighter SUSY states, radiative decays of
gauginos to a photon, decays to the Higgses, decays to gravitinos,
and R-parity violating decays to Standard Model particles. 
 The precise matrix elements  are used for the production and for the
 decay of sparticles, though helicity correlations are not taken into account.
 The mass spectrum of the SUSY particles at the electroweak scale 
 is either  calculated assuming a  common mass scale at the GUT scale 
 (Supergravity models or Gauge mediated SUSY breaking models), or the mass
 spectrum can be optionally provided by the user.
 The LSP can either be the neutralino, the sneutrino or the gravitino in
 R-parity conserving models,  or there can be 
 no LSP in the sense that all sparticles can decay to standard particles through
 R-Parity violation.  
 The initial state radiative corrections  take account of  
 $p_T / p_L$ effects  in the Structure Function formalism. QED final state
 radiation is implemented using the PHOTOS library. An optimised
 hadronisation interface to {\tt JETSET 7.4} is  provided, which also takes into
 account  the lifetime of sparticles.
 
\vskip 8pt

\leftline{\it Reasons for the new version}
\noindent
SUSYGEN has been  extensively used by the four LEP collaborations 
to simulate the expected signals on the way to the higher LEP
energies. The scope of version 1.0 of the program was limited in the following
way: only neutralino LSPs were implemented; gauge-mediated SUSY breaking
scenarios, i.e. the production and decay of gravitinos was not implemented; no
R-parity violation option was available; QED final state radiation was not
included, and none of the  Higgs production mechanisms were implemented. 

\vskip 8pt

\leftline{{\it Restrictions on the complexity of the problem} }
\noindent
The factorised treatment of the production and decay of sparticles 
through the appropriate matrix elements is  more accurate  than  the equivalent
treatment of generators like e.g ISAJET \cite{isajet} and SPYTHIA \cite{spythia},  where the
sparticle decays are described by phase space. Nevertheless
it is less accurate than programs which take helicity correlations into account,
as for example is done in SUSY-GRACE (SUSY23) \cite{susygrace}, or DFGT \cite{dfgt}. 
Squark hadronisation before decay is not implemented, and the hadronisation
of  squark decays via the  R-parity violating coupling $\lambda''$ is only 
 simulated using the independent fragmentation scheme. The decays to gluinos are
 at present not implemented. And three-body decays of sfermions to lighter
 sfermions are not included. 

\vskip 8pt
\leftline{{\it Typical running time:} 40 events/sec on a DEC ALPHA.} 
\vskip 8pt
\leftline{{\it Unusual features of the program:} none } 
\vskip 8pt
\leftline{{\it References} }
\noindent
[i] CERN Program Library, CN Division, CERN, Geneva.
\vskip 10pt\noindent
[ii] T.~Sj\"ostrand, Comp. Phys. Commun. {\bf 82} (1994) 74.
\vskip 10pt\noindent
[iii] F.~James, Comput. Phys. Commun. {\bf 79} (1994) 111.
\vskip 10pt\noindent
[iv] E.~Barberio and Z.~Was, Comp. Phys. Commun. {\bf 79} (1994) 291.
\vskip 10pt\noindent

\newpage
\tableofcontents
\newpage
\leftline{\Large{\bf LONG WRITE-UP}}
\vskip 15pt

\section{Introduction}
\pagenumbering{arabic}

In recent years   supersymmetry (SUSY)
has been  extensively used to chart the map of possible physics beyond 
the standard model (SM). This symmetry predicts the existence of 
 additional particles which differ from their standard model partners by
 half a unit of spin. The simplest model available is the
 Minimal Supersymmetric Standard Model (MSSM) \cite{moti,recent}
 which contains the minimal number of new particles and 
 interactions that are consistent with the
 SM gauge group. In the MSSM  all SM fermions have scalar
 SUSY partners (see Table~\ref{tablemssm}): the sleptons, sneutrinos and squarks. The
 SUSY equivalent of the gauge and Higgs bosons are the charginos and neutralinos
 (the gauginos), which are the mass eigenstates of the
 (${\tilde W^+}, {\tilde H^+}$) and (${\tilde \gamma},{\tilde Z},{\tilde
  H^0_1},{\tilde H^0_2}$) fields, respectively. 
 Within the  supersymmetric framework many theoretical questions
 of grand unified theories (GUTs) such as the hierarchy problem
 and the unification of couplings \cite{amaldi}  may be successfully addressed.

\begin{table}[hbt]
\begin{center}
\begin{tabular}{|lcclcc|}     \hline
&&&&&\\  
particle & & spin & sparticle & & spin \\
&&&&& \\ \hline
&&&&&\\
quark  & $q$ & 1/2 & squarks & $\tilde{q}_{L,R}$ & 0 \\
charged lepton & $l$ & 1/2 & charged sleptons & $\tilde{l}_{L,R}$ & 0 \\
neutrino  & $\nu$ & 1/2 &  sneutrino & $\tilde{\nu}$ & 0 \\
gluon & g & 1 & gluino & $\tilde{g}$ & 1/2 \\
photon & $\gamma$ & 1 & photino & $\tilde{\gamma}$ & 1/2 \\
&$Z^0$ & 1 & zino & $\tilde{Z}$ & 1/2 \\
neutral higgses & $h,H,A$ & 0 & neutral higgsinos & $\tilde{H}^{0}_{1,2}$ & 1/2 \\ 
& $W^\pm$ & 1 & wino & $\tilde{W^\pm}$ & 1/2 \\
charged higgs & $H^{\pm}$ & 0 & charged higgsino & $\tilde{H}^{\pm}$ & 1/2 \\ 
graviton & $G$ & 2 & gravitino & $\tilde{G}$ & 3/2 \\
&&&&& \\ \hline
&&&&& \\ 
\multicolumn{6}{|c|}{
$\tilde{W^\pm}$,$\tilde{H^\pm}$ mix to form 2 chargino 
mass eigenstates \cpmone,\cpmtwo}\\ 
\multicolumn{6}{|c|}{
$\tilde{\gamma}$, $\tilde{Z}$, $\tilde{H^0}_{1,2}$ mix to
form 4
neutralino mass eigenstates $\chi^0_1,\chi^0_2,\chi^0_3,\chi^0_4$}\\ 
\multicolumn{6}{|c|}{${\tilde t}_L,{\tilde t}_R$ (and ${\tilde b}, {\tilde
    \tau}$) mix to form the mass eigenstates ${\tilde t}_1, {\tilde t}_2$} \\ \hline
\end{tabular}
\caption{Particle-Sparticle correspondence in the MSSM}
\label{tablemssm}
\end{center}\end{table}

 The masses and couplings of the new supersymmetric states are  related by the 
 symmetry to those of the  SM states. If  supersymmetry were an exact
 symmetry,  the SUSY partners would be degenerate in mass with  their SM
 partners. Since no such states have been observed to date, supersymmetry must
 be broken at some higher energy scale. Two  supersymmetric breaking
 mechanisms have been studied in detail in the literature: gravity mediated
 supersymmetry breaking  (SUGRA) \cite{gramsb,gramsb1} and gauge mediated
 supersymmetry breaking (GMSB) \cite{gauge.mediated.SUSY,gaumsb}. 

 The first model (SUGRA) is inspired by supergravity theories, where 
 supersymmetry is broken at some very high scale, a "hidden sector", close to
 the Grand Unification  scale and is communicated to the  visible sector through
 gravitational interactions. If one assumes  universal soft breaking terms at
 the GUT scale,  the masses of the SUSY particles and their
 couplings can be calculated at the
 electroweak scale  through the evolution of the renormalisation group
 equations (RGEs). The model thus predicts the entire SUSY spectrum from a few
 parameters. 

 Recently the second model (GMSB), where the supersymmetry breaking occurs at a 
 relatively low scale (a few hundred TeV),   has received considerable 
 attention in the literature.  Here the messengers of the supersymmetric
 breaking are the gauge bosons. This model also predicts 
 mass relationships for the gauginos and the scalars, but the main difference to
 SUGRA models is that the {$\tilde G$} goldstino\footnote{Throughout this paper
 we will refer to the {$\tilde G$} as the gravitino.} -- the longitudinal component of
 the gravitino -- can be very light, and would thus be the 
 Lightest Supersymmetric Particle (LSP).

 Apart from the issue of supersymmetry breaking, a further complication arises
 when the SM is extended to incorporate SUSY. The most general interactions of
 the SM and SUSY  particles  invariant under the  $SU(3)_c\times SU(2)_L\times U(1)_Y$ gauge
 symmetry are those of the MSSM 
 plus the additional  superpotential terms \cite{rpsuper}
\begin{equation}
W_{\rpvm} =\lam_{ijk}L_iL_j\Ebar_k + \lam'_{ijk} L_iQ_j\Dbar_k
+\lam''_{ijk}\Ubar_i\Dbar_j \Dbar_k.
\label{eqrpv}
\end{equation}
Here $L$ ($Q$) are the lepton (quark) doublet superfields, and
$\Dbar,\Ubar$ ($\Ebar$) are the
down-like and up-like quark (lepton) singlet superfields, respectively;
$\lambda, \lambda', \lambda''$ are Yukawa couplings, and $i,j,k=1,2,3$ are
generation indices. The first two terms  in \eq{eqrpv} 
violate lepton number, while the last term violates baryon number. 
The simultaneous presence of the last two terms leads to
rapid proton decay, and the solution of this problem in the MSSM is to exclude all terms in
 \eq{eqrpv} by imposing the conservation of a discrete multiplicative
 quantum number, R-parity  \cite{rparity,rparity1}, defined as:
\begin{equation}
R_p=(-1)^{3B+L+2S}
\label{rpv.def}
\end{equation}
Here $B$  denotes the baryon number, $L$ the lepton number and $S$ the spin of
the particle state.  However, this
 solution  is not unique, and a number of models \cite{rpv.models} predict
 only a subset of the terms in (\ref{eqrpv}), thus protecting the proton from
 decay. These alternative solutions are denoted ``R-parity
 violation''. Theoretically there is no clear preference for models which impose
 R-parity conservation or violation \cite{rparity1}.  Experimentally the two cases
 correspond to vastly different signatures, as will be discussed
 below.

If R-parity is conserved one may expect two different classes of candidates for
the LSP, depending on whether the messengers of the supersymmetry breaking are 
gravitational or the gauge bosons: 
the neutralino (or the sneutrino), or the gravitino. In either case
the LSP is  stable, a consequence of R-parity conservation. 
If the neutralino (or the sneutrino) is the LSP  one expects events with
 large missing energy (\emiss) due to the neutralino escaping
detection. In the second case where the gravitino is the LSP, the possible
signatures depend on the decay rate of the heavier SUSY particles to the
gravitino  and on the nature of the NLSP (Next Lowest Supersymmetric
Particle). For example,  neutralino NLSPs  decay to an undetectable
gravitino and a photon, while sfermion NLSPs  decay to the corresponding 
fermion and the gravitino. Prompt decays to the gravitino 
would  produce the classical  missing energy signatures,
possibly with additional photons in the event; slow decays inside the detector
volume may  produce tracks with  kinks or displaced vertices; and  decays
outside the detector may also produce heavy stable charged particle
signatures if the NLSP is a charged particle. 

R-parity violation has two mayor consequences for collider phenomenology. Firstly, the
LSP is not stable and can decay to SM particles. And secondly, sparticles may be
produced singly. This opens up a  a whole new area of different
signatures, which in the simplest case may consist of acoplanar lepton pairs or
acoplanar jets, or in the more complex case of multi-lepton and/or  multi-jet  final
states, possibly with some missing energy. 

 SUSYGEN 2.2 is a Monte Carlo generator capable of simulating the 
 production and decay of all possible sparticles in the different scenarios
 described above, i.e. in 
 models  of different supersymmetry breaking (SUGRA or GMSB), and in models in which
 R-parity is either conserved  or violated. SUSYGEN is dedicated to 
 $e^+e^-$ colliders\footnote{The other three "general" SUSY programs are 
ISAJET \cite{isajet},  SPYTHIA \cite{spythia} and 
SUSY-GRACE (SUSY23) \cite{susygrace}. They have different areas of 
speciality: ISAJET and SPYTHIA treat both hadronic and $e^+e^-$
colliders, but they treat the sparticle decays through 
phase-space and they do not contain the full set of 
\rp decays, while SUSY-GRACE includes helicity
correlations, but does not incorporate all SUSY processes. It has no
\rpv option yet. See also \cite{altarelli} and references therein for other generators or
 programs related to SUSY.} where it has been extensively used (and debugged!) by the
 four LEP collaborations.  The program is based, for the most part, 
 on the formalisms of \cite{bartl} for the MSSM sparticle production and decays, 
 on \cite{ambros.gut.evo,carena} for SUGRA, on \cite{nano} for  GMSB,  and
 on  \cite{dreiner,dreiner.prod} for the \rpv production and decays.
 It uses the precise matrix elements for the production and
 for the decay of sparticles, although helicity effects are not taken into account.
 SUSYGEN incorporates initial and final state radiative corrections, and
 includes a  hadronisation interface to {\tt JETSET 7.4} \cite{jetset}. 
 
The outline of this paper is as follows: after reviewing and defining the MSSM
framework  in Section~\ref{mssm.frame}, 
the physics content of the generator is described in
Section~\ref{physics}. The structure of the program is outlined in Section~\ref{prog.des}, and
instructions on how to obtain, compile and run the program are given in
Section~\ref{runit}.  Conclusions and future plans are given in
Section~\ref{conclusions}.  The Appendix lists the particle and production code 
conventions, the gaugino mass matrix basis, 
the formulas used for the production and the decay of sparticles,
 common blocks and the  ntuple variables.
Finally a test run input and output listing is given.
\newpage

\section{The MSSM framework}\label{mssm.frame}
The MSSM is defined as the Supersymmetric extension of the SM
which  contains the minimal number of new particles and interactions that are
consistent with the SM gauge group. The gauge couplings of the SUSY particles
are equal to the SM gauge couplings. Experimental evidence shows that SUSY
must be broken, and the masses of the superpartners are therefore much heavier
than their SM counterparts. The exact nature of the Supersymmetry breaking mechanism 
 is unknown, but it can be parameterised by the following soft supersymmetric
 breaking terms \cite{gunion}:
\begin{itemize}
\item Gaugino masses: $M_i\lambda_i\lambda_i$, where $i=1,2,3$ for
$U(1),SU(2),SU(3)$ and $\lambda_i$ denotes the gaugino partner of
the corresponding gauge field.
\item Higgsino masses: $\mu\tilde{h}^0_{1}\tilde{h}^0_{2}$
and $\mu\tilde{h}^{\pm}_{1}\tilde{h}^{\pm}_{2}$, 
 where $\mu$ is the higgsino mass mixing parameter. 
\item
Scalar masses: \eg\ $m^2_{({\tilde t,\tilde
b})_L}({\tilde{t}_L}^\star\tilde{t}_L+{\tilde{b}_L}^\star\tilde{b}_L)+
m_{\tilde{t}_R}^2{\tilde{t}_R}^\star\tilde{t}_R+m_{\tilde{b}_R}^2{\tilde{b}_R}^\star\tilde{b}_R$,
plus similiar expressions for the first two generations and the sleptons and
sneutrinos.
\item `A' Left-Right trilinear mixing terms: \eg\ $A_t\lam_t
(\tilde{t}_LH_2^0-\tilde{b}_LH_2^+)\tilde{t}_R^\star$, plus similiar expressions
for $A_b$ and $A_{\tau}$.
\end{itemize}

In the ``unconstrained MSSM'' the above mass terms are considered to be free
parameters of the theory (Section~\ref{unconstrained}). The large number of unknowns can be substantially
reduced in ``constrained versions of the MSSM'' if a particular SUSY breaking
mechanism is employed. Two such mechanisms are implemented in SUSYGEN, SUGRA and
GMSB, and are discussed in Section~\ref{constrained}. One of the most important aspects for
collider phenomenology is the nature of the LSP, which determines the dominant 
decay modes of SUSY particles and hence their experimental signatures. This will
be discussed in Section~\ref{LSP}. 

\subsection{The unconstrained MSSM}\label{unconstrained}
The unconstrained MSSM corresponds in SUSYGEN to the following choices of
the input card ``MODES'' (see also  Section~\ref{input}): MODES=2, MODES=4 or MODES=5.  
\subsubsection{Gaugino Masses}\label{gaugino}
There are four spin 1/2 SUSY partners to the colorless neutral gauge fields and
higgs bosons, the neutral ``gauginos''. 
Their interaction eigenstates may be written as \cite{bartl}
\beq
 \psi_l^0=\left(-i\tilde{\gamma} ,-i\tilde{Z} , \cos\beta
 \tilde{H^0_1} -\sin\beta \tilde{H^0_2} , \sin\beta
 \tilde{H^0_1}  +\cos\beta \tilde{H^0_2}\right) ,
\eeq
 where $\tilde{\gamma},\tilde{Z},\tilde{H}^0_1,\tilde{H}^0_2$ denote the 
 the two component spinor fields of the photino, zino and the two neutral
 higgsinos, respectively;  $s_{\theta}$ and $c_{\theta}$ are the weak angle sine
 and cosine; and $\tan{\beta} = v_1 / v_2$, the ratio of the
 vacuum expectation values of the neutral Higgs fields. 
 The mass matrix of the neutral gauginos is given by 
\begin{equation}
M^0 = \left(\begin{array}{clcr}
  M_1 c^2_{\theta}+M_2 s^2_\theta & (M_2-M_1) c_\theta s_\theta & 0 & 0 \\
  (M_2-M_1) c_\theta s_\theta & M_2 c^2_\theta +M_1 s^2_\theta & m_Z & 0 \\
  0 & m_Z & \mu \sin2\beta & -\mu \cos2\beta \\
  0 & 0   & -\mu \cos2\beta & -\mu \sin2\beta
  \end{array}\right)
\label{mass.matrix}
\end{equation}
The physical mass eigenstates, the neutralinos ($\chi^0_k$), are defined by 
\beq
 \chi_k^0 =N_{kl}\psi_l^0
\eeq
and the neutralino mass eigenvalues ($m_{\chi^0_k}$) 
can be computed by diagonalising the mass matrix  \eq{mass.matrix} according to 
\beq
 m_{\chi^0_k}   \delta_{kl}= N_{km}N_{ln} M^0_{mn} .  
\label{neutralino.eigen}
\eeq

There are two SUSY partners to the W gauge boson and the charged Higgs
bosons. Their mass eigenstates are  the charginos ($\chi^+_k, \chi^-_k $), which
are admixtures of  their interaction eigenstates \cite{bartl}
\barr
 \psi_l^+&=&\left(-i\tilde{W}^+ , \tilde{H}^+_2 \right) \nonumber \\
 \psi_l^-&=&\left(-i\tilde{W}^- , \tilde{H}^-_1 \right) . 
\earr
The chargino mass matrix is given by 
\begin{equation}
M^c = 
\left(\begin{array}{cc}
  M_2 & \sqrt{2}m_W \sin\beta \\
  \sqrt{2}m_W \cos\beta & \mu
  \end{array}\right)
\label{mass.matrix.c}
\end{equation}
and the chargino mass eigenstates are given by 
\barr
 \chi_k^+ &=&V_{kl}\psi_l^+ \nonumber\\
 \chi_k^- &=&U_{kl}\psi_l^-       
\earr
The chargino mass eigenvalues ($m_{\chi^+_k}$) 
can be computed by diagonalising the mass matrix  \eq{mass.matrix.c} according to 
\beq
 m_{\chi^+_k}   \delta_{kl}= U^*_{km}V_{ln} M^0_{mn} .  
\label{chargino.eigen}
\eeq

The diagonalisation of the above two mass matrices gives 
four neutralino ($\chi^0_1,\chi^0_2,\chi^0_3,\chi^0_4$) and two chargino 
(\cpmone ,\cpmtwo) physical states. The masses and the couplings 
 of these states are determined by the eigenvalues and the eigenvectors of
 Eq.~(\ref{neutralino.eigen},\ref{chargino.eigen}).  They 
 depend on the three parameters ($M_2, \tan\beta, \mu$) in the case of 
charginos, and on the four parameters ($M_1, M_2, \tan\beta, \mu$)
in the case of  neutralinos. 
 Appendix~\ref{app.mass} gives more details on gaugino mixing and the different 
 neutralino bases used inside SUSYGEN.

 In order to simplify the parametric dependence a relationship between
 the gaugino masses is commonly assumed. 
 In the simplest possible picture the gaugino masses $M_1, M_2$ and $M_3$ 
 are assumed to be equal at the grand unification scale, an assumption which is
 natural in  the context of gravity 
 induced supersymmetry breaking.  The one-loop renormalisation group equations
 then allow one to calculate the evolution of the three couplings to the EW
 scale. This determines  the ``GUT relations'' for the gaugino mass terms \cite{gunion}:  
\barr
M_1 &=& \frac{5}{3} \tan^2{\theta_W} M_2    \label{gut.1} \\
M_3 &=& \frac{\alpha_S}{\alpha_{EM}} \sin^2{\theta_W} M_2  \label{gut.3}
\earr
Numerically this relationship (at the EW scale) is approximately given by
 $M_3:M_2:M_1 \simeq 7:2:1$. 
 The above relationship obtained in the context of SUGRA models is also valid
  for  GMSB models. 

However, the existence of string models or other models with non-universal masses 
 at the GUT scale \cite{nonuniversal} induces some scepticism on the general validity of
Eq.~(\ref{gut.1},\ref{gut.3}). By default SUSYGEN uses \eq{gut.1}, but the
ratio $M2:M1$ may be optionally  changed by the user.

The GUT relations  ensure that the
gluino (where $m_{\tilde g} = M_3$, neglecting QCD corrections)
 is normally much heavier than the SU(2) and U(1)
gauginos.  Their production is therefore typically inaccessible at LEP. 
However, in some models loop corrections to the gluino masses 
 predict light gluinos \cite{farrar.gluino}. And in GMSB models with two  effective SUSY
 breaking scales,  one for a colour-singlet and one for a colour non-singlet sector,  
 gluinos may also be light,  possibly even the LSP \cite{raby}. Light gluino
 scenarios are at present not implemented in SUSYGEN.

\subsubsection{The Gravitino Mass}

In GMSB models spontaneous SUSY breaking leads to a goldstone fermion, the
goldstino $\tilde{G}$ with very low mass.
 In local SUSY $\tilde{G}$ is the longitudinal component of the gravitino. 
 The $\tilde{G}$ mass is determined by $F$, the 
 characteristic scale of SUSY breaking, and  is given by \cite{gunion}:
\begin{eqnarray}
m_{\tilde{G}}=\frac{F}{\sqrt{3}M_{Planck}}=2.5 
\times (\frac{\sqrt{F}}{100 TeV})^2 eV
\label{mgrav}
\end{eqnarray}
For $\sqrt{F}\sim 10^8$ TeV -- the characteristic scale of 
string-motivated supergravity models --  the 
$\tilde{G}$ is fairly massive and has no consequences for LEP 
phenomenology. 
 In gauge mediated models $\sqrt{F}\sim 100-2000$ TeV, and  the $\tilde{G}$ 
 is the LSP. Note that the scale $\sqrt{F}$ also determines the coupling
 strength of the gravitino and thus enters in  the neutralino-gravitino 
 production cross section, and in the decay rate of the lifetime of the NLSP,
 which can only decay to the gravitino if R-parity is conserved.

\subsubsection{Third Generation Sfermion  Masses and Mixing Angles}
 The SUSY partners of the SM fermions are the sleptons ${\tilde l}_L, {\tilde l}_R$,
 the sneutrinos $\tilde \nu$ and the squarks ${\tilde q}_L, {\tilde q}_R$. The
 charged leptons and the quarks have two SUSY partners,  one for each
 lepton/quark  chirality.  The third generation sfermions ($\tilde t, \tilde b, \tilde
 \tau$)  are the most likely candidates for the lightest
 scalar quark or lepton states
  because of the potential for large mixing angles between
 the left and right handed states, and because of the large  third generation
 Yukawa couplings.  This may be seen from the stop and sbottom mass matrices
 \cite{bartl1} 
\begin{equation}
\begin{array}{c}
\left({\tilde t}_L {\tilde t}_R \right) 
\left(\begin{array}{cc}
  m_{{\tilde t}_L}^2+m_{t}^2+m_D^2 & m_{t} X_{t} \\
  m_{t} X_{t} & m_{{\tilde t}_R}^2+m_{t}^2+m_D^2
\end{array}\right)  
\left(\begin{array}{c}
  {\tilde t}_L \\
  {\tilde t}_R 
\end{array}\right) \\
% ---- now sbottom
\left({\tilde b}_L {\tilde b}_R \right) 
\left(\begin{array}{cc}
  m_{{\tilde b}_L}^2+m_{b}^2+m_D^2 & m_{b} X_{b} \\
  m_{b} X_{b} & m_{{\tilde b}_R}^2+m_{b}^2+m_D^2
\end{array}\right)  
\left(\begin{array}{c}
  {\tilde b}_L \\
  {\tilde b}_R 
\end{array}\right) \\
\end{array}
\label{stopmatrix}
\end{equation}
from which the mass eigenstates of the stop  (${\tilde t}_1, {\tilde
  t}_2$) and the sbottom (${\tilde b}_1, {\tilde  b}_2$) can be obtained by
  diagonalisation. Here 
\barr
X_t & =& A_t-\mu\cot\beta \\
 X_b &=& A_b -\mu\tan\beta \\
 m_D^2 &=& (I^3-Q\sin^2\theta_W)|\cos2\beta| m_Z^2 \label{d.terms}
\earr
  where $m_D$ are  the so-called ``D-terms'',  $I^3$ is the third component of isospin,
  and $Q$ is the electric charge.
  Note that for stops and sbottoms the off-diagonal terms in \eq{stopmatrix} are
  sizeable and lead to appreciable mixing between the left-right states. 
   This is
  not the case for the first two squark generations, where the analogous $m_q
  X_q$ terms are small since $m_q \sim 0$. Depending on $\tan\beta$ either the
  stop or the sbottom can be the lightest squark state.
  A similar argument applies for the   stau, which can be lighter than the
  selectron and the smuon due to mixing.

  SUSYGEN implements mixing between the left-right third generation
  sfermions, and the user can control the amount of mixing through the
  parameters $A_t, A_b, A_{\tau}$.
  In the unconstrained MSSM (MODES=2,4)  the user may
  alternatively  provide the explicit  mixing angles $\theta_{\tilde t}, \theta_{\tilde
  b}, \theta_{\tilde \tau}$ which parameterise the mass-eigenstates in terms of
  the left-right states:
 \barr
 {\tilde f}_1 &=& {\tilde f}_L \cos{\theta_{\tilde f}} + {\tilde f}_R
  \sin{\theta_{\tilde f}} \nonumber\\
 {\tilde f}_2 &=& -{\tilde f}_L \sin{\theta_{\tilde f}} + {\tilde f}_R
  \cos{\theta_{\tilde f}} \label{thetamix}
\earr

\subsubsection{The Higgs Mass}\label{higgs.masses}
The tree level Higgs masses only depend on two parameters: $m_A$, the 
mass of the CP-odd Higgs  and $\tan\beta$. The radiative corrections
\cite{haber} modify these predictions and introduce a dependence on the rest of
the supersymmetric mass spectrum. The RGE improved 
 formulae \cite{carena} implemented in SUSYGEN take into account most of these
 corrections. The Higgs masses show a strong dependence on  
the mixing of the stop $X_t$, 
increasing the theoretical upper limit on the  Higgs mass for larger 
values of  $X_t$.

\subsection{The constrained MSSM}\label{constrained}
The constrained MSSM corresponds in SUSYGEN to the following choices of
the input card ``MODES'' (see also  Section~\ref{input}): MODES=1 or  MODES=3.

\subsubsection{Mass Spectrum in SUGRA models}

In Supergravity inspired models the soft SUSY breaking parameters are assumed to be 
universal at the GUT scale, reducing the number of parameters to:
\begin{itemize}
\item
$m_0$, the common  mass   of the sfermions at the GUT scale. 
\item $M_2$, the SU(2) gaugino mass.
\item
$\mu$, the mass mixing parameter of the Higgs doublets.
\item
$\tan\beta$, the ratio of the vacuum expectation values of the
two Higgs doublets.
\item
$A_t,A_b,A_\tau$, the trilinear couplings in the Higgs sector. 
\end{itemize}
All above parameters are defined  at the EW scale except for $m_0$, 
which is defined at the GUT scale\footnote{Note that for example the   ISAJET program
\cite{isajet} defines the above parameters at the GUT scale.}.
 The sfermion masses $m_0$ are  evolved from the GUT scale to the EW scale
 according to the formulae given in Appendix B of reference
 \cite{ambros.gut.evo}. 
 Mixing of the third generation sfermions is
 taken into account according to \eq{stopmatrix} through the parameters
 $A_t, A_b$ and $A_\tau$. For certain sets of parameters the sfermion masses at
 the EW scale may turn out negative, and in this case a warning is issued and
 the program skips this particular parameter point.
 The Gaugino and the Higgs mass spectra are 
 calculated as discussed in Sections~\ref{gaugino},\ref{higgs.masses}. 

For the first two generations, and neglecting the D-terms \eq{d.terms},
the renormalisation of the scalar masses at e.g. $\tan\beta=1$ give
\begin{eqnarray}
({\tilde m}_R^2-m_0^2):({\tilde m}_L^2-m_0^2):({m}_{\tilde q}^2-m_0^2):M_2=0.22:0.75:6.40:1
\end{eqnarray}
displaying a clear mass  hierarchy, with the right sleptons (${\tilde m}_R$) being the
 lightest states, followed by a mass degenerate left slepton doublet (${\tilde m}_L$), 
 and the first and second generation squarks at higher masses. 

 Radiative corrections to the soft SUSY breaking  terms can 
 induce electroweak symmetry breaking, resulting in  further constrains on the
 parameters of SUGRA models. This is at present not implemented within SUSYGEN,
 although private versions of SUSYGEN interfaced to the code of \cite{boer}
 exist, and may be obtained from the authors.

\subsubsection{Mass Spectrum in GMSB models}
In  GMSB models  the scalar mass relationships are \cite{gunion}:
\begin{eqnarray}
m_{\tilde q}^2:{\tilde m}_L^2:{\tilde m}_R^2:M1=11.6:2.5:1.1:\sqrt{N_{5,10}}
\end{eqnarray}
Here $N_{5,10}$ is a number characterising the number of representations in the
messenger sector. Note that  the squarks are in principle very heavy in
GMSB models. The gravitino is the LSP, and its mass is given by
\eq{mgrav}. For $N_{5,10} = 1$ the neutralino is the NLSP, while for
$N_{5,10} \geq 2$  the $\tilde{l}_R$ is the NLSP. 
 The Gaugino and the Higgs mass spectra are calculated in the
 same way as in the SUGRA model. The following parameters determine the entire
 mass spectrum of the model: $N_{5,10}$, $\mu$, $M_2$, $\tan\beta$,
 $A_t,A_b,A_\tau$.

\subsection{The Nature of the LSP and the Sparticle Decay Modes}\label{LSP}

 SM particles have $R_p=+1$ while their SUSY partners have
 $R_p=-1$, which can  easily be seen from \eq{rpv.def}. 
 The multiplicative conservation of R-parity only allows vertices with
 an even number of SUSY particles, and the  LSP is therefore stable if
 R-parity is conserved. Cosmological arguments \cite{cosmological.args} then 
 require the LSP to be neutral, and the following R-parity conserving 
 LSP candidates are implemented in SUSYGEN:   the lightest neutralino, 
 the sneutrino or the gravitino.  The LSP is unstable and 
 can decay to SM particles 
 if R-parity is broken, and the above cosmological arguments do not apply. In
 this case the lightest chargino, the sleptons or the third generation squarks
 are also good LSP candidates. Gluino LSPs are at present not implemented in
 SUSYGEN.

 Note that only a subset of the LSP candidates may be realised in constrained
 versions of  the MSSM. In SUGRA models the lightest neutralino is the most
 likely LSP candidate, while in GMSB models the gravitino is the LSP. 

 All  sparticle decay modes relevant to LEP phenomenology are implemented with
 the exception of the three body decays of sfermions to lighter sfermions (which
 are only important in very specific  regions of parameter space, see for
 example \cite{stop.snu.decays,gmsb.selectron.decays}), and
 the decays to gluinos. Cascade decays to intermediate sparticle
 states are also fully taken into account. 

 SUSYGEN simulates the following set of R-parity conserving decay modes: 
\begin{itemize}
\item {Three body decays of gauginos to lighter gauginos, e.g. $\chi^+ \ra  f f'
  \chi^0$.}
\item {Two body decays of gauginos to sfermions, e.g. $\chi^+ \ra  l^+ {\tilde
      \nu}$.}
\item {Two body decays of gauginos to the Higgs, e.g. $\chi^0_k \ra H \chi^0_i$.}
\item {Radiative two body decays of neutralinos to a photon and a lighter
 neutralino, e.g. $\chi^0_k    \ra \gamma \chi^0_i$.}
\item {Two body decays of charginos to neutralinos, e.g. $\chi^+ \ra  W^+
    \chi^0$.}
\item {Two body decays of sfermions to gauginos, e.g. $\tilde f \ra f \chi$.}
\item {Two body decays to the gravitino, e.g. $\chi^0 \ra \gamma \tilde G$.}
\end{itemize}

The  R-parity violating superpotential  \eq{eqrpv} contains the 45
additional Yukawa couplings $\lambda_{ijk}$, $\lambda'_{ijk}$, $\lambda''_{ijk}$. 
 The $LL\bar{E}$ term corresponding to $\lambda_{ijk}$ is antisymmetric in $i,j$, and we take 
 $i<j$. The $\bar U \bar D \bar D$ term ($\lambda''_{ijk}$) is antisymmetric in
 $j,k$, therefore $j<k$. To avoid fast proton decay and evade other low energy
 constraints only a subset of the 45 couplings can be non-zero. It is usually
 assumed that the $\lambda$-couplings display a strong hierarchy
 \cite{rparity1}, in which case it is often sufficient to consider only one
 non-zero coupling, as is done in SUSYGEN.  The following set of R-parity
 violating decays  are included in SUSYGEN for the three operators $LL\bar{E},
 {LQ\bar{D}}$ and $\bar U \bar D \bar D$ and a single non-zero \rpv coupling:

\begin{itemize}
\item {Three body decays of gauginos to SM particles, e.g. $\chi^0 \ra l^+ l^-
    \nu$ for $\lambda_{ijk}\neq 0$.} 
\item {Two body decays of sfermions to SM particles, e.g. $\tilde{l}^+ \ra l^+ \nu$
    for $\lambda_{ijk}\neq 0$.}
\end{itemize}

\newpage
\section{Physics content}{\label{physics}}
This section provides an overview of the physics content of the
generator.  The implemented SUSY production mechanisms are described in
Section~\ref{prod}, and the sparticle decays in Section~\ref{decay}. 
The implementation of initial and final state radiative corrections are
discussed in Section~\ref{isr}, and  sparticle lifetime and the hadronisation
interface to JETSET are described in Sections~\ref{life} and \ref{hadro}.

\subsection{Sparticle Production}\label{prod}

\begin{figure}[htb]
\begin{center}
\epsfig{file=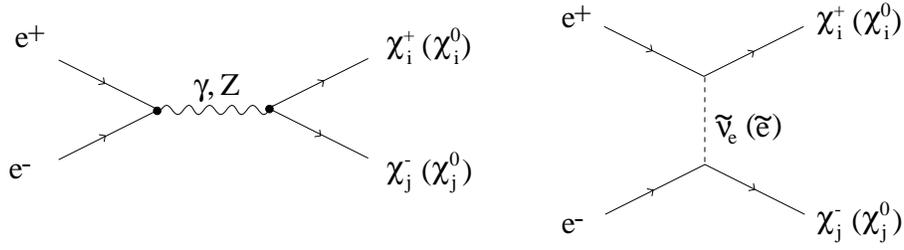,width=0.7\textwidth}
\caption{Chargino/neutralino production.}
\label{fig:fein1}
\end{center}
\end{figure}

The production mechanisms and their corresponding SUSYGEN process numbers are
summarised in Table~\ref{table.proc} in the Appendix. 
The diagrams for the s- and t-channel production of  charginos and neutralinos  (SUSYGEN
processes 1-13) are shown in Fig.~\ref{fig:fein1}. For large sneutrino and
slepton masses the s-channel diagram dominates, while for small sneutrino
(slepton) masses the t-channel contributions can be large, resulting in
destructive  (constructive) interference for chargino (neutralino) production
\cite{bartl}.
The cross sections therefore depend on the chargino and neutralino masses and their
couplings (and hence on $M_1, M_2, \mu$ and $\tan{\beta}$) as well as on the selectron and
electron sneutrino masses.

\begin{figure}[htb]
\begin{center}
\epsfig{file=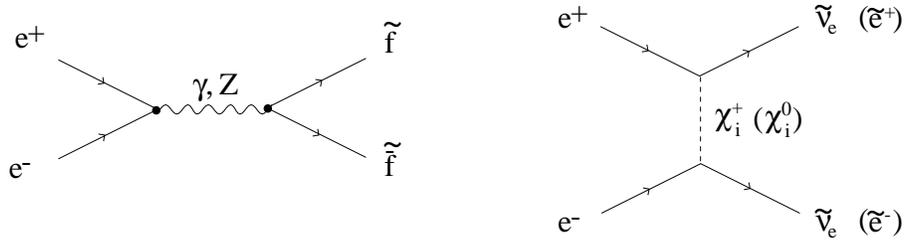,width=0.7\textwidth}
\caption{Sfermion production.}
\label{fig:fein2}
\end{center}
\end{figure}

Sfermions are produced in pairs (processes 14-35) through $\gamma,Z$ exchange in
the {\it s}-channel. The $\snue$\ and $\sel$\ are also produced
 with chargino and neutralino exchange in the {\it t}-channel \cite{bartl}. 
 Fig.~\ref{fig:fein2} shows the relevant diagrams. 
 The sfermion  cross sections  depend on the sfermion masses, and on the
 chargino or the neutralino masses and couplings in the case of  electron-sneutrino
 and selectron pair production,  respectively.   
 For the third generation sfermions ($\tilde \tau$, $\tilde t$ and $\tilde
 b$)  the left and right sfermion states can mix, and the 
 cross sections  also depend on the mixing angle $\theta_{\tilde f}$  \cite{bartl1}.

\begin{figure}[htb]
\begin{center}
\epsfig{file=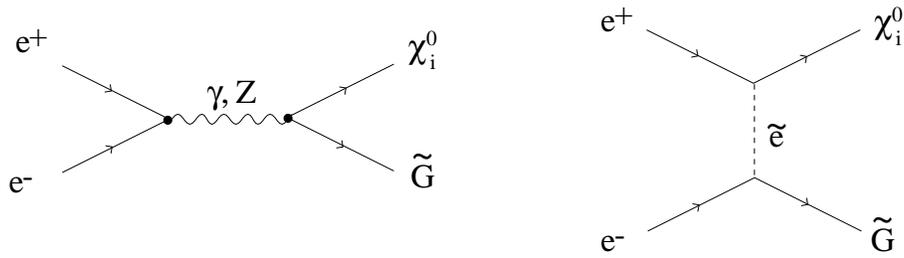,width=0.7\textwidth}
\caption{Single gravitino production.}
\label{fig:fein3}
\end{center}
\end{figure}

In GMSB models  single (processes 37-39) or double gravitino 
(not implemented in SUSYGEN) production becomes accessible  at LEP \cite{nano}. 
The corresponding feynman diagrams are shown in Fig.~\ref{fig:fein3}.

\begin{figure}[htb]
\begin{center}
\epsfig{file=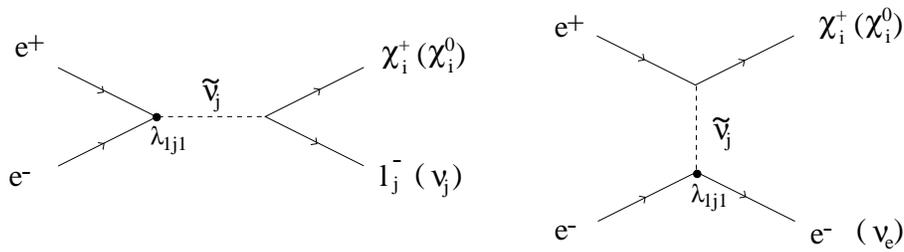,width=0.7\textwidth}
\caption{Sneutrino resonance production.}
\label{fig:fein4}
\end{center}
\end{figure}

In R-parity violating models sneutrinos can be produced singly via the $LL{\bar
  E}$ coupling at LEP,  either in  s-channel resonance
  \cite{resonant},  or in $\gamma e$  collisions \cite{egamma}. 
 At present only the resonant production of sneutrinos (processes 40-45) 
 (with non-zero couplings $\lambda_{121},\lambda_{131}$)
 and their subsequent decays  to
  ${\chi}^{\pm}_i l^{\mp}$ or ${\chi}^{0}_i\snu$ are
  implemented.  The corresponding  feynman diagrams are shown in Fig.~\ref{fig:fein4}.

Finally the 5 Higgs production processes ($hZ,HZ,hA,HA,H^+H^-$) are also
provided through an interface to PYTHIA. Here the RGE improved
 masses and mixings of Section~\ref{higgs.masses}  are used. 
This is the less developed part of the program, and
we suggest that more specialised programs  (e.g. see \cite{altarelli})  are used
for dedicated Higgs studies.

In summary the 50 production processes contained in SUSYGEN  
are the following:

\barr
{ e^+e^- \ra\left\{\begin{array}{l}
\chi_i \chi_j\\
{{\chi}^{\pm}}_i {{\chi}^{\mp}}_j \\
\tilde{f}\bar{\tilde{f}}\\
\chi_i \tilde{G} \\
{{\chi}^{\pm}}_i l^{\mp}\\
{{\chi}^{0}}_i {\nu}\\
hZ,HZ,hA,HA,H^+H^- 
\end{array}
\right.}
\label{eq:process}
\earr

%%%%%%%%%%%%%%%%%%%%%%%%%%%%%%%%%%%%%%%%%%%%%%%%%%%%%%%%%%%%%%%
\clearpage

 \subsection{Sparticle decays}\label{decay}
 Sparticles can decay in two modes: In the \rp conserving mode sparticles
 decay to lighter sparticles and SM particles (Sections \ref{sec:rpc.gaugino.decays},\ref{sec:rpc.sfermion.decays}). In the \rp
 violating mode sparticles can additionally decay to SM particles only (Sections \ref{sec:rpv.gaugino.decays},\ref{sec:rpv.sfermion.decays}).  
 
 \subsubsection{\rp Conserving Gaugino Decay Modes}
 \label{sec:rpc.gaugino.decays}

\begin{figure}[htb]
\begin{center}
\epsfig{file=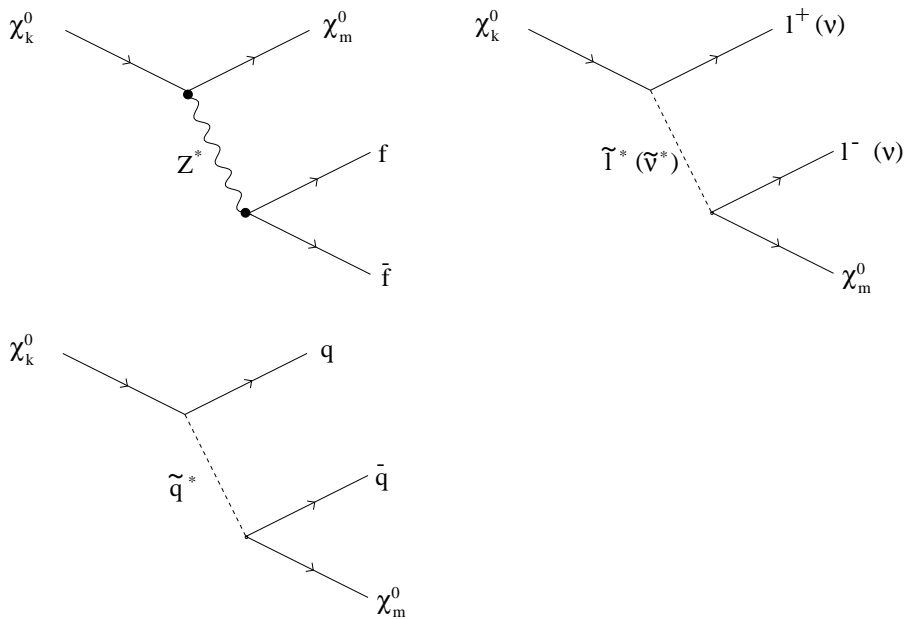,width=.7\textwidth}
\caption{\rp conserving neutralino three body decays.}
\label{fig:new1}
\end{center}
\end{figure}

\begin{figure}[htb]
\begin{center}
\epsfig{file=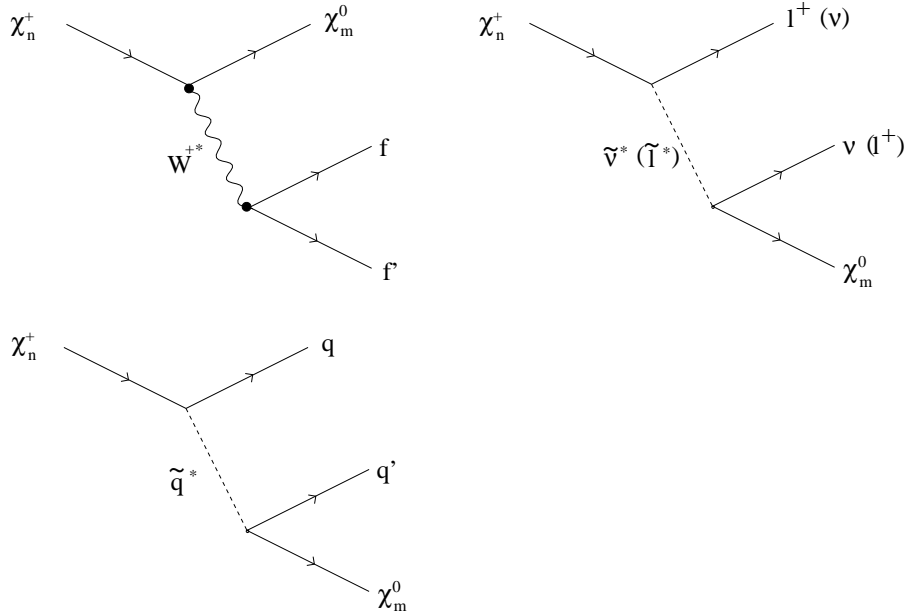,width=.7\textwidth}
\caption{\rp conserving chargino three body decays.}
\label{fig:new2}
\end{center}
\end{figure}

The charginos $\chi^+_n$ and the higher mass neutralinos ${{\tilde
 \chi}^0}_k$ (k=2,3,4) can decay to the three-body final states \cite{bartl}: 
\barr
\begin{array}{l}
{\chi^+_n \ra\left\{\begin{array}{l}
\chi^0_m l^+ \nu, \ 
\chi^0_m q {\bar q}', \\
\chi^+_m l^+ l^- , \
\chi^+_m q {\bar q}, \
\chi^+_m \nu {\bar \nu}
\end{array}
\right.}\\
\\
% \label{eq:rpccharg3decays}
% \earr
% \barr
{\chi^0_k \ra\left\{\begin{array}{l}
\chi^0_m l^+ l^- , \
\chi^0_m q {\bar q} , \
\chi^0_m \nu {\bar \nu} \\
{{\chi}^{\pm}}_m l^\mp {\bar \nu} , \
{{\chi}^\pm}_m q {\bar q}' 
\end{array}
\right.}
\end{array}
\label{eq:rpcgaug3decays}
\earr
Fig.~\ref{fig:new1} and \ref{fig:new2}  
show the feynman diagrams for the decays of \eq{eq:rpcgaug3decays}. 
 One can 
 in general distinguish between two regimes: 
In the first the scalar masses are much larger than the 
gaugino mass, 
and the decays occur through the {\it s}-channel to off-shell
gauge bosons and the lighter gauginos, e.g the 
$W^{*}\chi^0$, $Z^{*}\chi^+$
channels in the case of the chargino decays and the 
$W^{*}{{\chi}^{+}}$, $Z^{*}\chi^0$ in the case of the 
neutralino decays. In this case the different branching 
ratios are mostly determined from the decays of the off-shell $W^*$ and $Z^*$.
In the second regime the sfermions have masses  close to or below the gauginos, 
and the {\it u,t}-channels with ${\tilde l}^*$, ${\tilde \nu}^*$ or ${\tilde q}^*$
exchange  dominate and enhance the branching ratios to the corresponding  fermions. 
Particular care has been taken to take into account the masses of the final
fermions, so that e.g 
scenarios where one has "quasi-degenerate"  chargino and neutralino masses
\cite{perota} give the correct branching ratios and kinematics.

If the sfermion mass is below the gaugino mass, two-body decays of the gauginos
to the sfermions become dominant decay modes. This is implemented in SUSYGEN in
two ways: In the first mode -- which corresponds to the input card ``TWOB
FALSE'', see section~\ref{input} -- no distinction between  
three-body and two-body decays of gauginos is made. Instead the three-body calculation
is used, and the pole at $p_{\tilde f}^2 = M_{\tilde f}^2$ is avoided by the
inclusion of the sfermion width in the propagator\footnote{The
  propagator term $D=({{p_{\tilde f}^2-M_{\tilde f}^2}})^{-1}$ is
 replaced by $D=({{p_{\tilde f}^2-M_{\tilde f}^2} + iG_{\tilde f}})^{-1}$,
 where $G_{\tilde f} = \Gamma_{\tilde f} M_{\tilde f}$.}. Above threshold the
 propagator term forces the two-body kinematics, and the three-body decay is
 therefore  equivalent to two two-body decays. 
 In the second mode -- corresponding to the card ``TWOB TRUE'' --
 the two-body gaugino decays are calculated explicitly. The following two-body decays
 are implemented:
\barr
\begin{array}{l}
{\chi^+_n \ra\left\{\begin{array}{l}
 {\tilde f} f' \\
W^+ \chi^0_m
\end{array}
\right.}\\
\\
\chi^0_n \ra  {\tilde f} f
\end{array}
\label{eq:rpcgaug2decays}
\earr

In addition to the decay modes of Eq.~(\ref{eq:rpcgaug3decays})-(\ref{eq:rpcgaug2decays}) 
the gauginos can either decay to the Higgs \cite{hdecay}, 
 or in restricted regions of parameter space 
 radiatively via one loop diagrams \cite{ambros.gut.evo} 
to ${\chi}^{0}_{1} \gamma$. The radiative decays are particularly
important in  the supersymmetric limit $\tan(\beta) \ra 1$; $M,\mu \ra 0$, where
they are dominant.
\barr
\begin{array}{l}
{\chi^0_k \ra\left\{\begin{array}{l}
\chi^0_m H^0, \chi^0_m h, \chi^0_m A \\
\chi^0_m \gamma 
\end{array}
\right.}\\
\\
{\chi^+_n \ra \chi^0_m H^+}
\end{array}
\label{eq:rpcgaugraddecays}
\earr

In GMSB models the gravitino is the LSP, and
the gauginos can decay to~\cite{grav.ref}
\barr
\begin{array}{l}
{\chi^0_n \ra\left\{\begin{array}{l}
{{\tilde G}} \gamma, 
{{\tilde G}} Z, 
{{\tilde G}} H^0, {{\tilde G}} h, {{\tilde G}} A
\end{array}
\right.}\\
\\
{\chi^+_m \ra\left\{\begin{array}{l}
{{\tilde G}} W, 
{{\tilde G}} H^+
\end{array}
\right.}
\end{array}
\label{eq:rpcgauggmsbdecays}
\earr
The coupling of the gravitino  can be 
very small, and thus the decay ${\chi}^0_1 \ra {\tilde G} \gamma$ is
phenomenologically  the most interesting mode if it occurs inside the detector. 
 An isotropic two body decay has been assumed.

\subsubsection{\rp Conserving Sfermion Decay Modes}
\label{sec:rpc.sfermion.decays}

The sfermion decays to  a gaugino and a fermion 
 can also be separated into two cases. In the first case
the lightest neutralino (${\chi}_1^0$) is the only gaugino lighter than 
the  sfermion, and the typical decay mode is  
\beq
\tilde{f} \ra f {\chi}_1^0 
\eeq
with the
exception of the stop, where the decay $\tilde{t} \ra t {\chi}_1^0 $ is
kinematically inaccessible at LEP and therefore proceeds through the Cabbibo suppressed 
mode  $\tilde{t} \ra {\mathrm c} {\chi}_1^0$. In the second case heavier  
neutralinos $\chi^0_k$ $(k=2,3,4)$ or charginos $\chi^+_n$ 
may be lighter than the sfermion,  and the sfermion decays through a cascade of
gauginos:
\barr
\begin{array}{l}
{\tilde{l} (\tilde{\nu}) \ra\left\{\begin{array}{l}
l (\nu) \chi^0_k\\
\nu (l) \chi^+_n
\end{array}
\right.}\\
\\
{\tilde{q} \ra\left\{\begin{array}{l}
q \chi^0_k\\
q' \chi^+_n
\end{array}
\right.}
\end{array}
\label{eq:rpcsfermion2decays}
\earr

In GMSB models decays to gravitinos are also possible \cite{grav.ref}, although in 
practice only
the decay of the NLSP to the gravitino is of importance. The implemented decays
are:
\beq 
{\tilde f} \ra {\tilde G} f
\label{eq:rpcsfermiongravdecays}
\eeq

The above decays cover neutralino and gravitino LSP scenarios. If the
sneutrino is the LSP (or other sfermions are the LSP in \rp violating
models), sleptons and squarks can also decay via virtual $W^*, Z^*,
\chi^+*$ and $\chi^0*$ exchange to the sneutrino. At present
these decay modes are not  implemented in SUSYGEN.

\subsubsection{\rp violating Gaugino Decay Modes}
\label{sec:rpv.gaugino.decays}
\begin{figure}[htb]
\begin{center}
\epsfig{file=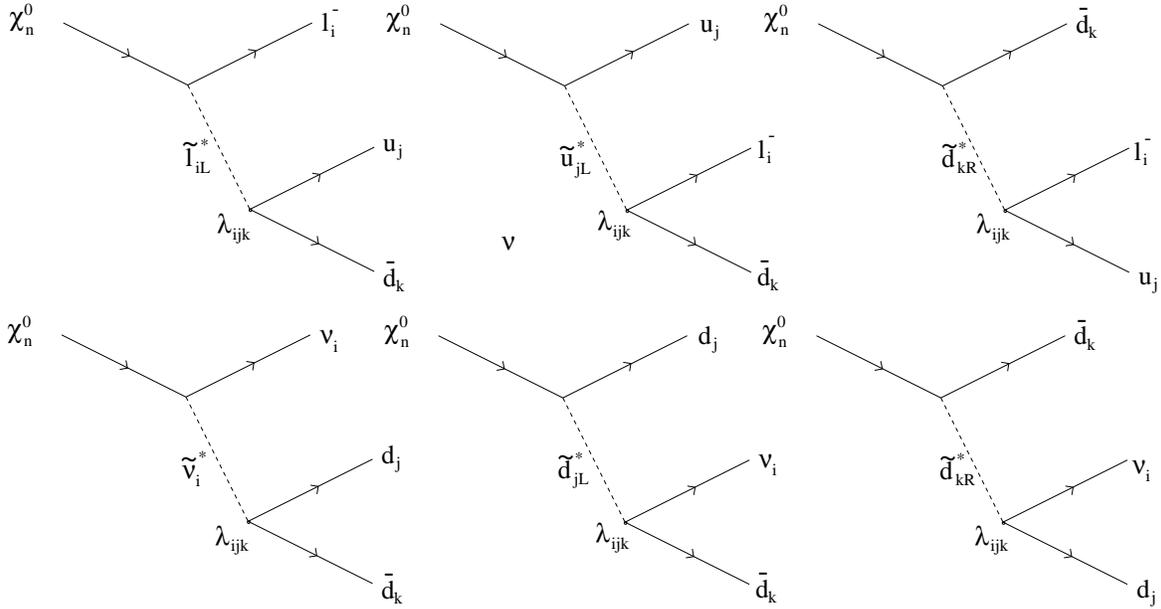,width=.9\textwidth}
\caption{R-parity violating neutralino decays for a dominant $LQ \bar D$ operator.}
\label{feyn3}
\end{center}
\end{figure}

SUSYGEN implements decays for a single non-zero R-parity violating coupling
$\lambda_{ijk}, \lambda'_{ijk}$ or $\lambda''_{ijk}$, where $i,j,k$ are
generation indices, and the three couplings correspond to the $LL\bar E$,
$LQ{\bar D}$ or ${\bar U}{\bar D}{\bar D}$ operators in \eq{eqrpv}. Neutralinos
can  decay to \cite{dreiner}:
\barr
\chi^0_n (LL\bar E) &\ra&\left\{\begin{array}{l}
l^-_i {\bar \nu}_j l^+_k ,
l^+_i {\nu}_j l^-_k , 
{\bar \nu}_i l^-_j l^+_k ,
{\nu}_i l^+_j l^-_k 
\end{array}
\right.
\label{eq:rpvneutlle3decays}\\
\chi^0_n (LQ{\bar D}) &\ra&\left\{\begin{array}{l}
l^-_i u_j {\bar d}_k ,
l^+_i {\bar u}_j {d}_k ,
{\nu}_i d_j {\bar d}_k , 
{\bar \nu}_i {\bar d}_j {d}_k
\end{array}
\right.
\label{eq:rpvneutlqd3decays}\\
\chi^0_n ({\bar U}{\bar D}{\bar D}) &\ra&\left\{\begin{array}{l}
u_i d_j d_k , 
{\bar u}_i {\bar d}_j {\bar d}_k 
\end{array}
\right.
\label{eq:rpvneutudd3decays}
\earr
 and Fig.~\ref{feyn3} show the corresponding  diagrams (for example) for the $LQ{\bar D}$
operator. Charginos can decay to~\cite{dreiner}:
\barr
\chi^+_n (LL \bar E) &\ra&\left\{\begin{array}{l}
\nu_i \nu_j e^+_k , 
e^+_i e^+_j e^-_k ,
e^+_i \nu_j \nu_k ,
\nu_i e^+_j \nu_k 
\end{array}
\right.
\label{eq:rpvcharglle3decays}\\
\chi^+_n (LQ{\bar D}) &\ra&\left\{\begin{array}{l}
\nu_i u_j {\bar d}_k , 
l^+_i {\bar d}_j d_k ,
l^+_i {\bar u}_j u_k ,
{\bar \nu}_i {\bar d}_j u_k
\end{array}
\right.
\label{eq:rpvcharglqd3decays}\\
\chi^+_n ({\bar U}{\bar D}{\bar D}) &\ra&\left\{\begin{array}{l}
{\bar d}_i {\bar d}_j {\bar d}_k ,
u_i u_j d_k ,
u_i d_j u_k
\end{array}
\right.
\label{eq:rpvchargudd3decays}
\earr
The above decays are particularly relevant when the chargino is the LSP.
 Charginos will normally decay to the
 neutralino via {\eq{eq:rpcgaug3decays}} if the neutralino is the LSP,
 in which case the decays of
 Eq.~({\ref{eq:rpvcharglle3decays}})-({\ref{eq:rpvchargudd3decays}}) 
 only dominate for large couplings $\lambda$ and when the exchanged sfermion mass is
 close to the mass of the chargino. 

\subsubsection{\rp violating Sfermion Decay Modes}
\label{sec:rpv.sfermion.decays}

Sfermions can decay directly to SM particles through a dominant $LL \bar E$ coupling via
\barr
{\tilde l^-}_{iL} &\ra& \nu_j l^-_k \nonumber \\ 
{\tilde l^-}_{jL} &\ra& \nu_i l^-_k \nonumber \\
{\tilde l^-}_{kR} &\ra& \nu_i l^-_j , \nu_j l^-_i \\
{\tilde \nu}_i &\ra& l^-_j l^+_k \nonumber \\ 
{\tilde \nu}_j &\ra& l^-_i l^+_k \nonumber
\earr
where ${\tilde l}_{iL}$ denotes a left-handed slepton of generation {\it i}. For
a dominant $LQ{\bar D}$ operator sfermions decay via 
\barr
{\tilde l^-}_{iL} & \ra & {\bar u}_j d_k \nonumber\\
{\tilde \nu}_{iL} & \ra & d_j {\bar d}_k \nonumber\\
{\tilde u}_{jL} & \ra & l^+_i d_k \\
{\tilde d}_{jL} & \ra & l^-_i u_k \nonumber\\
{\tilde d}_{kR} & \ra & l^-_i u_j \nonumber\\
{\tilde d}_{kR} & \ra & {\bar \nu}_i d_j \nonumber
\earr
and similarly for a dominant ${\bar U}{\bar D}{\bar D}$ operator 
\barr
{\tilde u}_{iR} & \ra & {\bar d}_j {\bar d}_k \nonumber\\
{\tilde d}_{jR} & \ra & {\bar u}_i {\bar d}_k \\
{\tilde d}_{kR} & \ra & {\bar u}_i {\bar d}_j .\nonumber
\earr
Note that SUSYGEN takes into account mixing between the left-right third
generation sfermion states. As already mentioned in Section
\ref{sec:rpc.sfermion.decays}, three-body decays
of sfermions to lighter sfermions  are not included in SUSYGEN. 
 These decays  are only of importance if the produced
 sfermion is the NLSP which cannot decay directly via a specific \rp violating
 coupling, and the LSP is another sfermion.

\subsection{Initial and Final State radiative corrections}\label{isr}

The initial state corrections are implemented using a factorised "radiator
formula" (REMT by Kleiss) where exponentiation of higher orders and a $P_T$ dependent
distribution have been implemented \cite{remt1}. They have been checked
against other initial state codes and found to agree at the percent 
level, once the prescriptions of the scales e.t.c were made to agree.

Final state radiative corrections are implemented within SUSYGEN in two ways: 
Firstly QED and
QCD corrections to quark final states are treated by JETSET (through a LUSHOW
call) in the hadronisation interface (see also section below). QED
Bremsstrahlung off leptons is implemented using the PHOTOS package
\cite{photos}. This allows an estimate of Bremsstrahlung effects in the
leading-logarithmic approximation up to O($\alpha^2$), 
including double emission
of photons.

\subsection{Lifetime}\label{life}
SUSYGEN calculates the lifetime of the sparticles from their decay rate, and
implements secondary vertices for long-lived sparticles accordingly. The
lifetime information is important for the \rp violating LSP decays when the
coupling strength $\lambda$ is small, and for the GMSB decays to the gravitino.
 The inclusion of lifetime effects can be optionally turned off.
 
\subsection{Hadronisation}\label{hadro}
SUSYGEN is fully interfaced to JETSET 7.4 \cite{jetset} which takes care of the
hadronisation of quarks. Here we describe  the details of the
interface and the hadronisation models used by SUSYGEN. 

Two fragmentation schemes can be chosen by the card {\it FRAG} 
(see also section~\ref{input}).  
In the independent fragmentation (IF) scheme the final state quarks are treated as independent
particles with respect to each other. There is no QCD radiation between
 the final state quarks, and the hadronisation process produces relatively
hard jets. Clearly this is an unphysical simplification, and the IF scheme
should only be used to compare and evaluate the effect of gluon radiation.
In the second scheme (the default) 
the final state quarks are evolved according to the Lund
Parton Shower model. A colour string is formed between colour singlet $q {\bar
q}$ states, and the two quarks are subsequently evolved in time/space, emitting
gluon radiation between the two quarks. The jets are
much softer owing to the gluon radiation between the quark states,
which produces additional soft particles. 

In the simplest case when there are only two final state quarks (for example in
the process $e^+ e^- \ra {\tilde b} {\tilde {\bar b}} \ra {\chi}^0 b
{\chi}^0 {\bar
b}$) there is no ambiguity in the colour string assignment. In the case when
quarks are produced by colourless sparticles a colour string is formed between
each of the decay products of the  colourless states. For example in the process $e^+
e^- \ra {\chi}^+ {\chi}^- \ra ({\chi}^0  q_1 q_2) ({\tilde
\chi}^0 q_3 q_4)$ a colour string is formed between the quarks $q_1 q_2$ and
another one between the quarks $q_3 q_4$. 

A further complication arises when the \rp violating couplings
$\lambda''$ produce triple quark final states (for example in the decay ${\tilde
\chi}^0 \ra q_1 q_2 q_3$). The triple quark vertices are not implemented in the
higher level user interface of JETSET, but are supported by the lower level JETSET
routines. Here we use a prescription kindly provided by Torbjoern
Sjoestrand \cite{torbjoern.priv} to implement this type of vertex:
The two quarks $q_1 q_2$ are connected by a double T-junction to a
diquark of flavour($q_1$,$q_2$) with zero momentum and the third quark $q_3$. JETSET
subsequently treats the parton shower evolution of the three quarks correctly,
but produces the error messages
\begin{verbatim}
     Advisory warning type 2 given after     1 LUEXEC calls:
     (LUPREP:) unphysical flavour combination
\end{verbatim}
which should be ignored.

At present the JETSET hadronisation interface described above has two
limitations: 
Firstly if squarks decay via the \rp violating coupling $\lambda''$ to two quarks (${\tilde
q}_1 \ra q_2 q_3$) the colour string connection between $q_2 q_3$ is not
supported by JETSET, and we have to resort to the independent fragmentation scheme
instead. Secondly, because the lifetime of the decay ${\tilde t} \ra {\chi}^0 c$
is larger than the hadronisation scale, stops should hadronise (to stop
mesons) prior to their decay. Stop meson hadronisation is not implemented
in SUSYGEN. 

\newpage
\section{Program description}\label{prog.des}
 The SUSYGEN program may be used in two different ways. Firstly as a standard MC
 program to generate MC events for one particular set of input SUSY parameters. 
 And secondly to scan  SUSY  parameter space in $M_2, \mu, \tan\beta$ and $m_0$,
 where SUSYGEN calculates the masses, cross sections and decay branching ratios
 for each point, and also optionally generates weighted signal events for each
 of the specified SUSY processes. In the ``SCAN'' mode the calculated SUSY
 parameters can also be  written to an ntuple, which can be subsequently
 analysed by the user.

 The user has the option to directly access the LUND common block within the
 program in the routine USER. This is particularly useful for feasibility
 studies in the SCAN mode, 
 where one might for example estimate detection efficiencies by
 applying a simple set of cuts inside the above routine. The calculated
 efficiencies are stored in the ntuple, one efficiency value for each of
 the generated SUSY signals. An example USER routine which selects hadronic
 events with large missing $E_T$ is provided with the release version of SUSYGEN.

 In the following the structure of the program and the main subroutines are
 described. A description of the common blocks is given in
 Appendix~\ref{commons}. Inputs and Outputs of the program are described in sections~\ref{input},\ref{output}.
 The program is divided into four stages. The first stage performs  general initialisations:
\begin{itemize}
\item reads the 
input cards in subroutine SCARDS 
using the CERN package FFREAD. The user  can define his/her own cards in subroutine
USER\_SCARDS.
\item then books  histograms in routine SBOOK and calls
NTUPLE\_INIT to define the SUSYGEN ntuple. 
\item initialises the sparticle names in SUPART and calls the routine
USER\_INIT for any other user-dependent initialisation.
\end{itemize}

The second stage  uses the input information to calculate the 
masses, mixings, branching ratios and cross sections\footnote{SUSYGEN
can operate in  SCAN mode, where a range of input values is scanned
(see input cards section). In this mode SUSYGEN loops over the 
input parameter range and recalculates the second stage parameters for 
each parameter space point.}. The routine SUSINI calls further subroutines:
\begin{itemize}
\item SUSANA for the calculation of the masses/mixings of the sparticles.
The routine SUSANA calls the routines:

\begin{itemize}
\item  SFERMION, which determines  the sfermion
masses either using the routine 
SMGUT, which computes the masses of the sfermions in the SUGRA model or the GMSB
model, or simply by taking each sfermion
mass from the input cards. The user can provide the mixings parameters of the
third generation sfermions through input cards,  or alternatively the sfermion mixing is
calculated in this routine according to Eq.~(\ref{stopmatrix},\ref{thetamix}).
\item  GAUGINO, which  computes the gaugino masses and mixing through the
diagonalisation of the corresponding matrices $M^0, M^c$.
\item  SUBH, which  computes
the Higgs masses; and PINTERF, which  passes the Higgs parameters to 
PYTHIA.
\item LEPLIM, which examines whether the specific point has already
been excluded by LEP~I searches. The imposition of this constraint
is selected through input cards and should be developed by the
user.
\item RPARINI, which initialises the calculations of the \rpv diagrams and
decay branching ratios.
\end{itemize}

\item BRANCH for the determination of the  widths and 
branching ratios of the sfermion and gaugino decays.
\begin{itemize}
\item First the $R_p$-conserving two body decay widths 
of sfermions and gauginos are  calculated.
\item Routine WICONST calculates the 
 3-body decay constants and the function WSC calculates the partial  
decay rates.
\item Routines NTONPH and NTOXH calculate the gaugino decays to 
 a photon or a Higgs.
\item In 
routine INTERF the widths of the 3 previous steps  are stored in the 
common block SSMODE. The nomenclature of ISAJET 7.03 \cite{isajet} 
has been retained for comparison purposes.
Then the gauge mediated and \rpv decays 
are calculated and stored in the same common block
 in the routines MGMDECAY and RPARDECAY, respectively.
 Finally the total sparticle widths and branching ratios 
 are determined by summing up the partial decay rates in  routine SSNORM.
\end{itemize}

\item The total cross sections are calculated in the routines CHARGI-HIGINIT
  (see table~\ref{sigmas}). The initial state radiation corrections are calculated in
  routine REMT1, and routine PROCINIT stores the cross sections and other 
  information for each of the 50 processes in the common blocks PROC1, PROC2.
\end{itemize}

\begin{table}[hbt]
\begin{center}
\begin{tabular}{l|l|l}     \hline
Production  & Total cross section & Differential cross section ($d\sigma/dt$) \\
mechanism   & Routine name & Routine name \\
\hline
$\chi^{+}_i\chi^{-}_j$ &  CHARGI  & GENCHAR \\
$\chi^0_i\chi^0_j $   &  PHOTI   & GENPHO\\
$\sel\sel$            &   -  & GENSEL \\
$\sell\selr$          &   -  & GENSELRS \\
$\smu\smu, \stau\stau, \squ\squ $ & - & GENSMUS \\
$\snu_e\snu_e$        &   - & GENSNUE \\
$\snu\snu$            &  - & GENSNU \\
$\tilde{G}\chi^0_i$   & - &   GENPHOG \\
$\nu\chi^0_i$         & - &  RPVZINT \\
$l^{\mp}\chi^{\pm}_i$ & - & RPVWINT \\
$hZ, HZ, H^+H^-, $  &  HIGINIT  &  PYTHIA \\
$HA, hA$           &   HIGINIT + SIGHA  &  PYTHIA \\
\end{tabular}
\caption{Subroutine names of the total and differential cross-sections.}
\label{sigmas}
\end{center}\end{table}

In the third stage ``weighted'' MC events are generated, i.e. one of the
selected  production processes is chosen with a probability proportional to its
relative cross section. The routine SUSEVE then loads the corresponding
variables from the commons PROC1,PROC2 and performs the following tasks:
\begin{itemize}
\item Picks  a value for $\cos\theta$ distributed according to the differential 
cross section ($d\sigma/dt$) for the process (see also table \ref{sigmas}) and
creates the  4-vectors for the final states of the hard process. 
\item The decay channel is determined in the routine DECABR from the branching
  ratios of the sparticles, and the two-body or three-body decay 4-vectors are
  calculated in the  routines SMBOD2 and SMBOD3, respectively. 
 In SMBOD3 the differential decay rate $d\Gamma/dsdt$  is sampled in $s$ and
  $t$, and the angular distributions of the decay products determined accordingly.
\item The decay products from the previous step are presented again to DECABR
  and decayed, until no more unstable SUSY particles are left.
\item The routines REMT3 and FINRAD\_PHOTOS perform the adjustments due to
 initial and final state radiation.
\item The above 4-vectors are interfaced to LUND in the subroutine 
SFRAGMENT where they fragment, hadronise and decay. 
\end{itemize}

While still in the third stage, inside the event loop, the routines
USER\_WRITE and USER are called where the user has the options to 
write an event out and/or analyse it directly  by accessing the LUND common
LUJETS.

At the end of the event loop -- while still inside the SCAN loop --
 the  calculated parameters (masses, BRs, 
cross-sections) and the rough detection efficiencies calculated in 
the routine USER are stored in an ntuple by the routine NTUPLE\_FILL.

In the fourth stage the routines SUSEND and USER\_END perform general and
user-specific closing functions.

\newpage
\section{Setting up and running SUSYGEN}\label{runit}
The program is supplied as a single Fortran file which can be compiled using
standard FORTRAN77. The release version may be obtained from the 
CPC Program Library, and the  most recent version can be obtained from 
% \newline
% \begin{verbatim}
http://lyohp5.in2p3.fr/delphi/katsan/susygen.html.
% \end{verbatim}
The program has to be linked to the standard CERNLIBs,
the JETSET7.4 library, PYTHIA and PHOTOS. 
This may be done by the following set of commands on UNIX:
\begin{verbatim}
f77 -g -w -static -c susygen.f
f77 -o susygen.exe  susygen.o /cern/pro/lib/libphotos.a \
/cern/pro/lib/libjetset74.a `cernlib mathlib packlib shift`
\end{verbatim}
and on VMS:
\begin{verbatim}
$ for/optimize/nodebug  susygen
$ cernlib jetset74,photos,packlib,genlib,kernlib
$ link/nodeb susygen,'lib$'
\end{verbatim}

To run the program one has to give appropriate input cards, 
which in the example below are assumed to be in the file 
 susygen.cards. To run SUSYGEN on UNIX:
\begin{verbatim}
susygen.exe < susygen.cards > susygen.log
\end{verbatim}
and on VMS:
\begin{verbatim}
$ define/user sys$input susygen.cards 
$ define/user sys$output susygen.log
$ run/nodeb susygen
\end{verbatim}
The Inputs and Outputs are discussed in the following.

\subsection{Input}\label{input}
\begin{figure}[hbt]
\begin{center}
\epsfig{figure=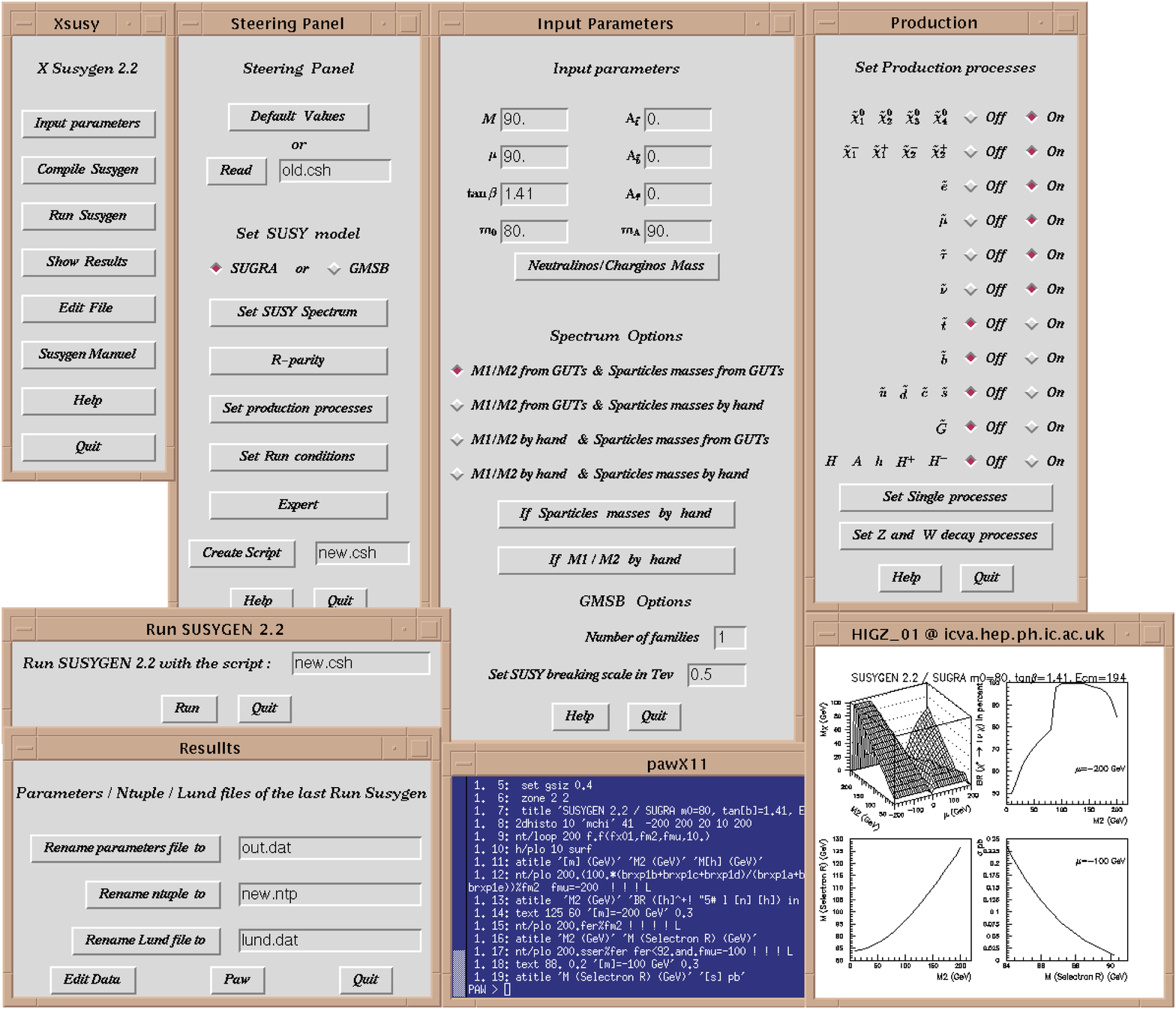,width=\textwidth}
\end{center}
\caption{The graphical user interface   {\tt Xsusy}.}\label{xsusy}
\end{figure}

The input to SUSYGEN consists of a single text file, the cards file, described
below.  Alternatively a  very advanced graphics X-interface, 
 {\tt Xsusy}, is also available for UNIX machines. The {\tt Xsusy} interface
allows input from pop-up menus rather than a cards file, and is extremely
user friendly. The SUSYGEN output may be analysed with {\tt PAW} and/or {\tt emacs} within
{\tt Xsusy}. Fig.~\ref{xsusy} shows a typical {\tt Xsusy}-session. 
{\tt Xsusy} is not part of the standard CPC distribution of SUSYGEN, but may be
obtained from 
% \newline
% \begin{verbatim}
http://lyohp5.in2p3.fr/delphi/katsan/susygen.html. 
% \end{verbatim}

In the absence of any of the cards  the default values given below are chosen.

\subsubsection{Set masses and mixings}
\begin{description}
\item[MODES 1] \hspace{.2in} MODES determines which input parameters will be
used to determine the supersymmetric masses and mixings:
\begin{itemize}
\item MODES=1 the gaugino/higgsino and sfermion spectrum is calculated
using the values M, $\mu$, $m_0$, $\tan\beta$, and A as input for the 
case of SUGRA models, and M, $\mu$, $\tan\beta$ and A for GMSB models.
\item MODES=2 the gaugino/higgsino spectrum is calculated
using the values M, $\mu$, $m_0$ and $\tan\beta$ as input, while the
the sfermion masses are taken from the input cards below.
\item MODES=3 the gaugino/higgsino spectrum is calculated
using the values M, $\mu$, $m_0$ and $\tan\beta$, but the "gaugino mass
relationship" does not relate $M_1$ to $M_2$ anymore, 
their relationship is set by the card RS below. 
The sfermion values are given according to the hierarchy proposed for
MODES=1.
\item MODES=4 the gaugino/higgsino spectrum is calculated
using the values M, $\mu$, $m_0$ and $\tan\beta$, but the "gaugino mass
relationship" does not relate $M_1$ to $M_2$ anymore, 
their relationship is set by the card RS below, while 
the sfermion values are taken from the input cards below.
\item MODES=5 TO BE USED WITH CARE. 
M below is the requested mass of the lightest neutralino. 
A corresponding M$_2$ may not exist for the given set of 
parameters, or very often double solutions are possible (e.g for positive $\mu$
and low $\tan\beta$).
\end{itemize}
\item[M]  90. \hspace{.2in} (M$_2$ the SU(2) gaugino mass in GeV,at the EW scale)
\item[mu] 90. \hspace{.2in} ($\mu$ in GeV, at the EW scale)
\item[m0] 90. \hspace{.2in} ($m_0$ in GeV, a common scalar mass at the GUT 
scale)
\item[tanb] 4.0 \hspace{.2in} ($\tan\beta=\upsilon_2/\upsilon_1$, at the weak 
scale)
\item[A] 0. 0. 0.     \hspace{.2in} (A$_t$, A$_b$, $A_{\tau}$ soft SUSY 
breaking trilinear couplings)
\item[RS] 1.    \hspace{.2in} RS is the gauge mass unification breaking scale
defined as $M_1 = M_2 \frac{5 \sin^2{\theta_W}}{3 \cos^2{\theta_W}} RS$)
\item[MA]  500. \hspace{.2in} (The mass of the A Higgs in GeV)
\item[MIX]    0 \hspace{.2in} If set to 1, permits sfermion mixing.
\item[C ----- ARE RELEVANT FOR MODES 2 and 4 ]
\item[MSQUARK] 1000. \hspace{.2in} (the mass of $\squ,m_{\squ}$)
\item[MLSTOP] 1000.  \hspace{.2in} (the mass of $\stp_L,m_{\stp_L}$)
\item[MRSTOP] 1000.  \hspace{.2in} (the mass of $\stp_R,m_{\stp_R}$)
\item[MLSEL]  1000.  \hspace{.2in} (the mass of $\sell,m_{\sell}$) 
\item[MRSEL]  1000.  \hspace{.2in} (the mass of $\selr,m_{\selr}$)
\item[MSNU]   1000.  \hspace{.2in} (the mass of $\snu,m_{\snu}$)
\item[MASMX] 0. 0. 0. 0. 0. 0. 
\hspace{.2in} When different from 0. sets
 the mass of $\tilde{t}_1, \tilde{b}_1, \tilde{\tau}_1$ 
$\tilde{t}_2$, $\tilde{b}_2$, $\tilde{\tau}_2$  respectively. 
\item[PHIMX] 0. 0. 0. \hspace{.2in} sets mixing 
 angle between $\tilde{t}_1$-$\tilde{t}_2$,
$\tilde{b}_1$-$\tilde{b}_2$ and $\tilde{\tau}_1$-$\tilde{\tau}_2$.
\end{description}

\subsubsection{Set production processes}
\begin{description}
\item[ZINO]          FALSE \hspace{.2in} (all pairs of $\chi^0$)
\item[WINO]          FALSE \hspace{.2in} (all pairs of $\chi^{\pm}$)
\item[SELECTRON]     FALSE \hspace{.2in} (all pairs of $\sel$)
\item[SMUON]         FALSE \hspace{.2in} (all pairs of $\smu$)
\item[STAU]          FALSE \hspace{.2in} (all pairs of $\stau$)
\item[SNU]           FALSE \hspace{.2in} (all pairs of $\snu$)
\item[SSTOP]         FALSE \hspace{.2in} (all pairs of $\stp$)
\item[SBOTTOM]       FALSE \hspace{.2in} (all pairs of $\sbt$)
\item[SQUARK]        FALSE \hspace{.2in} (all pairs of $\squ$ other than $\stp,\sbt$)
\item[HIGGS]         FALSE \hspace{.2in} (all higgses a la PYTHIA)
\item[GRAVITINO]     FALSE \hspace{.2in} ($\tilde{G}\chi^0$, when MGM is true)
\item[RSINGLE]       FALSE \hspace{.2in} ($\nu\chi^0$ and $\mu\chi^{\pm}$
or $\tau\chi^{\pm}$ through $\tilde{\nu}$ resonance when RPARITY is FALSE) 
\item[PROCSEL] 0 0 0 0 0 0 0 0 0 0 \\
Possibility to choose to produce any 10 processes.
The first word is the number of processes to be selected and the 
next ten give the process number (see process table in the  text). 
The corresponding generic card is turned TRUE. 
That is e.g ZINO is set to TRUE when 
one sets PROCSEL 1 2, asking to produce only the process $\chi_1^0\chi_2^0$.
\item[DECSEL] 1 1 1 1 1 1 1 1 1 1 1 1 1 1 1 1 1 1 1 1 1 1 1\\
Determines the modes of decay
of virtual Z and W. The first 4 determine whether the virtual Z$^*$ can decay to
$u\bar{u}$, $d\bar{d}$, $\nu\bar{\nu}$, $e^+e^-$ and the next two the decays
of W$^*$ to $u\bar{d}$ or $e^+\nu$. The next 12 determine the decays to the
two other families. The last 5 determine whether one wants to turn on/off the
decays to $\chi_i\gamma$, $\chi_i h$,$\chi_i H$, $\chi_i A$, 
$\chi^{\pm}_i H^{\mp}$. 
\end{description}

\subsubsection{Gauge mediated SUSY models}
\begin{description}
\item[MGM] FALSE \hspace{.2in} Switches on Production ($\chi {\tilde G}$)
of gravitinos and the decay modes to the gravitino in models of gauge mediated
supersymmetry breaking.
\item[LSUSY] 500. \hspace{.2in} Supersymmetry Breaking Scale 
$\sqrt{F}$ (in GeV) (gauge mediated SUSY models)
\end{description}

\subsubsection{\rp violating SUSY models}
\begin{description}
\item[RPARITY] TRUE \hspace{.2in} Is \rp conserved? 
\item[INDIC] o i j k \hspace{.2in} \rp violating coupling indices of the
single dominant coupling $\lambda_{ijk}$. The operators o=1,2,3 correspond to
the \rp violating  operators $L L {\bar E}$, $L Q {\bar D}$, ${\bar U} {\bar U} {\bar
D}$ respectively.
\item[LAMDA] 0.1 \hspace{.2in} Strength of the \rp violating coupling
$\lambda_{ijk}$.
\item[TWOB] FALSE \hspace{.2in} Switch on/off explicit calculation of the
gaugino two-body decays. The TWOB is automatically set to true if \rp is violated.
\end{description}

\subsubsection{Generation conditions}
\begin{description}
\item[ECM] 183. \hspace{.2in} (Center of mass energy).
\item[SUSEVENTS] 100.\hspace{.2in} (Total number of events to be produced. If
more than one processes are possible, they will be produced in percentages
corresponding to their respective cross-sections).
\item[ISR] 1  \hspace{.2in}  (Permit or not (0) initial state radiation).
\item[FSR] 1  \hspace{.2in}  (Permit or not (0) final state radiation).
\item[GENER] 0 \hspace{.2in} (Generate events(1) or not (0)).
\item[LIFE] TRUE \hspace{.2in} 
(Include effects of finite lifetime of sparticles).
\item[FRAG] 1 \hspace{.2in} (Fragmentation scheme. Normal string type
fragmentation (1) or independent fragmentation (2)).
\item[LCUT] 200. \hspace{.2in} (Decay length cut in cm, 
thereafter particles considered stable).
\end{description}

\subsubsection{Control switches}
\begin{description}
\item[DEBUG] 0 \hspace{.2in} (Print additional debug information. The higher the 
value, the more information is printed ($0 < $Debug$ < 31$)).
\item[LUWRIT] FALSE \hspace{.2in} (Write out events to FOR012.dat or not) .
\item[SCAN]   FALSE  \hspace{.2in} (Scan the values in M and $\mu$).
\item[ISCAN]  0 0 0 0  \hspace{.2in} (Number of scan values
in M$_2$, $\mu$, $m_0$ and $\tan\beta$). There is also the following option:
setting the relevant ISCAN value to -1, the program takes the first
corresponding value VSCx as a starting value, the second as ending value 
and the third as the step.
\item[VSC1] 0 \hspace{.2in}   (List of scan values in M$_2$, up to 10, 
or start, end, step values)
\item[VSC2] 0 \hspace{.2in}   (List of scan values in $\mu$, up to 10,
or start, end, step values)
\item[VSC3] 0 \hspace{.2in}   (List of scan values in $m_0$, up to 10,
or start, end, step values)
\item[VSC4] 0 \hspace{.2in}   (List of scan values in $\tan\beta$, up to 10,
or start, end, step values)
\item[LEPI]    FALSE  \hspace{.2in}  (Impose LEPI constraints or not).
\end{description}

\subsection{Output}\label{output}
Four output files are produced by SUSYGEN: 
\begin{itemize}
\item The file susygen.dat, which lists the sparticle masses, 
cross sections, branching ratios and  decay modes.
\item The file susygen.his. The ntuple with ID 200 
 in this file contains the calculated masses, cross
  sections and BRs.
\item The fortran file UNIT 12, e.g. ``fort.12'' on UNIX or ``FOR012.DAT'' on
  VMS, which contains the LUND common of the  generated events. See routine
  SXWRLU in the program for more information on the output format of this file.
\item The file susygen.log, which contains warning and error messages and a
  LUND listing of the generated  events (if the debug level is set to ``DEBUG n'',
  where $n>0$).
\end{itemize}

Ntuple 200 is a column-wise Ntuple, and contains one entry per SCAN
loop (SUSYGEN can scan parameter space in $\mu,M_2,\tan\beta,m_0$). The ntuple
variables are summarised in Tables~\ref{ntp1}-\ref{ntp12} in the Appendix.

\clearpage
\section{Conclusions and Future plans}\label{conclusions}

SUSYGEN 2.2 is a fast and versatile Monte-Carlo generator that has 
profited from the collective experience of the four LEP collaborations. Its 
current precision is sufficient for detailed studies at LEP. It permits
the study of gravity and gauge mediated models of SUSY breaking in models in
which R-parity is either conserved or violated.
 A total of  50 production processes are included, which are the 
 most important processes in the models studied above. SUSYGEN is 
 extremely user friendly and the code is transparent. It is maintained
 in CMZ, and  private developments can easily be  incorporated in the
 main body of the generator. 

There are a few things the authors would like to incorporate in the 
next version:
\begin{itemize}
\item Extend SUSYGEN to  hadron and $ep$ colliders.
\item Treatment of  helicity correlations. This development is 
well under way and should be implemented in a few months.
\item Include a light gluino scenario.
\item  Include single squark/sneutrino production 
 in \rpv processes where a radiated photon from one of the incoming
leptons interacts with the opposite lepton through a "resolved"
 quark or lepton component.
\item Include  three body decays of sfermions to lighter sfermions.
\item First order corrections to SUSY masses and cross-sections. 
 This development could be important in particularly difficult areas of
 the parameter space (e.g. $\tan\beta=1$ see \cite{king}), where 
 corrections of the oder  20\% are possible. 
\end{itemize}

Other more long-range plans include:

\begin{itemize}
\item A more precise Higgs treatment, which also  includes Higgs decays to SUSY particles.
\item Incorporation of  cosmological constraints  and other constraining measurements
 (e.g. $b\rightarrow s\gamma$ constraints) at an informative level.
\end{itemize}

\subsection*{Acknowledgments}
 We would  first of all like to thank S. Ambrosanio, M. Carena, H. Dreiner, 
 J.R. Espinosa, M. Lola, M. Quiros  and C. Wagner for their help, 
 ideas and the many lines of written code incorporated in SUSYGEN 2.2.
Furthermore we would like to thank R. Barbier for the {\tt Xsusy} interface, 
S. Melachroinos for providing the routines integrating the 
gaugino decay widths along one of the variables, 
S. Paiano and A. Perotta for providing the changes for the same gaugino
decay widths taking into account the masses of the product fermions,
R. Kleiss for the ISR code, 
T. Sjoestrand for his help with the JETSET interface and
D. Zerwas for his help with the FSR code.
Then we would like to thank the many users from all LEP collaborations
who provided us with comments, bug reports, bug fixes and their ideas. 
The interactive part of SUSYGEN is its best asset. So we would specially 
like to thank Y. Arnoud, L. Duflot, Y. Gao, M. Gruwe, K. Hultqvist, R.V. Kooten, 
I. Laktineh and  G. Wolf  who performed the above functions with 
patience and good humour.  P.M. would also  
like to thank J. Coles, J. Dann, G. Ganis, C. Hoffman, 
J. Nachtman, B. Orejudos, M. Schmitt, P. Van Gemmeren and M. Williams for the
countless hours  spent on chasing down numerous bugs and problems in the code.

\newpage
\appendix
{\leftline{\Large \bf APPENDIX}}

\section{Codes for sparticles and processes}\label{codes}

Table~\ref{table.part} lists the LUND KF codes which SUSYGEN assigns the sparticles 
in the LUND record. Note that {\it SusyProd} stands for the production process
of the sparticles, upper case $\tilde G$ is the Gravitino and lower case
$\tilde g$ is the gluino. Table~\ref{table.proc} lists the SUSY production 
process code conventions.

{\begin{table}[htb]
\begin{center}
\begin{tabular}{c|c|c|c}
LUND KF id & Sparticle & LUND KF id & Sparticle \\
\hline
\hline
41 & ${\tilde d}_{L}$& 61& ${\tilde b}_{R}$ \\
42 & ${\tilde u}_{L}$& 62& ${\tilde t}_{R}$ \\
43 & ${\tilde s}_{L}$& 63& ${\tilde b}_{1}$ \\
44 & ${\tilde c}_{L}$& 64& ${\tilde t}_{1}$ \\
45 & ${\tilde b}_{L}$& 65& $ {\tilde \tau}_{1}$ \\
46 & ${\tilde t}_{L}$& 66& $ {\tilde b}_{2}$ \\
47 & ${\tilde d}_{R}$& 67& ${\tilde t }_{2}$ \\
48 & ${\tilde u}_{R}$& 68& ${\tilde \tau}_{2}$ \\
49 & ${\tilde s}_{R}$& 69& $\tilde G$ \\
50 & ${\tilde c}_{R}$& 70& $\tilde g$ \\
51 & ${\tilde e}^{-}_{L}$ &71 & ${\chi}^{0}_{1}$ \\
52 & ${\tilde \nu_e}$  &72& ${\chi}^{0}_{2}$ \\
53 & ${\tilde \mu}^{-}_{L}$  &73& ${\chi}^{0}_{3}$ \\
54 & ${\tilde \nu_\mu}$  &74& ${\chi}^{0}_{4}$ \\
55 & ${\tilde \tau}^{-}_{L}$  &75& ${\chi}^{+}_{1}$ \\
56 & ${\tilde \nu_\tau}$  &76& ${\chi}^{+}_{2}$ \\
57 & ${\tilde e}^{-}_{R}$ &77 & ${\chi}^{-}_{1}$ \\
58 & ${\tilde \mu}^{-}_{R}$ &78 & ${\chi}^{-}_{2}$ \\
59 & ${\tilde \tau}^{-}_{R}$ &79 & SusyProd \\

\end{tabular}
\end{center}
\caption{\label{table.part} LUND KF labels and particle codes  (K(I,2) in the lund common).}
\end{table}}

{\begin{table}[htb]
\begin{center}
\begin{tabular}{c|c|c|c}
Process id & Process & Process id & Process \\
\hline
\hline
1  & ${\chi^0_1}{\chi^0_1}$ &26& $\tilde{t_1}\tilde{t_1}$ \\
2  & ${\chi^0_1}{\chi^0_2}$ &27& $\tilde{t_2}\tilde{t_2}$\\
3  & ${\chi^0_2}{\chi^0_2}$ &28& $\tilde{u_L}\tilde{u_L}$\\
4  & ${\chi^0_1}{\chi^0_3}$ &29& $\tilde{u_R}\tilde{u_R}$\\
5  & ${\chi^0_2}{\chi^0_3}$ &30& $\tilde{d_L}\tilde{d_L}$\\
6  & ${\chi^0_3}{\chi^0_3}$ &31& $\tilde{d_R}\tilde{d_R}$\\
7  & ${\chi^0_1}{\chi^0_4}$ &32& $\tilde{c_L}\tilde{c_L}$\\
8  & ${\chi^0_2}{\chi^0_4}$ &33& $\tilde{c_R}\tilde{c_R}$\\
9  & ${\chi^0_3}{\chi^0_4}$ &34& $\tilde{s_L}\tilde{s_L}$\\
10 & ${\chi^0_4}{\chi^0_4}$ &35& $\tilde{s_R}\tilde{s_R}$\\
11 & ${\chi^+_1}{\chi^-_1}$ &36& ${\chi^0_1}\tilde{G}$\\
12 & ${\chi^+_1}{\chi^-_2}$ &37& ${\chi^0_2}\tilde{G}$\\
13 & ${\chi^+_2}{\chi^-_2}$ &38& ${\chi^0_3}\tilde{G}$\\
14 & $\tilde{\nu_e}\tilde{\nu_e}$       &39& ${\chi^0_4}\tilde{G}$\\
15 & $\tilde{\nu_{\mu}}\tilde{\nu_{\mu}}$    &40& ${\chi^0_1}\nu$ \\
16 & $\tilde{\nu_{\tau}}\tilde{\nu_{\tau}}$  &41& ${\chi^0_2}\nu$ \\
17 & $\tilde{e_L}\tilde{e_L}$                &42& ${\chi^0_3}\nu$ \\
18 & $\tilde{e_R}\tilde{e_R}$                &43& ${\chi^0_4}\nu$ \\
19 & $\tilde{e_L}\tilde{e_R}$                &44& ${\chi^+_1}\l^-$ \\
20 & $\tilde{\mu_L}\tilde{\mu_L}$            &45& ${\chi^+_2}\l^-$ \\
21 & $\tilde{\mu_R}\tilde{\mu_R}$            &46& hZ\\
22 & $\tilde{\tau_L}\tilde{\tau_L}$          &47& HZ\\
23 & $\tilde{\tau_R}\tilde{\tau_R}$          &48& hA\\
24 & $\tilde{b_1}\tilde{b_1}$                &49& HA\\
25 & $\tilde{b_2}\tilde{b_2}$                &50& $H^+H^-$\\

\end{tabular}
\end{center}
\caption{\label{table.proc} SUSY production process codes.}
\end{table}}

\newpage

\section{Gaugino Diagonalisation}\label{app.mass}
\subsection{Mass matrices}
The neutralino mass matrix in the base 
\beq
\psi_l^0=\left(-i\tilde{\gamma} ,-i\tilde{Z} , \cos\beta
 \tilde{H^0_1} -\sin\beta \tilde{H^0_2} , \sin\beta
 \tilde{H^0_1}  +\cos\beta \tilde{H^0_2}\right) \nonumber
\eeq
is given in
section~\ref{unconstrained}. The mass eigenstates can be computed by
diagonalising \eq{neutralino.eigen}. The diagonalisation is done
numerically in SUSYGEN.
In some calculations (\rp violating processes and gauge mediated processes)
 we use the base \cite{gunhab}:
\beq
\psi_k^0= (-i\lambda_{\tilde{B}} ,-i\lambda_{\tilde{W_3}} , 
\psi_{H1}^1 , \psi_{H2}^2) 
\eeq
In this base the diagonalised neutralino coupling matrix 
$N'_{ij}$ is defined in terms of the matrix $N_{ij}$ as 
\barr
N'_{1j}&=&N_{1j}\cos(\theta_W)-N_{2j}\sin(\theta_W)\nonumber\\
N'_{2j}&=&N_{1j}\sin(\theta_W)+N_{2j}\cos(\theta_W)\nonumber\\
N'_{3j}&=&N_{3j}\cos(\beta)+N_{4j}\sin(\beta)\nonumber\\
N'_{4j}&=&-N_{3j}\sin(\beta)+N_{4j}\cos(\beta)
\label{eq:neut.diag.gh}
\earr

The chargino mass matrix is also given in section~\ref{unconstrained}. 
Here the chargino masses can be computed analytically:
\begin{eqnarray}
 M^{2}_{\chi^+_1,\chi^+_2} &=& \frac{1}{2}[M_2^2+\mu^2+2 m^2_W \\
 & &  \mp \sqrt{(M_2^2-\mu^2)^2+4 m^4_W 
  \cos2^2\beta +4 m^2_W (M_2^2 +\mu^2+ 2 M_2 \mu \sin2\beta)}] \nonumber
\end{eqnarray}
The chargino mixing matrices are given by  
\begin{eqnarray}
 U_{12} &=& U_{21}=\frac{\theta_1}{\sqrt2}\sqrt{1+\frac{M_2^2-\mu^2-2m_W^2 \cos
 2\beta }{W} }   \\
 U_{22} &=& -U_{11}=\frac{\theta_2}{\sqrt2}\sqrt{1-\frac{M_2^2-\mu^2-2m_W^2 \cos
 2\beta }{W} }  \\
 V_{21} &=& -V_{12}=\frac{\theta_3}{\sqrt2}\sqrt{1+\frac{M_2^2-\mu^2+2m_W^2 \cos
 2\beta }{W} }   \\  
 V_{22} &=& V_{11}=\frac{\theta_4}{\sqrt2}\sqrt{1-\frac{M_2^2-\mu^2+2m_W^2 \cos
 2\beta }{W} }   
\end{eqnarray}
where W is given by 
\begin{eqnarray}
 W=\sqrt{(M_2^2+\mu^2+2m_W^2)^2-4(M_2\mu-m_W^2 \sin2\beta)^2}
\end{eqnarray}
and the sign factors $\theta_i$ are shown in table ~\ref{tabt}. 
The  chargino and neutralino  mass eigenvalues $m_{\chi^+_k},m_{\chi^0_n}$ 
 can have positive or negative signs $\eta_{k}=\pm1, \eta_{n}=\pm1$, while the
 physical masses  are defined positively, e.g. 
 $M_{\chi^+_k}=m_{\chi^+_k}\eta_{k}$.

\begin{table}[htb]
\center{
\begin{tabular}{|l|c|r|}
\hline
           & $\tan\beta > 1$ & $\tan\beta < 1$ \\
\hline
 $\theta_1 $  & 1                & $\varepsilon_B $  \\
 $\theta_2 $  & $\varepsilon_B $ & 1         \\
 $\theta_3 $  & $\varepsilon_A $ & 1         \\
 $\theta_4 $  & 1                & $\varepsilon_A $ \\
\hline
\end{tabular}
\caption{Sign factors $\theta_i$, where $\varepsilon_A$=sign$(M_2\sin\beta+
 \mu\cos\beta) $ and $\varepsilon_B$=sign$(M_2\cos\beta+\mu\sin\beta)$.}
\label{tabt}
}
\end{table}

\clearpage
\section{Formulae}
In  the following we list the formulae which are used in SUSYGEN for the 
cross sections and the decay rates. The listing is given for completeness and
easier understanding of the code, together with the original references.

\subsection{Cross sections}

\subsubsection{Neutralino pair production (processes 1-10)}

Differential neutralino cross section (function GENPHO) \cite{bartl};

\begin{eqnarray}
\frac{d\sigma}{dt} &=& \frac{d\sigma}{dt}\mid_Z +\frac{d\sigma}{dt}|_{\sel}
   +\frac{d\sigma}{dt}|_{Z\sel}
\end{eqnarray}

\begin{eqnarray}
 \frac{d\sigma}{dt}\mid_{Z} &=& \frac{g^4}{16 \pi s^2 \cwq} 
 |D_z(s)|^2 |O_{ij}^{''L} |^2  (|L_e|^2+|R_e|^2) \\  
 & &  [(M_i^2-t)(M_j^2-t)+(M_i^2-u)(M_j^2-u) -2 \eta_i \eta_j M_i M_j s] 
\nonumber 
\end{eqnarray}
\begin{eqnarray}
\frac{d\sigma}{dt}|_{\sel} &=&
    \frac{g^4}{64 \pi s^2} \{ |f_{li}^L|^2 |f_{lj}^L|^2 
 [|D_{\sell}(t)|^2 (M_i^2-t)(M_j^2-t) \\
   & & +|D_{\sell}(u)|^2 (M_i^2-u)(M_j^2-u)
   -2 Re(D_{\sell}(t) D_{\sell}^{*}(u))\eta_i \eta_j M_i M_j s] \nonumber \\
              & & +|f_{li}^R|^2 |f_{lj}^R|^2
   [|D_{\selr}(t)|^2 (M_i^2-t)(M_j^2-t) \nonumber \\
   & & +|D_{\selr}(u)|^2(M_i^2-u)(M_j^2-u)
   -2 Re(D_{\selr}(t) D_{\selr}^{*}(u))\eta_i \eta_j M_i M_j s] \}
\nonumber  
\end{eqnarray}
\begin{eqnarray}
\frac{d\sigma}{dt}|_{Z \sel} &= &
    \frac{g^4}{16 \pi s^2 \cwd} Re(D_{Z}(s)) O_{ij}^{''L}  \\ 
  & & \{L_e f^{L}_{li} f^{L}_{lj} [D_{\sell}(t)((M_i^2-t)(M_j^2-t) 
   -\eta_i \eta_j M_i M_j s) \nonumber \\    
 & & +D_{\sell}(u)((M_i^2-u)(M_j^2-u) -\eta_i \eta_j M_i M_j s)] \nonumber \\
 & &  -R_e f^{R}_{li} f^{R}_{lj} 
   [D_{\selr}(t)((M_i^2-t)(M_j^2-t) -\eta_i \eta_j M_i M_j s) \nonumber \\  
 & & +D_{\selr}(u)((M_i^2-u)(M_j^2-u) -\eta_i \eta_j M_i M_j s)] \} 
     \nonumber 
\end{eqnarray}

where the Standard model couplings are;

\begin{eqnarray}
e^2 &=&4\pi\alpha \\
 g&=&\frac{e}{\sin\thw} \\
L_f&=& T_3 -Q \sin^2\thw \\
R_f&=& -Q \sin^2\thw \\
 \nonumber 
\end{eqnarray}

%%%%($T_3$(Q) is the isospin 3d coordinate(charge) of the each fermion)

$M_{i}$ and $\eta_{i}$ are neutralino masses and diagonalisation phases,
the couplings of the neutralino to sfermions are;

\begin{eqnarray}
 f_{li}^{L} &=& -\sqrt{2} [\frac{L_l}{\ctw}N_{i2} -\frac{R_l}{\sin\theta_W} 
  N_{i1}]  \\
 f_{li}^{R} &= & \sqrt{2} R_l[\frac{1}{\cos\thw} N_{i2}^{*}-\frac{1}
 {\sin\thw} N_{i1}^{*}] 
\end{eqnarray}

the neutralino couplings to Z are;
 
\begin{eqnarray}
O_{ij}^{''L}&=& -\frac{1}{2}(N_{i3}N_{j3}^{*}-N_{i4}N_{j4}^{*})\cos2\beta
 -\frac{1}{2}(N_{i3}N_{j4}^{*}+N_{i4}N_{j3}^{*})\sin2\beta \\
O_{ij}^{''R}&=& -O_{ij}^{''L*}
\end{eqnarray}

and the propagators are;

\begin{eqnarray}
D_Z(s) &=& (s-m_Z^2+im_W\Gamma_{z})^{-1} \\
D_{\sel_{L,R}}(x)&=& (x-m_{\sel_{L,R}}^2)^{-1} 
\end{eqnarray}

The integrated cross section (function PHOTI) is;
\begin{eqnarray}
\sigma_{tot} &=& \frac{2-\delta_{ij}}{2}(\sigma_Z+\sigma_{\sel}+\sigma_{Z\sel})
\end{eqnarray}
\begin{eqnarray}
\sigma_Z &=& \frac{g^4}{4\pi \cwq} |D_z(s)|^2 \frac{q}{\sqs}
  |O_{ij}^{''L}|^2 (|L_e|^2+|R_e|^2) [E_i E_j+\frac{q^2}{3}-\eta_i \eta_j
  M_i M_j] \nonumber \\
\end{eqnarray}

\begin{eqnarray}
 \sigma_{\sel} &  = &  \frac{g^4q}{16 \pi s \sqs}  \\  
  & &  \{ |f^{L}_{ei}|^2 |f^{L}_{ej}|^2
  [\frac{E_i E_j -s d_L +q^2}{s d_L^2 -q^2}+2+\frac{\sqs}{2 q}(1-2 d_L-\frac{
  \eta_i \eta_j M_i M_j}{s d_L}) ln|\frac{d_L+\frac{q}{\sqs}}{d_L-\frac{q}
   {\sqs}}|]  \nonumber \\
     & &   +|f^{R}_{ei}|^2 |f^{R}_{ej}|^2 [\frac{E_i E_j -s d_R+q^2}
{s d_R^2 -q^2}+2+\frac{
\sqs}{2 q} (1-2 d_R-\frac{
  \eta_i \eta_j M_i M_j}{s d_R})ln|\frac{d_R+\frac{q}{\sqs}}{d_R-\frac{q}
   {\sqs}}|] \} \nonumber 
\end{eqnarray}
\begin{eqnarray}
  \sigma_{Z\sel} & = &  -\frac{g^4 q }{8\pi \cwd  \sqs} 
 Re(D_z(s)) O^{''L}_{ij}  \\
  & & \{L_{e} f_{ei}^{L} f_{ej}^{L} [\frac{E_i E_j -s d_L (1-d_L) -\eta_i    
    \eta_j M_i M_j}{q \sqs} ln|\frac{d_L +\frac{q}{\sqs}}{d_L-\frac{q}{\sqs}}
  | +2 (1-d_L)]  \nonumber  \\ 
 & &  -R_{e} f_{ei}^{R}  f_{ej}^{R} [\frac{E_i E_j -s d_R (1-d_R) 
-\eta_i    
    \eta_j M_i M_j}{q \sqs} ln|\frac{d_R +\frac{q}{\sqs}}{d_R-\frac{q}{\sqs}}
    | +2 (1-d_R)]\} \nonumber 
\end{eqnarray}

the phase space factors are;

\begin{eqnarray}
d_{L,R} &=& \frac{1}{2s}(s+2m_{\sel_{L,R}}^2-M_i^2-M_j^2) \\
E_i &=& \sqrt{q^2+M_i^2}
\end{eqnarray}
where q is the CM momentum of the neutralinos.

\subsubsection{Chargino pair production (processes 11-13)}

The differential chargino cross section (function GENCHAR) is \cite{bartl};

\begin{eqnarray}
 \frac{d\sigma}{dt}&=&\frac{d\sigma}{dt}\mid_{\gamma}+\frac{d\sigma}{dt}\mid_{Z}
  \frac{d\sigma}{dt}\mid_{\snu} +\frac{d\sigma}{dt}\mid_{\gamma Z}
  \frac{d\sigma}{dt}\mid_{\gamma \snu} \frac{d\sigma}{dt}\mid_{Z \snu } 
\end{eqnarray}
\begin{eqnarray}
 \frac{d\sigma}{dt}\mid_{\gamma} &=& \frac{e^4 \delta_{ij}}{8 \pi s^4}
  [(M_i^2-u)(M_j^2-u)  +(M_i^2-t)(M_j^2-t) +2 M_i M_j s] 
\end{eqnarray}
\begin{eqnarray}
 \frac{d\sigma}{dt}\mid_{Z} &=& \frac{g^4 |D_Z(s)|^2}{32 \pi s^2 \cwq}
 \{(L_e^2+R_e^2)(|O_{ij}^{'L}|^2+|O_{ij}^{'R}|^2)
 [(M_i^2-u)(M_j^2-u)   \\
 & & +(M_i^2-t)(M_j^2-t)]+4(L_e^2+R_e^2)O_{ij}^{'L}O_{ij}^{'R}  
  \eta_i \eta_j M_i M_j s \nonumber \\  
 & & -(L_e^2-R_e^2)(|O_{ij}^{'L}|^2-|O_{ij}^{'R}|^2) 
  [(M_i^2-u)(M_j^2-u) - (M_i^2-t)(M_j^2-t)] \} \nonumber  
\end{eqnarray}
\begin{eqnarray}
 \frac{d\sigma}{dt}\mid_{\snu} &=& \frac{g^4 |D_{\snu}(t)|^2 }{64 \pi s^2}
 |V_{i1}|^2 |V_{j1}|^2 (M_i^2-t)(M_j^2-t) 
\end{eqnarray}
\begin{eqnarray}
 \frac{d\sigma}{dt}\mid_{\gamma Z} &=& \frac{e^2 g^2 Re(D_Z(s))\delta_{ij}}
 {16\pi s^3 \cwd} \{(L_e+R_e)(O_{ij}^{'L}+O_{ij}^{'R}) 
 [(M_i^2-u)(M_j^2-u) \nonumber \\
  & & +(M_i^2-t)(M_j^2-t) +2 M_iM_j s] 
  -(L_e-R_e)(O_{ij}^{'L}-O_{ij}^{'R})  \nonumber \\
  & & [(M_i^2-u)(M_j^2-u)-(M_i^2-t)(M_j^2-t)]\}  
\end{eqnarray}
\begin{eqnarray}
\frac{d\sigma}{dt}\mid_{\gamma\snu} &=& \frac{e^2 g^2 D_{\snu}(t) \delta_{ij}}
 {16\pi s^3 } |V_{i1}|^2 [(M_i^2-t)(M_j^2-t)+M_iM_j s]  
\end{eqnarray}
\begin{eqnarray}
 \frac{d\sigma}{dt}\mid_{Z \snu} &=&\frac{g^2 Re(D_{\snu}(t) D^*_Z(s))} 
 {16\pi s^2 \cwd} L_e V_{i1} V_{j1} [O_{ij}^{'L}(M_i^2-t)(M_j^2-t) 
 +O_{ij}^{'R} \eta_i \eta_j M_i M_j s]  \nonumber \\
\end{eqnarray}

where the chargino couplings to the Z are; 

\begin{eqnarray}
 O_{ij}^{'L} &=& -V_{i1} V_{j1}^{*} -\frac{1}{2} V_{i2} V_{j2}^{*}+\delta_{ij}
 \sin^2\thw  \\
 O_{ij}^{'R} &=& -U_{i1}^{*} U_{j1} -\frac{1}{2}  U_{i2}^{*} U_{j2}+\delta_{ij}
 \sin^2\thw  
\end{eqnarray}

and;

\begin{eqnarray}
 D_{\snu}(t) &=& (t-m_{\snu}^2)^{-1}  
\end{eqnarray}

The total cross section is;
\begin{eqnarray}
\sigma_{tot} = \sigma_{\gamma} +\sigma_{Z} +\sigma_{\snu} +\sigma_{\gamma Z}
  +\sigma_{\gamma \snu} +\sigma_{Z \snu} \
\end{eqnarray}

\begin{eqnarray}
 \sigma_{\gamma} &=& \frac{e^4 q \sqs \delta_{ij}}{2\pi s^3} (E_i E_j+
                   \frac{q^2}{3} +M_i M_j) 
\end{eqnarray}
\begin{eqnarray}
 \sigma_{Z} &=&  \frac{g^4 q|D_Z(s)|^2}{8\pi \cwq \sqs}
   [(|O_{ij}^{'L}|^2+|O_{ij}^{'R}|^2)
  (L_e^2+R_e^2)   \\
  & & (E_i E_j+\frac{q^2}{3})  +2(L_e^2+R_e^2)O_{ij}^{'L}
   O_{ij}^{'R} \eta_i\eta_j M_i M_j] \nonumber                 
\end{eqnarray}
\begin{eqnarray}
 \sigma_{\snu} &=&  \frac{g^4 |V_{i1}|^2 |V_{j1}|^2 q}{32\pi \sqs m_{\snu}^4}
  [\frac{E_i E_j +q^2 -q \sqs \frac{a_L}{b_L}}{a_L^2-b_L^2}+
 \frac{2 q^2}{b_L^2}  \\
  & & +\frac{1}{2 b_L^2}(q \sqs -\frac{2q^2 a_L}{b_L})ln|\frac{a_L+b_L}
     {a_L-b_L}|]  \nonumber 
\end{eqnarray}
\begin{eqnarray}
 \sigma_{\gamma Z} &=&  \frac{e^2 g^2 q \sqs}{4\pi \cwd s^2 } Re(D_Z(s))
  \delta_{ij} (L_e+R_e)(O_{ij}^{'L}+O_{ij}^{'R})  \\
  & & (E_i E_j+  \frac{q^2}{3} +M_i M_j) \nonumber 
\end{eqnarray}
\begin{eqnarray}
 \sigma_{\gamma \snu} &=& -\frac{e^2 g^2 |V_{i1}|^2 \delta_{ij}}{16\pi s^2}
 (h +M_i M_j ln|\frac{a_L+b_L}{a_L-b_L}|) 
\end{eqnarray}
\begin{eqnarray}
 \sigma_{Z \snu}  &=& -\frac{g^4}{16\pi s \cwd} V_{i1} V_{j1} Re(D_Z(s))
   L_e (O_{ij}^{'L} h +O_{ij}^{'R}\eta_i\eta_j M_i M_j
 ln|\frac{a_L+b_L}{a_L-b_L}|) \nonumber \\
\end{eqnarray}

where the phase space factors are;

\begin{eqnarray}
  a_L &=& \frac{2 m_{\snu}^2 +s -M_i^2 -M_j^2}{2 m_{\snu}^2} \\
  b_L &=& \frac{q\sqs}{m_{\snu}^2} \\
  h &=& 2 q\sqs -2 q^2 \frac{a_L}{b_L} +(E_iE_j +q^2 \frac{a_L^2}{b_L^2}
  -q \sqs \frac{a_L}{b_L}) ln|\frac{a_L+b_L}{a_L-b_L}|
\end{eqnarray}

\subsubsection{Sneutrino production (processes 14-16)}
The differential cross section (function GENSNUE) for $\snu_{e}$ is \cite{bartl};

\begin{eqnarray}
\frac{d\sigma}{dt}  &=& \frac{g^4}{64 \pi s^2} (ut-m_{\snu}^4)
 [\frac{L_e^2+R_e^2}{\cwq}|D_Z(s)|^2 +(\sum_{k=1}^{2} |V_{k1}|^2 D_k (t))^2 \\
 & &  +\frac{2 L_e}{\cwd} Re(D_Z(s)) \sum_{k=1}^{2} |V_{k1}|^2 D_k (t)]
  \nonumber 
\end{eqnarray}

where the {\it t} channel chargino propagators are;

\begin{eqnarray}
  D_k(t)&=& (t-M_k^2)^{-1} 
\end{eqnarray}

For $\snu$ other than $\snu_e$ (function GENSNU) only the first term
is retained.

\subsubsection{Selectron production (processes 17-19)}

The differential cross section (function GENSEL) is \cite{bartl};

\begin{eqnarray}
\frac{d\sigma}{dt}&=& \frac{d\sigma}{dt}\mid_{\gamma} +\frac{d\sigma}{dt}\mid_Z
  \frac{d\sigma}{dt}\mid_{\schi} +\frac{d\sigma}{dt}\mid_{\gamma Z}
  \frac{d\sigma}{dt}\mid_{\gamma \schi}+\frac{d\sigma}{dt}\mid_{Z \schi}
\end{eqnarray}

We obtain ;

(i) $ e^+ + e^- \longrightarrow \sel_1^+ + \sel_1^- $

\begin{eqnarray}
\frac{d\sigma}{dt}\mid_{\gamma}&=& \frac{e^4}{8 \pi s^4} (ut-m_{\sel_1}^4)
\end{eqnarray}

\begin{eqnarray}
\frac{d\sigma}{dt}\mid_Z  &=& \frac{g^4}{16\pi s^2 \cwq}(ut-
 m_{\sel_1}^4)|D_Z(s)|^2 (L_e \cos^2\phi +R_e \sin^2\phi)^2(L_e^2 +R_e^2)
 \nonumber \\
\end{eqnarray}

\begin{eqnarray}
\frac{d\sigma}{dt}\mid_{\schi} &=& \frac{g^4}{64 \pi s^2} \{(ut-m_{\sel_1}^4)
[\cos^4\phi (\sum_{k=1}^{4} D_k (t) |f_{ek}^L|^2)^2 
+\sin^4\phi (\sum_{k=1}^{4} D_k (t) |f_{ek}^R|^2)^2]  \nonumber\\
 & & +2s \sin^2\phi \cos^2\phi \sum_{k,l=1}^{4} M_k M_l \eta_k \eta_l D_k(t)
 D_l(t) f_{ek}^R f_{ek}^L f_{el}^R f_{el}^L \}  
\end{eqnarray}

\begin{eqnarray}
\frac{d\sigma}{dt}\mid_{\gamma Z} &=&\frac{e^2 g^2}{8 \pi s^3 \cwd}
 (ut-m_{\sel_1}^4)(L_e \cos^2\phi +R_e \sin^2\phi)(L_e +R_e) Re(D_Z(s)) \nonumber\\
\end{eqnarray}

\begin{eqnarray}
  \frac{d\sigma}{dt}\mid_{\gamma \schi}&=& \frac{e^2 g^2}{16\pi s^3}
 (ut-m_{\sel_1}^4)\sum_{k=1}^{4} D_k(t)(\cos^2\phi |f_{ek}^L|^2+
  \sin^2\phi |f_{ek}^R|^2)
\end{eqnarray}

\begin{eqnarray}
\frac{d\sigma}{dt}\mid_{Z \schi} &=&\frac{g^4}{16\pi s^2 \cwd}
 (ut-m_{\sel_1}^4)
  (L_e \cos^2\phi +R_e \sin^2\phi) Re(D_Z(s)) \nonumber \\
  & &\sum_{k=1}^{4} D_k(t) (L_e |f_{ek}^L|^2 \cos^2\phi +R_e |f_{ek}^R|^2 \sin^2
     \phi)  
\end{eqnarray}

where the {\it t} channel neutralino propagators are;

\begin{eqnarray}
  D_k(t)&=& (t-M_k^2)^{-1} 
\end{eqnarray}

and $\phi$ is the L/R mixing angle;

\begin{eqnarray}
 \sel_1 &=&\sel_L\cos\phi+\sel_R\sin\phi, \nonumber \\
 \sel_2 &=&\sel_R\cos\phi-\sel_L\sin\phi
\end{eqnarray}

(ii) $e^+ + e^- \longrightarrow \sel_1^+ + \sel_2^-$
(function GENSELR)

\begin{eqnarray}
 \frac{d\sigma}{dt}\mid_{\gamma}=\frac{d\sigma}{dt}\mid_{\gamma Z}
  =\frac{d\sigma}{dt}\mid_{\gamma \schi}=0 
\end{eqnarray}

\begin{eqnarray}
\frac{d\sigma}{dt}\mid_Z  &=& \frac{g^4}{16\pi s^2 \cwq}(ut-
  m_{\sel_1}^2 m_{\sel_2}^2 ) \sin^2\phi \cos^2\phi |D_Z(s)|^2 (L_e -R_e)^2 
 (L_e^2 +R_e^2)  \nonumber \\
\end{eqnarray}

\begin{eqnarray}
\frac{d\sigma}{dt}\mid_{\schi} &=& \frac{g^4}{64 \pi s^2}\{(ut-m_{\sel_1}^2 
 m_{\sel_2}^2)\cos^2\phi \sin^2\phi [(\sum_{k=1}^{4} D_k (t) 
  |f_{ek}^L|^2)^2  +  \\
 & & (\sum_{k=1}^{4} D_k (t) |f_{ek}^R|^2)^2]  
   + s (\sin^4\phi + \cos^4\phi) \sum_{k,l=1}^{4} M_k M_l \eta_k \eta_l 
 D_k(t) D_l(t) f_{ek}^R f_{ek}^L f_{el}^R f_{el}^L \} \nonumber 
\end{eqnarray}

\begin{eqnarray}
\frac{d\sigma}{dt}\mid_{Z \schi} &=&\frac{g^4}{16\pi s^2 \cwd}
(ut-m_{\sel_1}^2 m_{\sel_2}^2) \cos^2\phi \sin^2\phi (L_e -R_e) Re(D_Z(s))
\nonumber  \\
  & &\sum_{k=1}^{4} D_k(t) (L_e |f_{ek}^L|^2 -R_e |f_{ek}^R|^2)  
\end{eqnarray}

\subsubsection{Other charged sfermion cross sections (processes 20-35)}

The differential cross section (function GENSMU) is;

\begin{eqnarray}
\frac{d\sigma}{dt}&=& \frac{d\sigma}{dt}\mid_{\gamma} +\frac{d\sigma}{dt}\mid_Z
+\frac{d\sigma}{dt}\mid_{\gamma Z}
\end{eqnarray}

\begin{eqnarray}
\frac{d\sigma}{dt}\mid_{\gamma}&=& \frac{Q^2 e^2}{8 \pi s^4} (ut-m_{\sfe_1}^4)
\end{eqnarray}

\begin{eqnarray}
\frac{d\sigma}{dt}\mid_Z  &=& \frac{g^4}{16\pi s^2 \cwq}(ut-
 m_{\sfe_1}^4)|D_Z(s)|^2 (L_f \cos^2\phi +R_f \sin^2\phi)^2(L_e^2 +R_e^2)
\nonumber \\ 
\end{eqnarray}

\begin{eqnarray}
\frac{d\sigma}{dt}\mid_{\gamma Z} &=&\frac{Q^2 g^2}{8 \pi s^3 \cwd}
 (ut-m_{\sfe_1}^4)(L_f \cos^2\phi +R_f \sin^2\phi)(L_e +R_e) Re(D_Z(s))
\nonumber \\ 
\end{eqnarray}

The total cross section (function GENSMUS) is;

\begin{eqnarray}
\sigma&=& \sigma_{\gamma} +\sigma_Z +\sigma_{\gamma Z}
\end{eqnarray}

\begin{eqnarray}
\sigma_{\gamma}&=& \frac{Q^2 e^2}{8 \pi s} \frac{\beta^3}{6}
\end{eqnarray}

\begin{eqnarray}
\sigma_Z  &=& \frac{g^4 s}{16\pi \cwq} 
|D_Z(s)|^2 (L_f \cos^2\phi +R_f \sin^2\phi)^2(L_e^2 +R_e^2)\frac{\beta^3}{6}
\end{eqnarray}

\begin{eqnarray}
\sigma_{\gamma Z} &=&\frac{Q^2 g^2}{8 \pi \cwd}
 (L_f \cos^2\phi +R_f \sin^2\phi)(L_e +R_e) Re(D_Z(s)) \frac{\beta^3}{6}
\end{eqnarray}

Where $\beta=\frac{2q}{\sqrt{s}}$

For $\tilde{q}$ the above cross sections are multiplied by a QCD factor R;
\begin{eqnarray}
R=3(1+\frac{4\alpha_{s}}{6\pi}(\frac{\pi^2}{\beta}-(1+\beta)
(\frac{\pi^2}{2}-3)))
\end{eqnarray}

\subsubsection{Neutralino-Gravitino production (processes 36-39)}

The cross section is given by \cite{nano}
\label{sec:general}
\begin{equation}
{d\sigma\over dt}={1 \over 16\pi s^2}\,
{F(s,t,u)\over \Lambda_S^4}\ ,
\label{eq:sigma}
\end{equation}

$\Lambda_S^4= 6 (M_{PLANCK}m_{\tilde G})^2$ is the supersymmetry breaking
scale, as a function  of  the Planck mass and the mass of the gravitino.

\begin{equation}
F(s,t,u)=F_{\gamma\gamma}+F_{tt}+F_{uu}+F_{ZZ}
+F_{\gamma t}+F_{\gamma u}+F_{Zt}+F_{Zu}+F_{\gamma Z}\ ,
\label{eq:F}
\end{equation}
where
\begin{eqnarray}
F_{\gamma\gamma}&=&(N_{i1}e)^2\,{2s(s-m^2_\chi)(t^2+u^2)\over s^2}
\label{eq:Fgg}\\
F_{tt}&=&(X_R)^2\,{t^2\,(m^2_\chi-t)(-t)\over(t-m^2_{\tilde e_R})^2}
+(X_L)^2\,{t^2\,(m^2_\chi-t)(-t)\over(t-m^2_{\tilde e_L})^2}
\label{eq:Ftt}\\
F_{uu}&=&(X_R)^2\,{u^2\,(m^2_\chi-u)(-u)\over(u-m^2_{\tilde e_R})^2}
+(X_L)^2\,{u^2\,(m^2_\chi-u)(-u)\over(u-m^2_{\tilde e_L})^2}
\label{eq:Fuu}\\
F_{ZZ}&=&\left(N_{i2}\,{g\over\cos\theta_W}\right)^2\,{(R_e^2+L^2_e)\over2}\,
{2s(s-m^2_\chi)(t^2+u^2)\over (s-M^2_Z)^2+(\Gamma_Z M_Z)^2}
\label{eq:FZZ}\\
F_{\gamma t}&=&(N_{i1}\,eX_R){(-t)(2st^2)\over s(t-m^2_{\tilde e_R})}
-(N_{i1}eX_L){(-t)(2st^2)\over s(t-m^2_{\tilde e_L})}
\label{eq:Fgt}\\
F_{\gamma u}&=&(N_{i1}\,eX_R){(-u)(2su^2)\over s(u-m^2_{\tilde e_R})}
-(N_{i1}eX_L){(-u)(2su^2)\over s(u-m^2_{\tilde e_L})}
\label{eq:Fgu}\\
F_{Zt}&=&-\left(N_{i2}\,R_e X_R\,{g\over\cos\theta_W}\right)
{(-t)(2st^2)(s-M^2_Z)\over[(s-M^2_Z)^2+(\Gamma_Z M_Z)^2](t-m^2_{\tilde e_R})}
\nonumber\\
&&+\left(N_{i2}\,L_e X_L\,{g\over\cos\theta_W}\right)
{(-t)(2st^2)(s-M^2_Z)\over[(s-M^2_Z)^2+(\Gamma_Z M_Z)^2](t-m^2_{\tilde e_L})}
\label{eq:FZt}\\
F_{Zu}&=&-\left(N_{i2}\,R_e X_R\,{g\over\cos\theta_W}\right)
{(-u)(2su^2)(s-M^2_Z)\over[(s-M^2_Z)^2+(\Gamma_Z M_Z)^2](u-m^2_{\tilde e_R})}
\nonumber\\
&&+\left(N_{i2}\,L_e X_L\,{g\over\cos\theta_W}\right)
{(-u)(2su^2)(s-M^2_Z)\over[(s-M^2_Z)^2+(\Gamma_Z M_Z)^2](u-m^2_{\tilde e_L})}
\label{eq:FZu}\\
F_{\gamma Z}&=&-2\left(N_{i1}N_{i2}\,e\,{g\over\cos\theta_W}\right)\,
{(R_e+L_e)\over2}\,{2s(s-m^2_\chi)(t^2+u^2)(s-M^2_Z)
\over s[(s-M^2_Z)^2+(\Gamma_Z M_Z)^2]} \nonumber \\
\label{eq:FgZ}
\end{eqnarray}

and:

\begin{eqnarray}
X_R&=&N_{i1}\,e-N_{i2}\, {g R_e\over\cos\theta_W}\ \label{eq:XR}\\
X_L&=&-N_{i1}\,e-N_{i2}\,{g\over\cos\theta_W}\
L_e \label{eq:XL}
\end{eqnarray}

\subsubsection{S-channel Sneutrino resonance cross sections (processes 40-45)}

The \rp violating single sparticle production processes 40-45 are
characterised by a sneutrino s-channel resonance for a non-zero coupling 
 $\lambda_{121}$ or $\lambda_{131}$. The s-channel sneutrino 
process $e^+ e^- \ra {\tilde {\nu_j}}$ has three principal decay modes:
\begin{eqnarray}
e^+ e^- \ra {\tilde {\nu_j}} \ra e^+ e^-  \label{eq:eemod}\\
e^+ e^- \ra {\tilde {\nu_j}} \ra \chi^0 \nu_j  \label{eq:chimod}\\
e^+ e^- \ra {\tilde {\nu_j}} \ra \chi^+ l^-_j  \label{eq:chamod}
\end{eqnarray}
but only the single chargino/neutralino processes 
(\ref{eq:chimod}),(\ref{eq:chamod}) are  implemented in SUSYGEN.

\subsubsection{Single Chargino Production}
Ignoring contributions to the vertices of the MSSM from mass terms, we have two
channels present ($s$ and $t$) for chargino production \cite{dreiner.prod}:
\begin{eqnarray}
\sigma_{e^+ e^- \ra l_j^\mp\tilde{\chi}^\pm_m} = \frac{1}{64\pi s^2}
\left (
\frac{V_{m1} \lambda_{1j1} e}
{ \sin \theta_{W}} \right ) ^2
\; \left \{ \frac{C_{s}}{(s- m^2_{\tilde {\nu_j}})^2+\Gamn^2 m^2_{\tilde {\nu_j}}} + \right.
\frac{C_{t}}{(t-\mnue)^2} - \nonumber \\
\left.
\frac{C_{st} (s-m^2_{\tilde {\nu_j}})}
{[(s-m^2_{\tilde {\nu_j}})^2+\Gamn^2 m^2_{\tilde {\nu_j}}](t-\mnue)}
\right \}
\end{eqnarray}
where
\begin{eqnarray}
C_{s} & = & s ( s-\mwi), \quad C_{t} =  t ( t-\mwi) \nonumber \\
C_{st} & = & s ( s-\mwi) +t ( t-\mwi) - u ( u-\mwi) 
\end{eqnarray}
Here $s$, $t$ and $u$ are the Mandelstam variables
\beq
t=(e^--{\tilde\chi}^\pm)^2,\quad u=(e^+-{\tilde\chi}^\pm)^2, 
\eeq
and the width of the sneutrino is given by 
\beq
\Gamma_{\tilde {\nu_j}} = \Gamma_{1} + \Gamma_{2}
\eeq
where the \rp violating sneutrino decay rate  
$\Gamma_{1} \equiv \tilde{\nu_j} \ra e^{i}e^{k}= 
\frac{\lambda_{1j1}^{2}}{16 \pi}m_{\tilde{\nu}_{j}}$; and the \rp
conserving  decay rate
$\Gamma_{2}
\equiv \tilde{\nu}_{j} \ra \tilde{\chi}_{m}^{\pm} l^{\mp}_{j}$.

\subsubsection{Single Neutralino Production}
Ignoring mass terms the cross-section for this process is given by \cite{dreiner.prod}:
\begin{eqnarray}
\sigma_{e^+ e^- \ra \nu\tilde{\chi}^0_m} = \frac{1}{32\pi s^2}
\lambda^2_{1j1} 
\; \left \{ 
V_{1}^2 \frac{C_{s}}{(s-m^2_{\tilde {\nu_j}})^2+\Gamn^2 m^2_{\tilde {\nu_j}}} +
V_{2}^2 \frac{C_{t}}{(t-m^2_{{\tilde e}_{L}})^2}
+ V_{3}^2 \frac{C_{u}}{(u-m^2_{{\tilde e}_{R}})^2} \right.
\nonumber \\
- V_{1} V_{2}^{*}
\frac{C_{st} (s-m^2_{\tilde {\nu_j}})}
{[(s-m^2_{\tilde {\nu_j}})^2+\Gamn^2 m^2_{\tilde {\nu_j}}](t-m^2_{{\tilde e}_{L}})}
- V_{1} V_{3}^{*}
\frac{C_{su} (s-m^2_{\tilde {\nu_j}})}
{[(s-m^2_{\tilde {\nu_j}})^2+\Gamn^2 m^2_{\tilde {\nu_j}}](u-m^2_{{\tilde e}_{R}})}
\nonumber \\
- V_{2} V_{3}^{*}
\left. \frac{C_{tu}} {(t-m^2_{{\tilde e}_{L}}) (u-m^2_{{\tilde e}_{R}})}
\right \} \nonumber \\
\end{eqnarray}
where
\begin{eqnarray}
V_{1} & = & 
-\frac{1}{2} \frac{g}{\cos{\theta_{W}}}
N_{m 2}^{' *} \nonumber \\
V_{2} & = & 
-e N_{m 1}^{' *} - 
\frac{g}{\cos{\theta_{W}}}
\left( \frac{1}{2} -\sin{\theta_{W}}^2
\right ) N_{m 2}^{' *}
 \nonumber \\
V_{3} & = & 
-e N_{j m}^{' *} +
\frac{g \sin{\theta_{W}}^2 }{\cos{\theta_{W}}}
N_{m 2}^{' *} 
\nonumber \\
\end{eqnarray}

Here the functions $C_s,\,C_t,$ and $C_{st}$ are given as before, just replacing
the chargino mass $m_{{\tilde\chi}^\pm}$ by the neutralino mass $m_{{\tilde
\chi}^0}$. The new functions $C_{u}$, $C_{su}$ and $C_{ut}$ are given by 
\begin{eqnarray}
C_{u} & = &  u ( u-\mneu) \nonumber \\
C_{su} & = & s ( s-\mneu) -
t ( t-\mneu) +
u ( u-\mneu) \nonumber \\
C_{tu} & = & -s ( s-\mneu) +
t ( t-\mneu) +
u ( u-\mneu)  
\end{eqnarray}
And  the \rp conserving  decay rate $\Gamma_{2}
\equiv \tilde{\nu}_{j} \ra \tilde{\chi}_{m}^{0} \nu_{j}$. 

\subsection{Decay Rates}
\subsection{Gaugino-Higgsino three body decay rates}

 The general 3-body decay from a chargino/neutralino to a chargino/neutralino
and 2 fermions can be parametrised as \cite{bartl};

\begin{eqnarray}
 \Gamma(\schi_i{\longrightarrow}\schi_k +f \sfe ') &=& \frac{\alpha^2}{32\pi 
 \sin^4\theta _W M_i^3} \int d\bar{s} d\bar{t} (W_s+W_t+W_u+W_{tu}+W_{st}+
  W_{su}) \nonumber \\
\label{eq:tbd} 
\end{eqnarray}

\begin{eqnarray}
 W_s &=& \frac{1}{(\bar{s}-M_s^2)^2+\gs^2}\{A_s(M_i^2-\bar{t})(\bar{t}-M_k^2)\\
 & & + B_s(M_i^2-\bar{u})(\bar{u}-M_k^2)  
   + 2C_s \eta_i \eta_k M_i M_k \bar{s} \} \nonumber \\
 W_t &=& A_t^L \frac{(M_i^2-\bar{t})(\bar{t}-M_k^2)}{(\bar{t}-M_L^2)^2+\gl^2}+
   A_t^R \frac{(M_i^2-\bar{t})(\bar{t}-M_k^2)}{(\bar{t}-M_R^2)^2+\gr^2}  \\
 W_u &=& A_u^L \frac{(M_i^2-\bar{u})(\bar{u}-M_k^2)}{(\bar{u}-M_L^2)^2+\gl^2}+
   A_u^R \frac{(M_i^2-\bar{u})(\bar{u}-M_k^2)}{(\bar{u}-M_R^2)^2+\gr^2}  \\
 W_{tu} &=& 2 A^L_{tu} \eta_i \eta_k M_i M_k \bar{s}
 \frac{(\bar{t}-M_L^2)(\bar{u}-M_L^2)+ \gl^2}
 {[(\bar{t}-M_L^2)^2+\gl^2][(\bar{u}-M_L^2)^2+\gl^2]} \\
 & & + 2 A^R_{tu} \eta_i \eta_k M_i M_k \bar{s}
 \frac{(\bar{t}-M_R^2)(\bar{u}-M_R^2)+ \gr^2}
 {[(\bar{t}-M_R^2)^2+\gr^2][(\bar{u}-M_R^2)^2+\gr^2]} \nonumber \\
 W_{st} &=& 2 A^L_{st}(M_i^2-\bar{t})(\bar{t}-M_k^2)
 \frac{(\bar{t}-M_L^2)(\bar{s}-M_s^2)+\gl \gs}
 {[(\bar{t}-M_L^2)^2+\gl^2][(\bar{s}-M_s^2)^2+\gs^2]}  \\
 & & + 2 B^L_{st}\eta_i \eta_k M_i M_k \bar{s}
 \frac{(\bar{t}-M_L^2)(\bar{s}-M_s^2)+\gl \gs}
 {[(\bar{t}-M_L^2)^2+\gl^2][(\bar{s}-M_s^2)^2+\gs^2]} \nonumber \\
 & & +2 A^R_{st}(M_i^2-\bar{t})(\bar{t}-M_k^2)
 \frac{(\bar{t}-M_R^2)(\bar{s}-M_s^2)+\gr \gs}
 {[(\bar{t}-M_R^2)^2+\gr^2][(\bar{s}-M_s^2)^2+\gs^2]} \nonumber \\
 & & + 2 B^R_{st}\eta_i \eta_k M_i M_k \bar{s}
 \frac{(\bar{t}-M_R^2)(\bar{s}-M_s^2)+\gr \gs}
 {[(\bar{t}-M_R^2)^2+\gr^2][(\bar{s}-M_s^2)^2+\gs^2]} \nonumber \\
 W_{su} &=& 2 A^L_{su}(M_i^2-\bar{u})(\bar{u}-M_k^2)
 \frac{(\bar{u}-M_L^2)(\bar{s}-M_s^2)+\gl \gs}
 {[(\bar{u}-M_L^2)^2+\gl^2][(\bar{s}-M_s^2)^2+\gs^2]}  \\
 & & + 2 B^L_{su}\eta_i \eta_k M_i M_k \bar{s}
 \frac{(\bar{u}-M_L^2)(\bar{s}-M_s^2)+\gl \gs}
 {[(\bar{u}-M_L^2)^2+\gl^2][(\bar{s}-M_s^2)^2+\gs^2]} \nonumber \\
 & & +2 A^R_{su}(M_i^2-\bar{u})(\bar{u}-M_k^2)
 \frac{(\bar{u}-M_R^2)(\bar{s}-M_s^2)+\gr \gs}
 {[(\bar{u}-M_R^2)^2+\gr^2][(\bar{s}-M_s^2)^2+\gs^2]} \nonumber \\
 & & + 2 B^R_{su}\eta_i \eta_k M_i M_k \bar{s}
 \frac{(\bar{u}-M_R^2)(\bar{s}-M_s^2)+\gr \gs}
 {[(\bar{u}-M_R^2)^2+\gr^2][(\bar{s}-M_s^2)^2+\gs^2]} \nonumber 
\end{eqnarray}

  where the factors $G_i$ are; $\gs=\gsg M_s$, $\gl=\glg M_L$,
$\gr=\grg M_R$, $\gn=\gng M_{\snu}$, $\gmk=\gmkg M_k$, $\gml=\gmlg M_l$
and the chargino-neutralino couplings are;

\begin{eqnarray}
 O_{ij}^L &=& -(N_{i4}\cos\beta -N_{i3}\sin\beta)\frac{V^*_{j2}}{\sqrt{2}} +
    (N_{i1}\sin\theta_W +N_{i3}\cos\theta_W) V^*_{j1} \\
 O_{ij}^R &=&  (N^*_{i4}\sin\beta +N^*_{i3}\cos\beta)\frac{U_{j2}}{\sqrt{2}} +
    (N^*_{i1}\sin\theta_W +N^*_{i3}\cos\theta_W) U_{j1} 
\end{eqnarray}

and the tables ~\ref{tab:1}, ~\ref{tab:2}, ~\ref{tab:3} and ~\ref{tab:4} 
contain the definitions of the A, B and C parameters.

\begin{table}[htb]
\center{
\begin{tabular}{|l|c|r|l|}
\hline
$\schi_i^{+}{\longrightarrow}$ &$\schi_k^{0}l^+\nu$ &$\schi_k^{+}l^+l^-$ &
$\schi_k^{+}\nu\snu$ \\
\hline
$ A_{s}$ &$2(O_{ki}^L)^2 $ &$4\frac{(L_l O_{ki}^{'L})^2+(R_l O_{ki}^{'R})^2}
{\cwq}$ 
  &$\frac{(O_{ki}^{'L})^2}{\cwq}$  \\
$ B_{s}$ &$2(O_{ki}^R)^2 $ &$4\frac{(L_l O_{ki}^{'R})^2+(R_l O_{ki}^{'L})^2}
{\cwq}$ 
  &$\frac{(O_{ki}^{'R})^2}{\cwq}$   \\
$ C_{s}$ &$-2 O_{ki}^L O_{ki}^R $ &$-4\frac{(L_l^2 +R_l^2)O_{ki}^{'L}O_{ki}^
{'R}}{\cwq}$
  &$-\frac{O_{ki}^{'L} O_{ki}^{'R}}{\cwq}$    \\
$ A_{t}^{L}$  & $(f_{\nu k}^L V_{i1})^2$ & $(V_{k1}V_{i1})^2$  & 0   \\
$ A_{u}^{L}$  & $(f_{lk}^L U_{i1})^2 $   &  0  &$(U_{k1}U_{i1})^2$   \\
$ A_{tu}^{L}$ & $f_{\nu k}^L f_{lk}^L V_{i1} U_{i1} $  & 0     & 0  \\
$ A_{st}^{L}$ &$-\sqrt{2} f_{\nu k}^L V_{i1} O_{ki}^L$ 
  &$2\frac{V_{k1}V_{i1}O_{ki}^{'L}L_l}{\cwd}$& 0 \\
$ B_{st}^{L}$ &$\sqrt{2} f_{\nu k}^L V_{i1} O_{ki}^R$  
  &$-2\frac{V_{k1}V_{i1}O_{ki}^{'R}L_l}{\cwd}$ &0 \\
$ A_{su}^{L}$ &$\sqrt{2} f_{lk}^L U_{i1} O_{ki}^R $ &  0  
  &$-\frac{U_{k1}U_{i1}O_{ki}^{'R}}{\cwd}$ \\
$ B_{su}^{L}$ &$-\sqrt{2} f_{lk}^L U_{i1} O_{ki}^L$ &  0  
  &$\frac{U_{k1}U_{i1}O_{ki}^{'L}}{\cwd}$ \\
\hline
\end{tabular}
\caption{Coefficients for leptonic chargino decays}
\label{tab:1}
}
\end{table}

\begin{table}[htb]
\center{
\begin{tabular}{|l|c|r|l|}
\hline
$\schi_i^{+}{\longrightarrow}$ &$\schi_k^{0}u\tilde{d}$ &$\schi_k^{+}u\tilde{u}
$ &$\schi_k^{+}d\tilde{d}$ \\
\hline
$ A_{s}$ &$6(O_{ki}^L)^2 $ &$12\frac{(L_u O_{ki}^{'L})^2+(R_u O_{ki}^{'R})^2}
{\cwq}$ 
  &$12\frac{(L_d O_{ki}^{'L})^2+(R_d O_{ki}^{'R})^2}{\cwq}$   \\
$ B_{s}$ &$6(O_{ki}^R)^2 $ &$12\frac{(L_u O_{ki}^{'R})^2+(R_u O_{ki}^{'L})^2}
{\cwq}$ 
  &$12\frac{(L_d O_{ki}^{'R})^2+(R_d O_{ki}^{'L})^2}{\cwq}$    \\
$ C_{s}$ &$-12 O_{ki}^L O_{ki}^R $ &$-12\frac{(L_u^2 +R_u^2)O_{ki}^{'L}O_{ki}^
{'R}}{\cwq}$
  &$-12\frac{(L_d^2 +R_d^2)O_{ki}^{'L}O_{ki}^{'R}}{\cwq}$    \\
$ A_{t}^{L}$  & $3(f_{uk}^L V_{i1})^2$ & 0 &$3(V_{k1}V_{i1})^2$     \\
$ A_{u}^{L}$  & $3(f_{dk}^L U_{i1})^2$ &$3(U_{k1}U_{i1})^2$ & 0     \\
$ A_{tu}^{L}$ & $3f_{uk}^L f_{dk}^L V_{i1} U_{i1} $  & 0     & 0  \\
$ A_{st}^{L}$ &$-3\sqrt{2} f_{uk}^L V_{i1} O_{ki}^L$ 
  & 0 &$6\frac{V_{k1}V_{i1}O_{ki}^{'L}L_d}{\cwd}$ \\
$ B_{st}^{L}$ &$3\sqrt{2} f_{uk}^L V_{i1} O_{ki}^R$  
   &0 & $-6\frac{V_{k1}V_{i1}O_{ki}^{'R}L_d}{\cwd}$ \\
$ A_{su}^{L}$ &$3\sqrt{2} f_{dk}^L U_{i1} O_{ki}^R $   
  &$-6\frac{U_{k1}U_{i1}O_{ki}^{'R}L_u}{\cwd}$ &  0 \\
$ B_{su}^{L}$ &$-3\sqrt{2} f_{dk}^L U_{i1} O_{ki}^L$ 
  &$6\frac{U_{k1}U_{i1}O_{ki}^{'L}L_u}{\cwd}$ &  0  \\
\hline
\end{tabular}
\caption{Coefficients for hadronic chargino decays}
\label{tab:2}
}
\end{table}

\begin{table}[htb]
\center{
\begin{tabular}{|l|c|r|l|}
\hline
$\schi_i^{o}{\longrightarrow}$ &$\schi_k^{o}l^{+}l^{-}$ &$\schi_k^{o}\nu\snu 
$ &$\schi_k^{o}q\tilde{q}$ \\
\hline
$ A_{s}=B_{s}=C_{s}$ &$4\frac{(L_l^2 +R_l^2)O_{ki}^{''L}}{\cwq}$ &
$\frac{(O_{ki}^{''L})^2}{\cwq}$
  &$12\frac{(L_q^2 +R_q^2)O_{ki}^{''L}}{\cwq}$   \\
$ A_{t}^{L}=A_{u}^{L}=A_{tu}^{L}$  &$(f_{lk}^L f_{li}^L)^2$  &$(f_{\nu k}^L 
f_{\nu i}^L)^2$    
  &$3(f_{qk}^{L} f_{qi}^{L})^2 $  \\
$ A_{t}^{R}=A_{u}^{R}=A_{tu}^{R}$  &$(f_{lk}^R f_{li}^R)^2$  & 0 
  &$3(f_{qk}^{R} f_{qi}^{R})^2 $  \\
$ A_{st}^{L}=B_{st}^{L}=A_{su}^{L}=B_{su}^{L}$ &$2\frac{f_{lk}^L f_{li}^L 
O_{ki}^{''L}L_l}{\cwd}$
  &$\frac{f_{\nu k}^L f_{\nu i}^L O_{ki}^{''L}}{\cwd}$
  &$6\frac{f_{qk}^L f_{qi}^L O_{ki}^{''L}L_q}{\cwd}$ \\
$ A_{st}^{R}=B_{st}^{R}=A_{su}^{R}=B_{su}^{R}$ &$2\frac{f_{lk}^R f_{li}^R 
O_{ki}^{''R}R_l}{\cwd}$
  &$ 0 $
  &$6\frac{f_{qk}^R f_{qi}^R O_{ki}^{''R}R_q}{\cwd}$ \\
\hline
\end{tabular}
\caption{Coefficients for $\schi_i^{o}{\longrightarrow}\schi_k^{o}f\tilde
{f^{'}}$}
\label{tab:3}
}
\end{table}

\begin{table}[htb]
\center{
\begin{tabular}{|l|c|r|}
\hline
$\schi_i^{o}{\longrightarrow}$ &$\schi_k^{+}l^{-}\snu$ &$\schi_k^{+}u\tilde{d}
$ \\
\hline
$ A_{s}$ &$2(O_{ki}^L)^2 $ &$6(O_{ki}^L)^2 $ \\
$ B_{s}$ &$2(O_{ki}^R)^2 $ &$6(O_{ki}^R)^2 $  \\
$ C_{s}$ &$-2 O_{ki}^L O_{ki}^R $ &$-6 O_{ki}^L O_{ki}^R $  \\
$ A_{t}^{L}$  & $(f_{\nu i}^L V_{k1})^2$ &$3(f_{ui}^L V_{k1})^2$     \\
$ A_{u}^{L}$  & $(f_{li}^L U_{k1})^2$ &$3(f_{di}^L U_{k1})^2$     \\
$ A_{tu}^{L}$ & $f_{\nu i}^L f_{li}^L V_{k1} U_{k1}$  &$3f_{ui}^L f_{di}^L 
V_{k1} U_{k1}$ \\
$ A_{st}^{L}$ &$-\sqrt{2}f_{\nu i}^L V_{k1} O_{ik}^L$ 
  &$-3\sqrt{2}f_{ui}^L V_{k1} O_{ik}^L$\\
$ B_{st}^{L}$ &$\sqrt{2}f_{\nu i}^L V_{k1} O_{ik}^R$ 
  &$3\sqrt{2}f_{ui}^L V_{k1} O_{ik}^R$ \\
$ A_{su}^{L}$ &$\sqrt{2}f_{li}^L U_{k1} O_{ik}^R$ 
  &$3\sqrt{2}f_{di}^L U_{k1} O_{ik}^R$ \\
$ B_{su}^{L}$ &$-\sqrt{2}f_{li}^L U_{k1} O_{ik}^L$ 
  &$-3\sqrt{2}f_{di}^L U_{k1} O_{ik}^L$ \\
\hline
\end{tabular}
\caption{Coefficients for $\schi_i^{o}{\longrightarrow}\schi_k^{+}f\tilde
{f^{'}}$}
\label{tab:4}
}
\end{table}

\subsection{Gaugino two body and radiative decay rates}

The above three-body branching ratio formulas include the width of the
exchanged particles in the propagator terms, and therefore smoothly converge
to a two-body formula when above threshold. For example if the
sneutrino is lighter than the chargino and the neutralino is the LSP,
 the three-body decay rate $\Gamma(\chi^+ \ra l^+ \nu \chi^0)$ is the same as the
two-body decay rate $\Gamma(\chi^+ \ra  l^+ {\tilde \nu})$. Furthermore the
kinematics of the three-body decay is equivalent to the two-body decay
kinematics (once the sneutrino decay ${\tilde \nu} \ra \nu \chi^0$ has been
included). The procedure has been numerically  checked
with  the explicit 2-body calculations, and the branching ratios, 
partial decay widths and kinematics agree well. 

Sometimes it is useful to calculate the gaugino two-body decays explicitly (for
example for \rp violating models). The
gaugino decay rate to a lighter sfermion is  given by:
\barr
\Gamma(\chi \ra {\tilde f} f') &=& c_f \frac{1}{16 \pi m^3_\chi} 
 \sqrt{(m^2_\chi-(m_{\tilde f}+m_{f'})^2) (m^2_\chi-(m_{\tilde f}-m_{f'})^2)} \nonumber \\
& & \times {{\cal M}_{\chi {\tilde f} f'}} \label{gaug.2bd.master}
\earr
where $c_f=3$ is the colour factor for gaugino decays to squarks ($c_f=1$ for
sleptons and sneutrinos). And
\barr
M_{\chi {\tilde f} f'} &=& 2 \sqrt{(m^2_\chi-(m_{\tilde f}+m_{f'})^2)
(m^2_\chi-(m_{\tilde f}-m_{f'})^2) +4 m^2_\chi m^2_{f'}} \{a^2+b^2\} \nonumber \\
\earr
where the gaugino coupling to the sfermion is parametrised as
$i\{a(1+\gamma_5)+b(1-\gamma_5))\}$; and the constants a,b are for example given in
\cite{gunhab}, Fig.~22-24. Note that for gaugino decays to third generation sfermions mixing
must be taken into account by modifying the constants a,b and the sfermion mass appropriately.

The gaugino decay rate to a gauge boson (W or Z) is again given by
\eq{gaug.2bd.master}, and the coupling constants a,b are given in
\cite{haberkane} (Fig.~75).

The two-body decay rates for the neutralino decays to a lighter neutralino plus a Higgs
boson are given by \cite{gunion.2}:
\barr
\begin{array}{ll}
\Gamma(\chi_i \ra \chi_j H^0_k) = \frac{g^2
\lambda(M_{i},M_{j},M_{k})^{\frac{1}{2}}}{32 \pi
M^3_i} &[(F^2_{ijk}+F^2_{jik})(M^2_i+M^2_j-M^2_H)\nonumber \\
&  +4F_{ijk}F_{jik}\epsilon_i \epsilon_j \eta'_k M_i M_j]
\end{array}
\earr
where k=1,2,3 correspond to the three neutral Higgs $h,H,A$; $\eta'_k=1$ for
$k=1,2$ and $\eta'_k=-1$ for $k=3$; the factors $\epsilon_i$ stand for the sign
of the neutralino mass; the kinematic function $\lambda$ is defined by
\barr
\lambda(a,b,c) = a^2+b^2-c^2-4a^2b^2
\earr
and 
\barr
F_{ijk} &=& \frac{c_k}{2 sin{\beta}}[N'_{3i}N'_{2j} + N'_{3j}N'_{2i} -
\tan{\theta_W}(N'_{3i}N'_{1j}+N'_{3j}N'_{1i})] \nonumber \\
& &+ \frac{d_k}{2 m_W \sin{\beta}}[M N'_{2i}N'_{2j}+ M' N'_{1i}N'_{1j} -
\mu(N'_{3i}N'_{4j}+N'_{4i}N'_{3j})]
\earr
where $M,M',\mu$ are the usual neutralino mixing parameters, and the constants $c_k,d_k$ are given by 
\barr
c_1=\sin(\beta-\alpha) &,& d_1=-\sin(\alpha) \nonumber\\
c_2=\cos(\beta-\alpha) &,& d_2=\cos(\alpha) \nonumber\\
c_3=\cos(2\beta) &,& d_3=\cos(\beta). 
\earr

The two-body decay rates for the neutralino decays to a lighter chargino plus a
charged Higgs
boson are given by \cite{gunion.2}:
\barr
\Gamma(\chi_i \ra \chi^\pm_j H^\mp) = \frac{g^2
\lambda(M_{i},M_{j},M_{H^\pm})^{\frac{1}{2}}}{32 \pi M^3_i} 
[(F_L^2+F_R^2)(M^2_i+M^2_j-M^2_H)+4F_L F_R \epsilon_i\epsilon_j M_i M_j]
\nonumber \\
\earr
where 
\barr
F_L&=&\cos{\beta}[N'_{4i}V_{j1}+\frac{1}{\sqrt{2}}(N'_{2i}+N'_{1i}\tan{\theta_W})V_{j2}]
\nonumber \\
F_R&=&\sin{\beta}[N'_{3i}U_{j1}-\frac{1}{\sqrt{2}}(N'_{2i}+N'_{1i}\tan{\theta_W})U_{j2}].
\earr

Finally the radiative two-body decays occurring through loop-diagrams
have  the general form:
\barr
\Gamma(\schi^0_j\rightarrow\schi^0_i\gamma)=
\frac{g^2_{\schi^0_j\schi^0_i\gamma}(M_j^2-M^2_i)^3}{8\pi M_j^5}
\earr

The expressions for $g^2_{\schi^0_j\schi^0_i\gamma}$ can be found in \cite{ambros.gut.evo}.

\clearpage

\subsection{Sfermion decay rates}

%------------------------------------------------------------------------
\noi The most general sfermion decay widths  are: 
\beq
  \Gamma (\sf_{i}\to f\nt_{k}) =
  \frac{g^{2}\lambda^{\onehf}(\msf{i}^{2}, m_{f}^{2}, \mnt{k}^{2})}
       {16\pi\msf{i}^{3}} \,
  \left[ (a_{ik}^{2} + b_{ik}^{2}) (\msf{i}^{2} - m_{f}^{2} - \mnt{k}^{2}) -
         4 a_{ik}b_{ik}m_{f}\mnt{k} \right]
  \label{eq:ntwidth}
\eeq
and
\beq
  \Gamma (\sf_{i}\to f^{\prime}\ch_{j}) =
  \frac{g^{2}\lambda^{\onehf}(\msf{i}^{2}, m_{f^{\prime}}^{2}, \mch{j}^{2})}
       {16\pi\msf{i}^{3}}\,
  \left[ (l_{ij}^{2} + k_{ij}^{2}) 
         (\msf{i}^{2} - m_{f^{\prime}}^{2} - \mch{j}^{2}) -
         4\, l_{ij}k_{ij}m_{f^{\prime}}\mch{j} \right]
  \label{eq:chwidth}
\eeq
where $\lambda (x,y,z) = x^{2}+y^{2}+z^{2}-2xy-2xz-2yz$. \\

and where: 
\beq
  {a_{1k}^{f} \choose a_{2k}^{f}} = 
    \left(\baa{ll} \cth_{\sf} & \sth_{\sf} \\ 
                  -\sth_{\sf} & \cth_{\sf} \eaa\right)\:
  {h_{Lk}^{f} \choose f_{Rk}^{f}},
  \hspace{4mm}
  {b_{1k}^{f} \choose b_{2k}^{f}} = 
    \left(\baa{ll} \cth_{\sf} & \sth_{\sf} \\ 
                  -\sth_{\sf} & \cth_{\sf} \eaa\right)\:
  {f_{Lk}^{f} \choose h_{Rk}^{f}},  
  \label{eq:copmix}
\eeq

\noi
\beq \begin{array}{ll}
  h_{Lk}^{t} = Y_{t} \left( 
               N_{k3}\sin\b - N_{k4}\cos\b \right),  &
  f_{Lk}^{t} = -\frac{2\sqrt{2}}{3} \sin\tW N_{k1} - \sqrt{2}\,
        (\onehf - \twoth\sin^{2}\tW ) \frac{N_{k2}}{\cos\tW}, \\
  h_{Rk}^{t} = Y_{t} \left( 
               N_{k3}\sin\b - N_{k4}\cos\b \right),  &
  f_{Rk}^{t} = -\frac{2\sqrt{2}}{3} \sin\tW 
               (\tan\tW N_{k2} - N_{k1}), 
  \label{eq:stntcop}
\end{array} \eeq

\noi
\beq \begin{array}{ll}
  h_{Lk}^{b} = -Y_{b} \left( 
               N_{k3}\cos\b + N_{k4}\sin\b \right), &
  f_{Lk}^{b} = \frac{\sqrt{2}}{3}\sin\tW N_{k1} +
     \sqrt{2}\,(\onehf - \oneth\sin^{2}\tW ) \frac{N_{k2}}{\cos\tW}, 
     \hspace{5mm} \\
  h_{Rk}^{b} = -Y_{b} \left( 
               N_{k3}\cos\b + N_{k4}\sin\b   \right),  &
  f_{Rk}^{b} = \frac{\sqrt{2}}{3}\sin\tW 
                   (\tan\tW N_{k2} - N_{k1}),                             
  \label{eq:sbntcop}
\end{array} \eeq

\noi
\beq \begin{array}{ll}
  h_{Lk}^{\tau} = -Y_{\tau} \left(  
                  N_{k3}\cos\b + N_{k4}\sin\b \right),  &
  f_{Lk}^{\tau} = \sqrt{2}\sin\tW N_{k1} +
              \sqrt{2}\,(\onehf - \sin^{2}\tW ) \frac{N_{k2}}{\cos\tW}, 
              \hspace{7.5mm} \\ 
  h_{Rk}^{\tau} = -Y_{\tau} \left( 
                  N_{k3}\cos\b + N_{k4}\sin\b   \right),  &   
  f_{Rk}^{\tau} = \sqrt{2}\sin\tW (\tan\tW N_{k2} - N_{k1}),                                     
  \label{eq:slntcop}
\end{array} \eeq

\noi for the sfermion-fermion-neutralino interaction, and
%\begin{small}
\beq \begin{array}{ll}
 l_{1j}^{t} =       -V_{j1}\cth_{\st} +
              Y_{t}\,V_{j2}\sth_{\st}, \hspace{6mm} &
 k_{1j}^{t} = Y_{b}\,U_{j2}\cth_{\st}, \\
 l_{2j}^{t} =        V_{j1}\sth_{\st} + 
              Y_{t}\,V_{j2}\cth_{\st}, \hspace{6mm} &
 k_{2j}^{t} = -\,Y_{b}\,U_{j2}\sth_{\st}, 
  \label{eq:stchcop}
\end{array} \eeq
\beq \begin{array}{ll}
 l_{1j}^{b} =       -U_{j1}\cth_{\sb} +
              Y_{b}\,U_{j2}\sth_{\sb}, \hspace{4.8mm} &
 k_{1j}^{b} = Y_{t}\,V_{j2}\cth_{\sb},\\
 l_{2j}^{b} =        U_{j1}\sth_{\sb} + 
              Y_{b}\,U_{j2}\cth_{\sb}, \hspace{4.8mm} &
 k_{2j}^{b} = -\,Y_{t}\,V_{j2}\sth_{\sb}, 
  \label{eq:sbchcop}
\end{array} \eeq
\beq \begin{array}{ll}
 l_{1j}^{\tau} =          -U_{j1}\cth_{\stau} +
                 Y_{\tau}\,U_{j2}\sth_{\stau}, \hspace{3.6mm} &
 k_{1j}^{\tau} = 0,\hspace{21.5mm}\\
 l_{2j}^{\tau} =           U_{j1}\sth_{\sb} +
                 Y_{\tau}\,U_{j2}\cth_{\stau}, \hspace{3.6mm} &
 k_{2j}^{\tau} = 0, 
  \label{eq:slchcop}
\end{array} \eeq

\noi for the sfermion-fermion-chargino interaction. 

$Y_{f}$ denote the Yukawa couplings:
\beq
  Y_{t} = m_{t}/(\sqrt{2}\,m_{W}\sin\b), \hspace{6mm} 
  Y_{b} = m_{b}/(\sqrt{2}\,m_{W}\cos\b),\hspace{6mm}
  Y_{\tau} = m_{\tau}/(\sqrt{2}\,m_{W}\cos\b).
  \label{eq:yukcop}
\eeq

For the first two generations the Yukawa couplings and the mixing are
 negligible.

For the decay rate of the process  $\stp\rightarrow c \schi^0$ we use:
\begin{eqnarray}
 \Gamma({\tilde t}^{L,R}{\longrightarrow}c +\schi_i^0) &=&
  3\cdot 10^{-10} m_{\tilde t}^{L,R} (1 - \frac{\mxo^2}{(m_{\tilde t}^{L,R})^2})^2 
\end{eqnarray}

\subsection{Two Body Decays of Sparticles to the Gravitino}

The decay width of the neutralinos to the Gravitino can be written 
as \cite{grav.ref}
\barr
& &\Gamma ({{\tilde \chi}^0_i} \rightarrow \gamma {\tilde G} ) =
{\kappa_{i\gamma} \over 8 \pi}{m_{{{\tilde \chi}^0_i}}^5 \over 
\Lambda_S^4}
\label{ngammaGdecay}
\\
& &\Gamma ({{\tilde \chi}^0_i} \rightarrow Z{\tilde G} ) =
{2\kappa_{iZ_T}+ \kappa_{iZ_L}\over 16 \pi}
{m_{{{\tilde \chi}^0_i}}^5  \over \Lambda_S^4}
\left ( 1- {m_Z^2\over m_{{{\tilde \chi}^0_i}}^2}\right )^4
\label{nzGdecay-eq}\\
& & \Gamma ({{\tilde \chi}^0_i} \rightarrow \varphi{\tilde G} ) =
{\kappa_{i\varphi} \over 16 \pi}{m_{{{\tilde \chi}^0_i}}^5 \over 
\Lambda_S^4}
\left ( 1- {m_\varphi^2\over m_{{{\tilde \chi}^0_i}}^2}\right )^4
\label{nhGdecay-eq}
\earr
where \barr
& &\kappa_{i\gamma} = |N'_{i1}\cos\theta_W + N'_{i2}\sin \theta_W |^2 \nonumber
\\
& &\kappa_{iZ_T} = |N'_{i1}\sin\theta_W - N'_{i2} \cos\theta_W |^2 \nonumber
\\
& &\kappa_{iZ_L} = |N'_{i3} \cos \beta -  N'_{i4}\sin\beta|^2 \nonumber
\\
& &\kappa_{ih^0} = |N'_{i3}\sin\alpha -N'_{i4}\cos\alpha |^2 \nonumber
\\
& &\kappa_{iH^0} = |N'_{i3}\cos\alpha +N'_{i4}\sin\alpha |^2 \nonumber
\\
& &\kappa_{iA^0} = |N'_{i3}\sin\beta +N'_{i4}\cos\beta |^2
\earr
and  $\varphi= (h^0,H^0,A^0)$ is any of the neutral Higgs scalar bosons.
The couplings $N'_{ij}$ are given in \eq{eq:neut.diag.gh}. The chargino
decay widths into gravitino final states are given by 
\barr
& &\Gamma ({{\tilde \chi}^+_i} \rightarrow W^+{\tilde G} ) =
{2\kappa_{iW_T}+ \kappa_{iW_L}\over 16 \pi}
{m_{{{\tilde \chi}^+_i}}^5  \over \Lambda_S^4}
\left ( 1- {m_W^2\over m_{{{\tilde \chi}^+_i}}^2}\right )^4
\\
& &\Gamma ({{\tilde \chi}^+_i} \rightarrow H^+{\tilde G} ) =
{\kappa_{iH^+}\over 16 \pi}
{m_{{{\tilde \chi}^+_i}}^5  \over \Lambda_S^4}
\left ( 1- {m_{H^+}^2\over m_{{{\tilde \chi}^+_i}}^2}\right )^4
\earr
with
\barr
& &\kappa_{iW_T} = {1\over 2}\left ( | V_{i1} |^2 + |U_{i1} |^2 \right )
\nonumber\\
& &\kappa_{iW_L} = |V_{i2}|^2 \sin^2 \beta + |U_{i2}|^2 \cos^2\beta
\nonumber\\
& &\kappa_{iH^+} = |V_{i2}|^2 \cos^2 \beta + |U_{i2}|^2 \sin^2\beta
\earr
and the widths of the 2-body slepton decays into gravitinos are given by 
\barr
\Gamma({\tilde f} \rightarrow f {\tilde G}) =
{m_{\tilde f}^5\over 8 \pi \Lambda_S^4}
\earr

\subsection{$R_p$ violating  two body decays}

For the  \rp violating Yukawa coupling $\lambda''_{ijk}$ the
decay rate of a squark ${\tilde q}_1$ of flavour $i$ into the final states
${\tilde q}_{1i} \ra {\bar q}_j {\bar q}_k$
is given by 
\barr
\Gamma({\tilde q}_{1i} \ra {\bar q}_j {\bar q}_k) = \frac {{\lambda''_{ijk}}^2
  \sin^2{\theta_{{\tilde q}}} |p_i|}{8\pi M_{{\tilde q}_{1i}}} 2\dpr{p_j}{p_k}
\earr
where $\theta_{{\tilde q}}$ is the mixing angle of the squarks ${\tilde
  q}_{Li},{\tilde q}_{Ri}$ which
form the mass eigenstates ${\tilde q}_{1i},{\tilde q}_{2i}$. Here
\barr
  |p_i|&=&
{((M_{{\tilde q}_{1i}}^2-(M_{qj}+M_{qk})^2)(M_{{\tilde
      q}_{1i}}^2-(M_{qj}-M_{qk})^2))^{\frac{1}{2}}} /({2M_{{\tilde q}_{1i}}})
\nonumber \\
\dpr{p_j}{p_k}&=&((M_{qi}^2+|p_i|^2)(M_{qj}^2+|p_i|^2))^{\frac{1}{2}} +
|p_i|^2
\earr
The decay rate for ${\tilde q}_{2i} \ra {\bar q}_j {\bar q}_k$ can be
simply obtained from the above formula by replacing $\sin{\theta_{{\tilde q}}}
\ra \cos{\theta_{{\tilde q}}}$ and ${\tilde q}_{1i} \ra {\tilde q}_{2i}$. The
decay rates for the other operators are 
analogous, except that the mixing term $\sin{\theta}$ has to be replaced by
$\cos{\theta}$ for those sfermions which are part of the SU(2) doublet superfields (the
$L_{i,j}$ and $Q_j$ fields in \eq{eqrpv}). Note also that the result has to be
multiplied by a colour factor $c_f = 3$ for the slepton decays via $L_i Q_j {\bar D_k}$.

\subsection{$R_p$ violating  three body decays}

Neutralinos can decay through  \rp violating couplings 
to SM particles via
Eq.~(\ref{eq:rpvneutlle3decays})-(\ref{eq:rpvneutudd3decays}).
The Matrix element 
squared for the decay ${{\tilde \chi}^0_l}\ra e_i u_j {\bar d}_k$ of the
operator $L_i Q_j {\bar D}_k$ is given by \cite{dreiner}: 

\begin{eqnarray}
 |{\cal M}({{\tilde \chi}^0_l}\ra e_i u_j {\bar d}_k)|^2&=8c_fg^2\lam_{ijk}'^2 \left\{ 
\right.&  \\
 & \hspace{-2.5cm} D({\tilde u}_j)^2  e_i\cdot d_k&\hspace{-1cm}
 [(a(u_j)^2 + b(u)^2 )    {{\tilde \chi}^0_l}\cdot u_j 
                 +   2 a(u_j) b(u) m_{u_j} M_{{{\tilde \chi}^0_l}}   \, ] \nonumber \\
& \mbox{}\hspace{-2.5cm} + D({\tilde d}_k)^2 e_i\cdot u_j & \hspace{-1cm}
[ (a(d_k)^2 + b({\bar d})^2)   {{\tilde \chi}^0_l}\cdot d_k-
                    2 a(d_k)b({\bar d}) m_{d_k}  M_{{{\tilde \chi}^0_l}}\, ]\nonumber \\
&  \mbox{}\hspace{-2.5cm} + D({\tilde e}_j)^2\hfil  u_j\cdot d_k
&\hspace{-1cm} [(a(e_i)^2 + b(e)^2)
          {{\tilde \chi}^0_l}\cdot e_i   + 2 a(e_i)b(e)  m_{e_i}  M_{{{\tilde \chi}^0_l}}\, ] \nonumber \\
 &\mbox{}\hspace{-2.5cm} -D({\tilde e}_i)D({\tilde u}_j) &
\hspace{-1cm}[ a(u_j)a(e_i)
m_{e_i}m_{u_j}  {{\tilde \chi}^0_l}\cdot d_k 
          + a(u_j)  b(e)  m_{u_j}  M_{{{\tilde \chi}^0_l}} e_i\cdot d_k  \nonumber \\
  && \hspace{-1cm}       + a(e_i)  b(u)  m_{e_i}  M_{{{\tilde \chi}^0_l}}  u_j\cdot d_k 
          + b(e)  b(u) \, {\cal G}(u_j,{{\tilde \chi}^0_l},e_i,d_k) ] \nonumber \\
 &\mbox{}\hspace{-2.5cm}  -D({\tilde u}_j) D({\tilde d}_k)&
\hspace{-1cm} [a(u_j) a(d_k) m_{u_j}m_{d_k}  {{\tilde \chi}^0_l}\cdot e_i
 - a(u_j)  b({\bar d})   m_{u_j}  M_{{{\tilde \chi}^0_l}}  e_i\cdot d_k  \nonumber \\
 && \hspace{-1cm} + a(d_k)  b(u)  m_{d_k}  M_{{{\tilde \chi}^0_l}}  e_i\cdot u_j 
          - b(u)  b({\bar d}) \, {\cal G}(u_j,{{\tilde \chi}^0_l},d_k,e_i)]  \nonumber \\
 &\mbox{}  \hspace{-2.5cm}    -D({\tilde e}_i)D({\tilde d}_k)& 
\hspace{-1cm} [-a(e_i)b({\bar d})
  m_{e_i} M_{{{\tilde \chi}^0_l}}  u_j\cdot d_k + a(e_i)  a(d_k)  m_{e_i}m_{d_k} 
{{\tilde \chi}^0_l}\cdot u_j   \nonumber \\
 && \hspace{-1cm}\left. + a(d_k)  b(e)  m_{d_k}  M_{{{\tilde \chi}^0_l}}  e_i\cdot u_j  
          - b(e)  b({\bar d})  \, {\cal G}({{\tilde \chi}^0_l},e_i,u_j,d_k) ]  \nonumber \right\}
\end{eqnarray}
where the function ${\cal
G}(a,b,c,d)= \dpr{a}{b}\dpr{c}{d}-\dpr{a}{c}\dpr{b}{d}+ \dpr{a}{d} \dpr{b}{c}$.
Here $c_f=3$ is the colour factor and $g$ is the weak coupling constant
$g={\frac {e} {\sin{\theta_w}}}$. 
We have denoted the 4-momenta of the initial and final state particles 
by their particle symbols. The functions $D(p_i)$ denote the propagators squared 
for particle $p_i$ and are given by
\beq
D({\tilde p}_i)^{-1}= M_{{\tilde \chi}^0_l}^2 + m_{p_i}^2-2{{\tilde \chi}^0_l}\cdot p_i - {\tilde m}^2_{p_i}.
\eeq
The coupling constants $a(p_i),\,b(p)$ are given in
Table~\ref{NeutCouplings}. 

\begin{table}
\begin{center}
\begin{tabular}{|c|c|c|}\hline
& $a(p_i)$&$b(p_i)$ \\ \hline
$e_i$ &$ \frac{N'_{3l}}{2 M_W\cos\beta}m_{e_i}$&$ -\frac{1}{2} (N'_{2l}
 + \tan\theta_W N'_{1l})$\\ && \\
${\bar e}_i$ &$\frac{N'_{3l}}{2 M_W \cos\beta}\,m_{e_i}$&$-\tan\theta_W
N'_{1l}$ \\ && \\
$\nu_i$ &$0$&$\frac{1}{2}(N'_{2l}- \tan\theta_W N'_{1l})$ \\ && \\
$u_i$ & $\frac{N'_{4l}}{2M_W\sin\beta} m_{u_j}$ &$ \frac{1}{2}(N'_{2l}
 + {\frac{1}{3}}\tan\theta_W N'_{1l})$ \\  && \\
$\ubar_i$ &$\frac{N'_{4l}}{2M_W\sin\beta} m_{u_j}$&$\frac{2}{3}\tan\theta_W
N'_{1l}$ \\  && \\
$d_i$ &$ \frac{N'_{3l}}{2M_W\cos\beta}  m_{d_k}$&$ 
-\frac{1}{2}(N'_{2l} - \frac{1}{3}\tan\theta_W  N'_{1l})$ \\  && \\
$\dbar_i$ & $ \frac{N'_{3l}}{2M_W\cos\beta}  m_{d_k} $ &$- \frac{1}{3} \tan
\theta_W N'_{1l}$ \\ \hline
\end{tabular}
\end{center}
\caption{\label{NeutCouplings} The Coupling constants $a(p_i)$ and $b(p)$ used in
the LSP decay calculation for a neutralino ${\tilde \chi}^0_l$.}
\end{table}
The amplitude squared of the decay to the neutrino, ${{\tilde \chi}^0_l}\ra\nu_i d_j 
\dbar_k$, can be obtained from
the above result by a set of simple transformations of the
4-momenta, the propagator functions $D$ and the couplings $a(p_i),\,b(p)$ : 
$e_i\ra\nu_i$, $u_j\ra d_j$.

The result for the operators $L_iL_j{\bar E}_k$ is completely analogous,
except the colour factor $c_f=1$. The matrix  element squared for the operator 
${\bar U}_i{\bar D}_j{\bar D}_k$ is 
\barr
 |{\cal M}({{\tilde \chi}^0_l}\ra {\bar u}_i{\bar d}_j{\bar d}_k)|^2 &= 8c_f g^2\lam''^2
\left\{\right.& \\ 
   & \hspace{-2.5cm}     D({\tilde u}_i)^2 {\bar d}_j\cdot {\bar d}_k&
\hspace{-1cm}  [ ( b(u)^2 +
a(u_i)^2 ) {{\tilde \chi}^0_l}\cdot {\bar u}_i   -2  b(u) a(u_i) m_{u_i} M_{{\tilde \chi}^0_l} ]
\nonumber \\
   &   \hspace{-2.5cm} +D({\tilde d}_j)^2  \ubar_i\cdot\dbar_k&\hspace{-1cm}
[ (a(d_j)^2 + b(d)^2)
{{\tilde \chi}^0_l}\cdot\dbar_j   -2  b(d)  a(d_j) m_{d_j} M_{{\tilde \chi}^0_l} ]
\nonumber \\
   &\hspace{-2.5cm} +D({\tilde d}_k)^2 \ubar_i\cdot\dbar_j&\hspace{-1cm}[ 
( a(\dbar_k)^2 
+ b(\dbar)^2 ) {{\tilde \chi}^0_l}\cdot\dbar_k -2  b(\dbar) a(\dbar_k) m_{d_k}M_{{\tilde \chi}^0_l}]
\nonumber \\
   &\hspace{-2.5cm} + D({\tilde u}_i)D({\tilde d}_k)&\hspace{-1cm}[ 
 a(\ubar_i)a(\dbar_j)
\,{\cal G}(\ubar_i,{{\tilde \chi}^0_l},\dbar_j,\dbar_k) + b(\ubar)  b(\dbar) m_{u_i}m_{d_k} 
{{\tilde \chi}^0_l}\cdot\dbar_k \nonumber \\
&&\hspace{-1cm}   - b(\ubar)  a(\dbar_j) m_{u_i} M_{{\tilde \chi}^0_l} \dbar_j\cdot\dbar_k  
  -b(\dbar)  a(\ubar_i) m_{d_j} M_{{\tilde \chi}^0_l} \ubar_i\cdot\dbar_k   ] 
\nonumber \\
   &\hspace{-2.5cm}   + D({\tilde u}_i)  D({\tilde d}_k)&  \hspace{-1cm}
[ a(\ubar_i)a(\dbar_k)\,{\cal G}(u_i,{{\tilde \chi}^0_l},d_k,d_j)+ b(\ubar)  b(\dbar) 
m_{u_i}m_{d_k} {{\tilde \chi}^0_l}\cdot\dbar_j    \nonumber \\
   && \hspace{-1cm}-b(\ubar)  a(\dbar_k)m_{u_i} M_{{\tilde \chi}^0_l} \dbar_j\cdot\dbar_k   
      -b(\dbar) a(\ubar_i) m_{d_k} M_{{\tilde \chi}^0_l} \ubar_i\cdot\dbar_j  ]
\nonumber \\
   &\hspace{-2.5cm} + D({\tilde d}_j) D({\tilde d}_k)&\hspace{-1cm}
 [ a(\dbar_j)a(\dbar_k) \,{\cal G}(d_j,{{\tilde \chi}^0_l},d_k,u_i)
+ b(\dbar) b(\dbar) m_{d_j} m_{d_k}  {{\tilde \chi}^0_l}\cdot\ubar_i 
\nonumber \\
   && \hspace{-1cm}-b(\dbar)  a(\dbar_k) m_{d_j} M_{{\tilde \chi}^0_l}  
\ubar_i\cdot\dbar_k 
    \left.     -b(\dbar)  a(\dbar_j) m_{d_k} M_{{\tilde \chi}^0_l}\ubar_i\cdot\dbar_j  ]
\right\} \nonumber
\earr
Here the colour factor $c_f=6$.
The partial decay rates for the neutralino decay modes is then given by 
\barr
\begin{array}{lll}
\Gamma_{{\tilde \chi}^0_l} (LL{\bar E}) & = & \int \frac{1}{2\pi^3}\frac{1}{16
M_{{\tilde \chi}^0}}(2|{\cal M}({{\tilde \chi}^0_l}\ra e_i {\bar \nu}_j {\bar e}_k)|^2+2|{\cal
M}({{\tilde \chi}^0_l}\ra {\bar \nu}_i e_j {\bar e}_k)|^2) 
dE_idE_j \\
\Gamma_{{\tilde \chi}^0} (LQ{\bar D}) & = & \int \frac{1}{2\pi^3}\frac{1}{16
M_{{\tilde \chi}^0}}(2|{\cal M}({{\tilde \chi}^0_l}\ra e_i u_j {\bar d}_k)|^2+2|{\cal
M}({{\tilde \chi}^0_l}\ra\nu_i d_j \dbar_k)|^2) 
dE_idE_j \\
\Gamma_{{\tilde \chi}^0} ({\bar U}{\bar D}{\bar D}) & = & \int \frac{1}{2\pi^3}\frac{1}{16
M_{{\tilde \chi}^0}}2|{\cal M}({{\tilde \chi}^0_l}\ra {\bar u}_i{\bar d}_j{\bar d}_k)|^2 
dE_idE_j 
\end{array}
\label{gammaneut}
\earr
where the matrix element squares are multiplied by a factor of two 
since the LSP is a Majorana fermion and can decay to the
conjugate final states.

Charginos can decay via
Eq.~(\ref{eq:rpvcharglle3decays})-(\ref{eq:rpvchargudd3decays}), 
and the matrix elements squared for the $L_i Q_j {\bar D}_k$ operator can be
written as \cite{dreiner}:  
\barr
\label{eq:matrix1}
|{\cal M}({{\tilde\chi}^+_l} \ra \nu_i u_j {\bar d}_k)|^2 &=& 4c_fg^2{\lam'}^2 \left[
\frac{\alpha_R^2}{R^2({\tilde e}_{iL})} \dpr{{{\tilde \chi}^+_l}}{\nu_i} 
                     \dpr{u_j}{{\bar d}_k} \right. \\
&&+ \frac{\dpr{\nu_i}{{\bar d}_k}}{R^2({\tilde d}_{jL})}
\left\{(\beta_L^2+\beta_R^2)\dpr{{{{\tilde \chi}^+_l}}}{u_j}
+2{\cal R}e(\beta_L\beta_R^*m_{uj}M_{{{{\tilde \chi}^+_l}}})\right\}\nonumber \\
&&\left.-{\cal R}e\left\{
 \frac{\alpha_R}{R({\tilde e}_{iL})R({\tilde d}_{jL})}\left(\beta_L^*
m_{uj} M_{{{{\tilde \chi}^+_l}}}
\dpr{u_j}{{\bar d}_k}+\beta_R^*{\cal G}(p,\nu_i,{\bar d}_k,u_j)\right)
\right\} \right]\nonumber
\\
\label{eq:matrix2}
|{\cal M}({{\tilde\chi}^+_l} \ra {\bar e}_i {\bar d}_j d_k)|^2 &=&  
4c_fg^2{\lam'}^2 \left[\frac{\dpr{d_j}{{\bar d}_k}}{R^2({\tilde \nu}_{iL})}
\left\{(\gamma_L^2+\gamma_R^2)\dpr{{{{\tilde \chi}^+_l}}}{e_i} 
+2{\cal R}e(\gamma_L\gamma_R^* m_{ei}M_{{{{\tilde \chi}^+_l}}}
)\right\} \nonumber \right.\\
&& + \frac{\dpr{e_i}{{\bar d}_k}}{R^2({\tilde u}_{jL})}
\left\{(\delta_L^2+\delta_R^2)\dpr{{{{\tilde \chi}^+_l}}}{d_j} 
+2{\cal R}e(\delta_L\delta_R^* m_{dj}M_{{{{\tilde \chi}^+_l}}})
\right\}\nonumber \\
&&\left.-\frac{1}{R({\tilde \nu}_{iL})R({\tilde u}_{jL})} {\cal R}e\left\{
\gamma_L \delta_L^*{\cal G}({{{\tilde \chi}^+_l}},e_i,{\bar d}_k,d_j)
+\gamma_L \delta_R^* m_{dj}M_{{{{\tilde \chi}^+_l}}}\dpr{e_i}{{\bar d}_k} \right.\right.
\nonumber \\
&& \left.\left.
+\gamma_R \delta_L^* m_{ei}M_{{{{\tilde \chi}^+_l}}}\dpr{d_j}{{\bar d}_k}
+\gamma_R \delta_R^* m_{ei}m_{dj}\dpr{{{{\tilde \chi}^+_l}}}{{\bar d}_k} \right\}\right]
\\
\label{eq:matrix3}
|{\cal M}({{\tilde\chi}^+_l} \ra {\bar e}_i {\bar u}_j u_k)|^2 &=& 
\frac{2c_f{\lam'}^2g^2m_{dk}^2 |U_{l2}|^2}{M_W^2\cos^2\beta R^2({\tilde
d}_{kR})}\dpr{e_i}{u_j}
\dpr{{{{\tilde \chi}^+_l}}}{u_k}\\
\label{eq:matrix4}
|{\cal M}({{\tilde\chi}^+_l} \ra {\bar \nu}_i {\bar d}_j u_k)|^2
&=& \frac{2c_f{\lam'}^2g^2m_{dk}^2|U_{l2}|^2}{M_W^2\cos^2\beta R^2({\tilde
d}_{kR})}\dpr{\nu_i}{d_j} \dpr{{{{\tilde \chi}^+_l}}}{u_k}
\earr
where the couplings $\alpha_{L,R},\beta_{L,R},\gamma_{L,R}$ and
$\delta_{L,R}$ are 
\barr
\begin{array}{llll}
\alpha_L&=&0, \quad &\alpha_R=-iU_{l1} \nonumber \\
\beta_L&=&\frac{im_{uj}V^*_{l2}}{\sqrt{2}M_W\sin\beta}, &\beta_R=\alpha_R\nonumber\\
\gamma_L&=&iV^*_{l1}, &\gamma_R=-\frac{igm_{ei}U_{l2}}
{\sqrt{2}M_W\cos\beta} \\
\delta_L&=&\gamma_L, &\delta_R= -\frac{igm_{dj}U_{l2}}
{\sqrt{2}M_W\cos\beta}
\end{array}
\earr
The final state momenta are denoted by the particle
symbols. $M_{{{{\tilde \chi}^+_l}}}$ is the chargino mass and $m_{ei,dj,dk}$ are the
final state fermion masses. $c_f=3$ is the colour factor. 
The propagator terms $R(p)$ are defined by
\barr
\begin{array}{llll}
R({\tilde e}_{iL}) &=&(\chi^+-\nu_i)^2-{\tilde m}^2_{eiL},\quad&
R({\tilde d}_{jL}) =(\chi^+- u_j)^2-{\tilde m}^2_{djL},\nonumber\\
R({\tilde \nu}_{iL}) &=&(\chi^+-e_i)^2-{\tilde m}^2_{\nu iL}, &
R({\tilde u}_{jL}) =(\chi^+-d_j)^2-{\tilde m}^2_{ujL},\nonumber\\
R({\tilde d}_{kR}) &=&(\chi^+-u_k)^2-{\tilde m}^2_{dkR},&
R({\tilde e}_{kR}) =(\chi^+-\nu_k)^2-{\tilde m}^2_{ekR} \\
R({\tilde u}_{iR}) &=&(\chi^+-{\bar d}_i)^2-{\tilde m}^2_{uiR},&
R({\tilde d}_{jR}) =(\chi^+-{u}_j)^2-{\tilde m}^2_{djR}.
\end{array}
\earr
For the operator $L_iL_j{\bar E}_k$ the matrix elements squared are 
\barr
\label{eq:llematrix1}
|{\cal M}({{\tilde\chi}^+_l} \ra \nu_i \nu_j {\bar e}_k)|^2 &=& 4g^2{\lam}^2 \left[
\frac{\alpha_R^2}{R^2({\tilde e}_{iL})} \dpr{{{\tilde \chi}^+_l}}{\nu_i} 
                     \dpr{\nu_j}{{\bar e}_k} 
+ \frac{\beta_R^2}{R^2({\tilde e}_{jL})}
\dpr{{{{\tilde \chi}^+_l}}}{\nu_j}\dpr{\nu_i}{{\bar e}_k}\right.\nonumber \\
&&\left.-{\cal R}e\left\{
 \frac{ \alpha_R\beta_R^* } { R({\tilde e}_{iL})R({\tilde e}_{jL}) }
{\cal G}(p,\nu_i,{\bar e}_k,\nu_j)
\right\} \right] \\
\label{eq:llematrix2}
|{\cal M}({{\tilde \chi}^+_l} \ra {\bar e}_i {\bar e}_j e_k)|^2 &=&  
4g^2{\lam}^2 \left[\frac{\dpr{e_j}{{\bar e}_k}}{R^2({\tilde \nu}_{iL})}
\left\{(\gamma_L^2+\gamma_R^2)\dpr{{{{\tilde \chi}^+_l}}}{e_i} 
+2{\cal R}e(\gamma_L\gamma_R^* m_{ei}M_{{{{\tilde \chi}^+_l}}}
)\right\} \nonumber \right.\\
&& + \frac{\dpr{e_i}{{\bar e}_k}}{R^2({\tilde \nu}_{jL})}
\left\{(\delta_L^2+\delta_R^2)\dpr{{{{\tilde \chi}^+_l}}}{e_j} 
+2{\cal R}e(\delta_L\delta_R^* m_{ej}M_{{{{\tilde \chi}^+_l}}})
\right\}\nonumber \\
&&\left.-\frac{1}{R({\tilde \nu}_{iL})R({\tilde \nu}_{jL})} {\cal R}e\left\{
\gamma_L \delta_L^*{\cal G}({{{\tilde \chi}^+_l}},e_i,{\bar e}_k,e_j)
+\gamma_L \delta_R^* m_{ej}M_{{{{\tilde \chi}^+_l}}}\dpr{e_i}{{\bar e}_k} \right.\right.
\nonumber \\
&& \left.\left.
+\gamma_R \delta_L^* m_{ei}M_{{{{\tilde \chi}^+_l}}}\dpr{e_j}{{\bar e}_k}
+\gamma_R \delta_R^* m_{ei}m_{ej}\dpr{{{{\tilde \chi}^+_l}}}{{\bar e}_k}
\right\}\right]\\
|{\cal M}({{\tilde\chi}^+_l} \ra {\bar e}_i \nu_j \nu_k)|^2 &=& 4g^2{\lam}^2 \left[
\frac{\epsilon_R^2}{R^2({\tilde e}_{kR})} \dpr{{{\tilde \chi}^+_l}}{\nu_k} 
                     \dpr{{\bar e}_i}{\nu_j}\right]\\
|{\cal M}({{\tilde\chi}^+_l} \ra \nu_i {\bar e}_j  \nu_k)|^2 &=& 4g^2{\lam}^2 \left[
\frac{\epsilon_R^2}{R^2({\tilde e}_{kR})} \dpr{{{\tilde \chi}^+_l}}{\nu_k} 
                     \dpr{\nu_i}{{\bar e}_j}\right]
\earr
where $\alpha,\beta,\gamma,\delta$ are given as above except that in $\delta_R$
$m_{dj}$ is replaced by $m_{ej}$ and $\beta_L=0$ because of vanishing neutrino
mass, and $\epsilon_R=-\frac{igm_{ek}U_{l2}} {\sqrt{2}M_W\cos\beta}$.  
For the operator ${\bar U}{\bar D}{\bar D}$ the matrix elements squared are 
\barr
|{\cal M}({{\tilde\chi}^+_l} \ra {\bar d}_i {\bar d}_j {\bar d}_k)|^2 &=& 
4c_fg^2{\lam}^2 \left[\frac{\beta_R^2}{R^2({\tilde u}_{iR})} \dpr{{{\tilde
\chi}^+}}{{\bar d}_i} \dpr{{\bar d}_j}{{\bar d}_k}\right]\\
|{\cal M}({{\tilde\chi}^+_l} \ra u_i u_j d_k)|^2 &=& 
4c_fg^2{\lam}^2 \left[\frac{\gamma_R^2}{R^2({\tilde d}_{jR})} \dpr{{{\tilde
\chi}^+}}{{u}_j} \dpr{{u}_i}{{d}_k}\right]\\
|{\cal M}({{\tilde\chi}^+_l} \ra u_i d_j u_k)|^2 &=& 
4c_fg^2{\lam}^2 \left[\frac{\delta_R^2}{R^2({\tilde d}_{kR})} \dpr{{{\tilde
\chi}^+}}{{u}_k} \dpr{{u}_i}{{d}_j}\right]
\earr
$\beta_R,\gamma_R,\delta_R$ are given by 
\barr
\begin{array}{lll}
\beta_R&=&-\frac{im_{ui}V^*_{l2}}{\sqrt{2}M_W\sin\beta}, \gamma_R=-\frac{igm_{dj}U_{l2}}
{\sqrt{2}M_W\cos\beta}\\
\delta_R&=&-\frac{igm_{dk}U_{l2}}
{\sqrt{2}M_W\cos\beta} \\
\end{array}
\earr
and the colour factor $c_f=6$.
The partial decay rate for these modes is then given by 
\barr
\begin{array}{llll}
\Gamma_{{\tilde \chi}^+_l} (LL{\bar E}) & = & \int \frac{1}{2\pi^3}\frac{1}{16
M_{{\tilde \chi}^+_l}}(&|{\cal M}({{\tilde\chi}^+_l} \ra \nu_i \nu_j {\bar
e}_k)|^2+|{\cal M}({{\tilde \chi}^+_l} \ra {\bar e}_i {\bar e}_j
e_k)|^2+\\
& & &
|{\cal M}({{\tilde\chi}^+_l} \ra {\bar e}_i \nu_j \nu_k)|^2+|{\cal
M}({{\tilde\chi}^+_l} \ra \nu_i {\bar e}_j  \nu_k)|^2)
dE_idE_j \\
\Gamma_{{\tilde \chi}^+_l} (LQ{\bar D}) & = & \int \frac{1}{2\pi^3}\frac{1}{16
M_{{\tilde \chi}^+_l}}(&|{\cal M}({{\tilde\chi}^+_l} \ra \nu_i u_j {\bar
d}_k)|^2+|{\cal M}({{\tilde\chi}^+_l} \ra {\bar e}_i {\bar d}_j
d_k)|^2+\\
& & &|{\cal M}({{\tilde\chi}^+_l} \ra {\bar e}_i {\bar u}_j u_k)|^2+|{\cal
M}({{\tilde\chi}^+_l} \ra {\bar \nu}_i {\bar d}_j u_k)|^2)
dE_idE_j \\
\Gamma_{{\tilde \chi}^+_l} ({\bar U}{\bar D}{\bar D}) & = & \int \frac{1}{2\pi^3}\frac{1}{16
M_{{\tilde \chi}^+_l}}(&|{\cal M}({{\tilde\chi}^+_l} \ra {\bar d}_i {\bar d}_j {\bar
d}_k))|^2+|{\cal M}({{\tilde\chi}^+_l} \ra u_i u_j d_k)|^2+\\
&&&|{\cal
M}({{\tilde\chi}^+_l} \ra u_i d_j u_k)|^2)
dE_idE_j 
\end{array}
\label{gammacharg}
\earr

\clearpage

\section{Common Blocks}\label{commons}
In the following the common blocks are listed in  alphabetical order.

\begin{itemize}
\item COMMON/ CHAMIX/ OIJL(2,2), OIJR(2,2), V(2,2), U(2,2), FM(2), ETA(2)
\item COMMON/ CHANEU/ OIJLP(4,2), OIJRP(4,2)
\begin{description}
\item[OIJL, OIJR] :  chargino couplings to Z.
\item[V, U] :   diagonalising matrices of chargino..
\item[FM]  :  positive chargino masses.
\item[ETA] :  sign of diagonal values.
\item[OIJLP, OIJRP]:  chargino couplings to neutralinos.
\end{description}

\item COMMON/ CONST/ IDBG, IGENER, IRAD, FRAD: (Input cards DEBUG, GENER, ISR
and FSR)

\item COMMON/ COUPLS/ A,B: Coupling constants used in the \rp violating gaugino
decay calculation.

\item COMMON/ DECSEL/ IDECSEL(23),IPROCSEL(11): (Input cards DECSEL,PROCSEL)

\item COMMON/ DOTPROD/PP: Dot-products used in the \rp violating gaugino
decay calculation.

\item COMMON/ FINDEX/ FMPR1, FMPR2, XCROST, APRO
\begin{description}
\item[FMPR1, FMPR2] : masses of the 2 current produced sparticles.
\item[XCROST] : current cross section.
\item[APRO] : maximum differential amplitude.
\end{description}

\item COMMON/ GAUG2BDECS/ IJKGAUG(MAXGAUG2BODDECS,3),NGAUG
\item COMMON/ GAUG2BDECS2/ GIJKGAUG(MAXGAUG2BODDECS)
\begin{description}
\item[NGAUG] : Number of \rp conserving 2-body gaugino decay modes.
\item[IJKGAUG(n,i)] : i=1...3 gives the initial(i=1) and final state particles
(i=2-3) in the 2-body decay.
\item[GIJKGAUG] : Width of 2-particle decay.
\end{description}

\item COMMON/ GAUGM/ MASSNEU,MASSCHA : Neutralino and chargino mass used in current
\rp violating gaugino calculation.

\item COMMON/ HABKANE/XM(i,j) : Mixing matrix $N'_{ij}$ defined by \eq{eq:neut.diag.gh}.

\item  COMMON/ HIGGSES/MLH,MA,MHH,MHPM,SINA,COSA : Higgs masses and mixing angles.

\item COMMON/ INDEXX/ INDEX, INDEX1, INDEX2, NEVT 
\begin{description}
\item[INDEX] is an internal variable determining whether the program 
currently processes 
neutralino (INDEX=1), chargino (INDEX=2) sfermion (INDEX=3),
gravitino (INDEX=4), single sparticle production (INDEX=5), or
Higgs (INDEX=5).
\item[INDEX1, INDEX2] are the index code of the specific type of 2 
sparticles produced. 
\end{description}

\item COMMON/ INDICES/I,IJ,IJC,IJF,IJF\_MIX 
\begin{description}
\item[I(4)] : Input Card INDIC
\item[IJ(i=1..3,j=1..4)] : Decay product(i) of \rp violating neutralino(j) decay.
\item[IJC(i=1..3,j=1..2)] : Decay product(i) of \rp violating chargino(j) decay.
\item[IJF(n=1..68,c=1..2,fs=1..2)] : Decay products of direct \rp violating
sfermion decays. n=left-handed sfermion decaying into the final
state  fs=1..2. There are up to c=1..2 decay modes.
\item[IJF\_MIX] : Same as IJF, except that the n=right-handed sfermion.
\end{description}

\item      COMMON/ ISR/ QK(4) (4-vector of ISR $\gamma$).

\item COMMON/ KINEM/ FLUM, ECM, S, ROOTS, T, Q, Q2, EN(2)
\begin{description}
\item[FLUM, ECM, S, ROOTS]: ${\it L}, E_{CM}, s, \sqrt{s} $ (
 {\it L} is the luminosity, $E_{CM}$ is the nominal energy, {\it s} and 
{\it $\sqrt{s}$} change when initial state radiation is permitted).
\item[T, Q, Q2, EN]: ${\it t}, q, q^2$ and energies of 2 products.
\end{description}

\item COMMON/ LAMDA/XLAMA : \rp violating coupling strength $\lambda$.

\item COMMON/ LIFE/LIFETIME : Input Card LIFE.

\item COMMON/ LSUSY/XLSUSY : Spontaneous Supersymmetry breaking scale $\lambda_{SUSY}$. Used to
calculate the mass of the gravitino.

\item COMMON/ LUSHOWERS/LISTQMX
\item COMMON/ LUSHOWERS2/LISTSHOWERIND,NSHOWERS
\begin{description}
\item[LISTQMX(i),LISTSHOWERIND(i,j)] : Give the invariant mass and the index of
the particles(j) of the i-th shower which are to be processed  by JETSET routine LUSHOW.
\item[NSHOWERS] : Number of showers to process. 
\end{description}

\item COMMON/ MDS/MODES,MIX : Input Cards MODES and MIX.

\item COMMON/ MSSM/ TANB, SINB, COSB, FMGAUG, FMR, FM0, ATRI(3)
\begin{description}
\item[TANB, SINB, COSB]  : $\tan\beta=\upsilon_2/\upsilon_1$.
\item[FMGAUG] :  $M_2$ gaugino mass.
\item[FMR] :     Higgs mixing term $\mu$.
\item[FM0] :  common scalar mass $m_0$.
\item[ATRI] :  3 trilinear couplings for stop, sbottom and stau
respectively.
\end{description}

\item COMMON/ NEUMIX/ ZR(4,4), WAS(4), ESA(4), VOIJL(4,4), 
 VOIJR(4,4), GFIR(4,4), GFIL(4,4)
\begin{description}
\item[ZR] :  neutralino diagonalising matrix.
\item[ESA] :  phase factors of eigenvalues .
\item[WAS] :  absolute masses in ascending order.
\item[VOIJL, VOIJR] :  neutralino couplings   to Z.
\item[GFIL, GFIR] : neutralino couplings to $\ssup, \sdn, \snu, \sel$.
\end{description}

\item COMMON/ REORDER/ ISPA(12,2), KL(2,18), KLAP(2,18), IDECS(12,2) 
 (Correspondence matrices for decays and equivalence with LUND)
\begin{description}
\item[ISPA]: LUND codes of left/right sfermions (18 possible).
\item[KL]: standard particles accompanying a gaugino in 3-body decay .
\item[KLAP]: IDs of {\it t,u} exchanged particles.
\item[IDECS]: mapping of SUSYGEN standard particle convention to the  
 LUND ones.
\end{description}

\item  COMMON/ RESA/RES,RES1A,RES2A : 
\begin{description}
\item[RES] : Total Decay rate of \rp violating neutralino decay.
\item[RES1A,RES2A] : Partial Decay widths of \rp violating
neutralino decays to final states of type (1),(2).
\end{description}

\item COMMON/ RKEY/rgmaum, rgmaur, rgm0, rgtanb, rgatri, rgma, rfmsq, 
rfmstopl, rfmstopr, rfmsell, rfmselr, rfmsnu, rfmglu, recm, rflum
\begin{description}
\item[rgmaum,..,rfmglu] : The single precision version of the variables in the
COMMON STEER.
\item[recm,rflum,rvscan] : CM-Energy, Luminosity in single precision.
\end{description}

\item COMMON/ RPARI/RPAR,MGM
\begin{description}
\item[RPAR] : Logical, TRUE $\ra$ \rp is conserved.
\item[MGM] : Logical, TRUE $\ra$  Gauge Mediated SUSY gravitino decays enabled.
\end{description}

\item COMMON/ RWIDTHS/GIJ,GIJC,GIJF,GIJF\_MIX : Decay rates of the \rp
violating gaugino and sfermion decays. See also COMMON INDICES.

\item COMMON/ SDECAY/ DAS, DBS, DCS, DATL, DAUL, 
 DATUL, DASTL, DBSTL, DASUL, DBSUL, DATR, DAUR, 
 DATUR, DASTR, DBSTR, DASUR, DBSUR, XDEC(17,64)  
\item DIMENSION CURENT(17) 
\item EQUIVALENCE (CURENT(1),DAS)
\item DAS etc correspond to the 17 variables that characterise 
    gaugino 3-body decay. There are 64 different patterns;
\begin{enumerate}
\item 6x4 neutralino-neutralino        to uu, dd, vv, ll.
\item 8x4 chargino-neutralino          to ud, lv.
\item 1x8 chargino-chargino            uu,dd, vv, ll.
\end{enumerate}

\item COMMON/ SM/ FMW, FMZ, GAMMAZ, SINW, COSW, ALPHA, E2, G2, PI, TWOPI,
      FLC(12), FRC(12), GMS(12), ECHAR(12)
\begin{description}
\item[FMW] : W mass (80.2GeV).
\item[FMZ] : Z mass  (91.19GeV).
\item[GAMMAZ] : Z width (2.497GeV).
\item[SINW, COSW] : $\sin\theta_W (0.231243)$, $\cos\theta_W$ Weinberg.
\item[PI, ALPHA, e2,  G2] : $\pi$, $\frac{1}{128}$, $e^2=4\pi\alpha$, 
$g^2=\frac{e^2}{\sin^2\theta_{\mathrm W}}$.
\item[FLC] : $T_3-Q\sin^2{\theta_W}$, left SM couplings.
\item[FRC]:     $-Q\sin^2{\theta_W}$, right SM couplings.
\item[GMS]: masses of standard particles in  the following order
 (up, down, neutrino, electron) $\times 3$ for the three families.
\item[ECHAR]: charge of standard particles.
\end{description}

\item COMMON/ SPARC/ ZINO, WINO, SELE, SMUO, STAU, SNU, SQUA, STOPA, SBOTA 
(Input cards ZINO etc)

\item COMMON/ SPARTCL/ FMAL(12), FMAR(12), RATQA(12), FGAMC(12),   
 FGAMCR(12), COSMI(12), FM1(12), FM2(12) 
\begin{description}
\item[FMAL(12), FMAR(12)] : Left/right masses of sparticles.
\item[RATQA(12)] : charge.
\item[FGAMC(12), FGAMCR(12)] : left/right coupling to Z.
\item[COSMI(12)] : cosine of mixing angle ($\cos\phi$).
\item[FM1(12), FM2(12)] : masses of the 1,2 combinations for the 12 sfermions. 
\item All the above are in the following order: 
$\tilde{u}, \tilde{d}, \tilde{\nu}, \tilde{e}$ $\times 3$ generations.
\end{description}

\item COMMON/ SSCALE/RSCALE : Input Cards RS.

\item COMMON/ SSMODE/ NSSMOD, ISSMOD(MXSS), JSSMOD(5,MXSS)  
\item COMMON/ SSMOD1/ GSSMOD(MXSS), BSSMOD(MXSS)
(ISAJET type commons, for comparison)
\begin{description}
\item[MXSS]       : maximum number of modes.
\item[NSSMOD]     : number of modes.
\item[ISSMOD]     : initial particle.
\item[JSSMOD]     : final particles.
\item[GSSMOD]     : width.
\item[BSSMOD]     : branching ratio.
\end{description}

\item COMMON/ STEER/ GMAUM, GMAUR, GM0, GTANB, GATRI,
 FMSQ, FMSTOPL, FMSTOPR, FMSELL, FMSELR, FMSNU, FMGLU \\ 
(M, $\mu$, $m_0$, $\tan\beta$, A, $m_{\squ}$, $m_{\stp_L}$, $m_{\stp_R}$, 
$m_{\sell}$, $m_{\selr}$, $m_{\snu}$, $m_{\tilde{g}}$)

\item COMMON/ STOPMIX/STOP1,SBOT1,STAU1,STOP2,SBOT2,STAU2,
PHIMIX1,PHIMIX2,PHIMIX3 : Input cards MASMX

\item COMMON/ STR/ WRT, SCAN, LEPI (Input cards LUWRIT, SCAN, LEPI)

\item COMMON/ SUSY/TANTHW,COSB,SINB : $\tan{\theta_W}, \cos{\beta}, \sin{\beta}$,
where $\beta$ is the ratio of the neutral Higgs vevs.

\item COMMON/ TWOBODIES/ TWOB : Input cards TWOB

\item COMMON/ UBRA/ NDECA(-80:80)
\begin{description}
\item[NDECA] : number of decay nodes open to the particle with corresponding ID.
\end{description}
\item COMMON/ UBRA1/ BRTOT(2, 50, -80:80)
\begin{description}
\item[BRTOT] : i)pointer to SSMODE, ii)integrated branching ratios.
\end{description}

\item COMMON/ WIDTHS/ GW, WIDFL(12), WIDFR(12), FMELTW, FMERTW,
 FMELUW, FMERUW, LINDA(18,6,6) 
\begin{description} 
\item[GW] : The product of Z (or W) mass and width ($GeV^2$).  
\item[WIDFL, WIDFR] : Sparticle widths (GeV). 
\item[FMELTW/FMERTW, FMELUW/FMERUW] : The products of sparticle masses 
 and widths ($GeV^2$). 
\item[LINDA] : (Internal use). 
\end{description}

\item COMMON/ VARIABLES/ FMS, FMI, FMK, FML1, FML2, ETAI, ETAK,   
 BRSPA(6,48), LIND(6,6,6), BRGAUG(23,6,6), FMELT, FMERT, FMELU, FMERU \\ 
All variables relate to the R-parity conserving gaugino decays except for BRSPA,
 which relates to the R-parity conserving sfermion decays.
\begin{description} 
\item[FMS]:  is the s channel exchanged particle W or Z.
\item[FMI, ETAI] : is the mass,$\eta$ of the father.
\item[FMK, ETAK,] : is the mass,$\eta$ of the son.
\item[FML1, FML2] :  are the accompanying partons.
\item[FMELT/FELU, FMERT/FMERU] : are the sparticles exchanged in {\it u,t} 
 channels.
\end{description}
\begin{itemize}
\item LIND(i, j, k) are the permitted transitions;
\begin{description}
\item[i]: is one of the 6 patterns uu, dd, ll, vv, ud, lv.
\item[j]: is one of the son gauginos.
\item[k]: is one of the father gauginos.
\end{description}

\item BRGAUG(i, j, k) are the integrated differential widths;
\begin{description}
\item[i]: is one of the 18 patterns (uu, dd, vv, ll, ud, lv) $\times 3$ 
 generations.
\item[j]: is one of the son gauginos.
\item[k]: is one of the father gauginos.
\end{description}

\item BRSPA(i,j) are the partial sparticle widths;
\begin{description}
\item[i]: is the index of the daugther gaugino.
\item[j]: is the index of the father sfermion.
\end{description}  
\end{itemize}

\item COMMON/ XCROS/ XGAUG(8), XETA(8) 
\begin{description}
\item[XGAUG] are the masses of the 4 neutralinos and the 2 charginos $\chi^\pm_{1,2}$.
\item[XETA] the $\eta=\pm 1$ phase factors.
\end{description}

\end{itemize}

\clearpage

\section{Ntuple variables}
\begin{table}[htb]
\begin{center}    
\begin{tabular}{l|l||l|l||l|l}     
 Variable & Description& Variable & Description &Variable & Description \\
\hline
 fECM & $\sqrt{s}$    & fM2 & $M_2$      & fmu & $\mu$ \\
  fm0 & $m_0$  & ftanb & $\tan\beta$      & fA  & $A_t$ \\
  fmA & $m_A$ & & & & \\
\end{tabular} 
\end{center} 
\caption { Input parameters - block ``COL1''.}\label{ntp1}
\end{table}

\begin{table}[htb]
\begin{center}    
\begin{tabular}{l|l||l|l||l|l}     
 Variable & Description& Variable & Description& Variable & Description \\
\hline
 fx01 & $M(\chi^0_1)$  & fx02 & $M(\chi^0_2)$ & fx03 & $M(\chi^0_3)$ \\
 fx04 & $M(\chi^0_4)$ &
 fx1p & $M(\chi^+_1)$ & fx2p & $M(\chi^+_2)$ \\ 
 fupl & $M({\tilde u}_L)$ & fdownl & $M({\tilde d}_L)$ &
 fneu & $M({\tilde \nu})$ \\
 fel & $M({\tilde e}_L)$  &
 fupr & $M({\tilde u}_R)$ & fdownr & $M({\tilde d}_L)$ \\
 fer & $M({\tilde e}_R)$ & ftop1 & $M({\tilde t}_1)$ &
 fbot1 & $M({\tilde b}_1)$ \\
 ftau1 & $M({\tilde \tau}_1)$ &
 ftop2 & $M({\tilde t}_2)$& fbot2 & $M({\tilde b}_2)$ \\
 ftau2 & $M({\tilde \tau}_2)$ & fmh1  & $M(h)$  &
 fmh2  & $M(H)$ \\
 fmhpc & $M(H^\pm)$ & & & & 
\end{tabular} 
\end{center} 
\caption { Masses - block ``COL1''.}\label{ntp2}
\end{table}

\begin{table}[htb]
\begin{center}    
\begin{tabular}{l|l||l|l||l|l}     
 Variable & Description& Variable & Description& Variable & Description \\ \hline
 zx01a & $N_{11}$ &
 zx01b & $N_{21}$ &
 zx01c & $N_{31}$ \\
 zx01d & $N_{41}$ &
 zx02a & $N_{12}$ &
 zx02b & $N_{22}$ \\
 zx02c & $N_{32}$ &
 zx02d & $N_{42}$ &
 zx03a & $N_{13}$ \\
 zx03b & $N_{23}$ &
 zx03c & $N_{33}$ &
 zx03d & $N_{43}$ \\
 zx04a & $N_{14}$ &
 zx04b & $N_{24}$ &
 zx04c & $N_{34}$ \\
 zx04d & $N_{44}$ &
 u11 & $U_{11}$ &
 u12 & $U_{12}$ \\
 v11 & $V_{11}$ &
 v12 & $V_{12}$ & & 
\end{tabular} 
\end{center} 
\caption {Neutralino/Chargino eigenvectors - block ``COL3''.}\label{ntp4}
\end{table} 

\begin{table}[htb]
\begin{center}    
\begin{tabular}{l|l||l|l||l|l}     
 Variable & Description& Variable & Description& Variable & Description \\ \hline
 sx1x1 & $\sigma(\chi^0_1 \chi^0_1)$ & sx1x2 & $\sigma(\chi^0_1 \chi^0_2)$ &
         sx2x2 & $\sigma(\chi^0_2 \chi^0_2)$ \\
         sx1x3 & $\sigma(\chi^0_1 \chi^0_3)$ &
         sx2x3 & $\sigma(\chi^0_2 \chi^0_3)$ &
         sx3x3 & $\sigma(\chi^0_3 \chi^0_3)$ \\
         sx1x4 & $\sigma(\chi^0_1 \chi^0_4)$ &
         sx2x4 & $\sigma(\chi^0_2 \chi^0_4)$ &
         sx3x4 & $\sigma(\chi^0_3 \chi^0_4)$ \\
         sx4x4 & $\sigma(\chi^0_4 \chi^0_4)$ &
         sx1x1p & $\sigma(\chi^+_1 \chi^+_1)$ &
         sx1x2p & $\sigma(\chi^+_1 \chi^+_2)$ \\
         sx2x2p & $\sigma(\chi^+_2 \chi^+_2)$ &
         snue   & $\sigma({\tilde \nu}_e {\tilde \nu}_e)$  &
         snutot & $\sigma({\tilde \nu}_e {\tilde \nu}_e)$ + $\sigma({\tilde
 \nu}_\mu {\tilde \nu}_\mu)$ \\
        & & & & & + $\sigma({\tilde \nu}_\tau {\tilde \nu}_\tau)$ \\
         ssel & $\sigma({\tilde e}_L {\tilde e}_L)$  &
         sser & $\sigma({\tilde e}_R {\tilde e}_R)$  &
         sselser & $\sigma({\tilde e}_L {\tilde e}_R)$  \\
         smul & $\sigma({\tilde \mu}_L {\tilde \mu}_L)$  &
         smur & $\sigma({\tilde \mu}_R {\tilde \mu}_R)$  &
         stau1 & $\sigma({\tilde \tau}_1 {\tilde \tau}_1)$  \\
         stau2 & $\sigma({\tilde \tau}_2 {\tilde \tau}_2)$  &
         sbt1 & $\sigma({\tilde b}_1 {\tilde b}_1)$  &
         sbt2 & $\sigma({\tilde b}_2 {\tilde b}_2)$  \\
         stp1 & $\sigma({\tilde t}_1 {\tilde t}_1)$  &
         stp2 & $\sigma({\tilde t}_2 {\tilde t}_2)$  &
         sqlr & $\sum_{i=1}^{4} \sigma({\tilde q^i}_L {\tilde
 q^i}_L)+\sigma({\tilde q^i}_R {\tilde q^i}_R)$  \\
         sgx1 & $\sigma(\chi^0_1 {\tilde G})$ &
         sgx2 & $\sigma(\chi^0_2 {\tilde G})$ &
         sgx3 & $\sigma(\chi^0_3 {\tilde G})$ \\
         sgx4 & $\sigma(\chi^0_4 {\tilde G})$ &
         srpvx1 & $\sigma(\chi^0_1 { \nu})$ &
         srpvx2 & $\sigma(\chi^0_2 { \nu})$ \\
         srpvx3 & $\sigma(\chi^0_3 { \nu})$ &
         srpvx4 & $\sigma(\chi^0_4 { \nu})$ &
         srpvx1p & $\sigma(\chi^+_1 {l^-})$ \\
         srpvx2p & $\sigma(\chi^+_2 {l^-})$ &
         szh1 & $\sigma(hZ)$ &
         szh2  & $\sigma(HZ)$ \\
         sah1 & $\sigma(hA)$ &
         sah2 & $\sigma(HA)$ &
         shphp & $\sigma(H^+ H^-)$ 
\end{tabular} 
\end{center} 
\caption {  Cross sections in pb - block ``COL2''.}\label{ntp3}
\end{table}

\begin{table}[htb]
\begin{center}    
\begin{tabular}{l|l||l|l||l|l}     
 Variable & Description& Variable & Description& Variable & Description \\ \hline
 sx01 & $\Gamma_{R_p}(\chi^0_1)$ &
 sx02 & $\Gamma_{R_p}(\chi^0_2)$ &
 sx03 & $\Gamma_{R_p}(\chi^0_3)$ \\
 sx04 & $\Gamma_{R_p}(\chi^0_4)$ &
 sxp1 & $\Gamma_{R_p}(\chi^+_1)$ &
 sxp2 & $\Gamma_{R_p}(\chi^+_2)$ \\
 sul  & $\Gamma_{R_p}({\tilde u}_L)$ &
 sdl    & $\Gamma_{R_p}({\tilde d}_L)$ &
 snel  & $\Gamma_{R_p}({\tilde \nu}_e)$ \\
 sel  & $\Gamma_{R_p}({\tilde e}_L)$ &
 scl  & $\Gamma_{R_p}({\tilde c}_L)$ &
 ssl  & $\Gamma_{R_p}({\tilde s}_L)$ \\
 snmel  & $\Gamma_{R_p}({\tilde \nu}_\mu)$ &
 sml  & $\Gamma_{R_p}({\tilde \mu}_L)$ &
 stl  & $\Gamma_{R_p}({\tilde t}_1)$ \\
 sbl  & $\Gamma_{R_p}({\tilde b}_1)$ &
 sntl  & $\Gamma_{R_p}({\tilde \nu}_\tau)$ &
 stal  & $\Gamma_{R_p}({\tilde \tau}_1)$ \\
 sur  & $\Gamma_{R_p}({\tilde u}_R)$ &
 sdr  & $\Gamma_{R_p}({\tilde d}_R)$ &
 sner  & -  \\
 ser  & $\Gamma_{R_p}({\tilde e}_R)$ &
 scr  & $\Gamma_{R_p}({\tilde c}_R)$ &
 ssr  & $\Gamma_{R_p}({\tilde s}_R)$ \\
 snmer  & -  &
 smr  & $\Gamma_{R_p}({\tilde \mu}_R)$ &
 str  & $\Gamma_{R_p}({\tilde t}_2)$ \\
 sbr  & $\Gamma_{R_p}({\tilde b}_2)$ &
 sntr  & -  &
 star  & $\Gamma_{R_p}({\tilde \tau}_2$
\end{tabular} 
\end{center} 
\caption {R-parity conserving widths ($\Gamma_{R_p}$) of sparticles in GeV - block ``COL4''. Does not include GMSB decays to the gravitino.}\label{ntp5}
\end{table}

\begin{table}[htb]
\begin{center}    
\begin{tabular}{l|l||l|l||l|l}     
 Variable & Description& Variable & Description& Variable & Description \\
 \hline
 rx01 & $\Gamma_{\rpvm}(\chi^0_1)$ &
 rx02& $\Gamma_{\rpvm}(\chi^0_2)$ &
 rx03& $\Gamma_{\rpvm}(\chi^0_3)$ \\
 rx04& $\Gamma_{\rpvm}(\chi^0_4)$ &
 rxp1& $\Gamma_{\rpvm}(\chi^+_1)$ &
 rxp2& $\Gamma_{\rpvm}(\chi^+_2)$ \\
 rul& $\Gamma_{\rpvm}({\tilde u}_L)$ &
 rdl& $\Gamma_{\rpvm}({\tilde d}_L)$ &
 rnel& $\Gamma_{\rpvm}({\tilde \nu}_e)$ \\
 rel& $\Gamma_{\rpvm}({\tilde e}_L)$ &
 rcl& $\Gamma_{\rpvm}({\tilde c}_L)$ &
 rsl& $\Gamma_{\rpvm}({\tilde s}_L)$ \\
 rnmel& $\Gamma_{\rpvm}({\tilde \nu}_\mu)$ &
 rml& $\Gamma_{\rpvm}({\tilde \mu}_L)$ &
 rtl& $\Gamma_{\rpvm}({\tilde t}_1)$ \\
 rbl& $\Gamma_{\rpvm}({\tilde b}_1)$ &
 rntl& $\Gamma_{\rpvm}({\tilde \nu}_\tau)$ &
 rtal& $\Gamma_{\rpvm}({\tilde \tau}_1)$ \\
 rur& $\Gamma_{\rpvm}({\tilde u}_R)$ &
 rdr& $\Gamma_{\rpvm}({\tilde d}_R)$ &
 rner& -  \\
 rer& $\Gamma_{\rpvm}({\tilde e}_R)$ &
 rcr& $\Gamma_{\rpvm}({\tilde c}_R)$ &
 rsr& $\Gamma_{\rpvm}({\tilde s}_R)$ \\
 rnmer& -  &
 rmr& $\Gamma_{\rpvm}({\tilde \mu}_R)$ &
 rtr& $\Gamma_{\rpvm}({\tilde t}_2)$ \\
 rbr& $\Gamma_{\rpvm}({\tilde b}_2)$ &
 rntr& -  &
 rtar& $\Gamma_{\rpvm}({\tilde \tau}_2)$ 
\end{tabular}
\end{center} 
\caption {R-parity violating widths ($\Gamma_{\rpvm}$) of sparticles in GeV -  block ``COL5''.} \label{ntp6}
\end{table}

\begin{table}[htb]
\begin{center}    
\begin{tabular}{l|l||l|l||l|l}     
 Variable & Description& Variable & Description& Variable & Description \\
 \hline
 rx01 & $\Gamma_{GMSB}(\chi^0_1)$ &
 rx02& $\Gamma_{GMSB}(\chi^0_2)$ &
 rx03& $\Gamma_{GMSB}(\chi^0_3)$ \\
 rx04& $\Gamma_{GMSB}(\chi^0_4)$ &
 rxp1& $\Gamma_{GMSB}(\chi^+_1)$ &
 rxp2& $\Gamma_{GMSB}(\chi^+_2)$ \\
 rul& $\Gamma_{GMSB}({\tilde u}_L)$ &
 rdl& $\Gamma_{GMSB}({\tilde d}_L)$ &
 rnel& $\Gamma_{GMSB}({\tilde \nu}_e)$ \\
 rel& $\Gamma_{GMSB}({\tilde e}_L)$ &
 rcl& $\Gamma_{GMSB}({\tilde c}_L)$ &
 rsl& $\Gamma_{GMSB}({\tilde s}_L)$ \\
 rnmel& $\Gamma_{GMSB}({\tilde \nu}_\mu)$ &
 rml& $\Gamma_{GMSB}({\tilde \mu}_L)$ &
 rtl& $\Gamma_{GMSB}({\tilde t}_1)$ \\
 rbl& $\Gamma_{GMSB}({\tilde b}_1)$ &
 rntl& $\Gamma_{GMSB}({\tilde \nu}_\tau)$ &
 rtal& $\Gamma_{GMSB}({\tilde \tau}_1)$ \\
 rur& $\Gamma_{GMSB}({\tilde u}_R)$ &
 rdr& $\Gamma_{GMSB}({\tilde d}_R)$ &
 rner& -  \\
 rer& $\Gamma_{GMSB}({\tilde e}_R)$ &
 rcr& $\Gamma_{GMSB}({\tilde c}_R)$ &
 rsr& $\Gamma_{GMSB}({\tilde s}_R)$ \\
 rnmer& -  &
 rmr& $\Gamma_{GMSB}({\tilde \mu}_R)$ &
 rtr& $\Gamma_{GMSB}({\tilde t}_2)$ \\
 rbr& $\Gamma_{GMSB}({\tilde b}_2)$ &
 rntr& -  &
 rtar& $\Gamma_{GMSB}({\tilde \tau}_2)$
\end{tabular}
\end{center} 
\caption {GMSB widths ($\Gamma_{R_{GMSB}}$) of sparticles in GeV -  block ``COL6''.} \label{ntp7}
\end{table}

\begin{table}[htb]
\begin{center}    
\begin{tabular}{l|l||l|l||l|l}     
 Variable & Description& Variable & Description& Variable & Description \\
 \hline
 rx01 & $\Gamma(\chi^0_1)$ &
 rx02& $\Gamma(\chi^0_2)$ &
 rx03& $\Gamma(\chi^0_3)$ \\
 rx04& $\Gamma(\chi^0_4)$ &
 rxp1& $\Gamma(\chi^+_1)$ &
 rxp2& $\Gamma(\chi^+_2)$ \\
 rul& $\Gamma({\tilde u}_L)$ &
 rdl& $\Gamma({\tilde d}_L)$ &
 rnel& $\Gamma({\tilde \nu}_e)$ \\
 rel& $\Gamma({\tilde e}_L)$ &
 rcl& $\Gamma({\tilde c}_L)$ &
 rsl& $\Gamma({\tilde s}_L)$ \\
 rnmel& $\Gamma({\tilde \nu}_\mu)$ &
 rml& $\Gamma({\tilde \mu}_L)$ &
 rtl& $\Gamma({\tilde t}_1)$ \\
 rbl& $\Gamma({\tilde b}_1)$ &
 rntl& $\Gamma({\tilde \nu}_\tau)$ &
 rtal& $\Gamma({\tilde \tau}_1)$ \\
 rur& $\Gamma({\tilde u}_R)$ &
 rdr& $\Gamma({\tilde d}_R)$ &
 rner& -  \\
 rer& $\Gamma({\tilde e}_R)$ &
 rcr& $\Gamma({\tilde c}_R)$ &
 rsr& $\Gamma({\tilde s}_R)$ \\
 rnmer& -  &
 rmr& $\Gamma({\tilde \mu}_R)$ &
 rtr& $\Gamma({\tilde t}_2)$ \\
 rbr& $\Gamma({\tilde b}_2)$ &
 rntr& -  &
 rtar& $\Gamma({\tilde \tau}_2)$
\end{tabular}
\end{center} 
\caption {Total  widths ($\Gamma =
 \Gamma_{R_{p}}+\Gamma_{\rpvm}+\Gamma_{R_{GMSB}}$) of sparticles in GeV - block
 ``COL7''. This block is only present in the GMSB or $\rpvm$ modes.}\label{ntp8}
\end{table}

\begin{table}[htb]
\begin{center}    
\begin{tabular}{l|l||l|l}     
 Variable & Description& Variable & Description \\
 \hline
 brupx01 & $d\Gamma({\tilde u}_L \ra u \chi^0_1)$   &
 brupx02  & $d\Gamma({\tilde u}_L \ra u \chi^0_2)$ \\
 brupxp1  & $d\Gamma({\tilde u}_L \ra d \chi^+_1)$ &
 brdownx01 & $d\Gamma({\tilde d}_L \ra d \chi^0_1)$ \\
 brdownx02 & $d\Gamma({\tilde d}_L \ra d \chi^0_2)$ &
 brdownxp1 & $d\Gamma({\tilde d}_L \ra u \chi^-_1)$ \\
 brneux01 & $d\Gamma({\tilde \nu}_e \ra \nu_e \chi^0_1)$ &
 brneux02 & $d\Gamma({\tilde \nu}_e \ra \nu_e \chi^0_1)$ \\
 brneuxp1 & $d\Gamma({\tilde \nu}_e \ra e^\mp \chi^\pm_1)$ &
 brelx01 & $d\Gamma({\tilde e}_L \ra e \chi^0_1)$ \\
 brelx02& $d\Gamma({\tilde e}_L \ra e \chi^0_2)$ &
 brelxp1& $d\Gamma({\tilde e}_L \ra \nu_e \chi^+_1)$ \\
 brstx01& $d\Gamma({\tilde t}_1 \ra c (or t) \chi^0_1)$ &
 brstx02& $d\Gamma({\tilde t}_1 \ra c (or t) \chi^0_1)$ \\
 brstxp1& $d\Gamma({\tilde t}_1 \ra b \chi^+_1)$ &
 brbotx01& $d\Gamma({\tilde b}_1 \ra b \chi^0_1)$ \\
 brbotx02& $d\Gamma({\tilde b}_1 \ra b \chi^0_2)$ &
 brbotxp1& $d\Gamma({\tilde b}_1 \ra c (or t) \chi^-_2)$ \\
 brntx01 & $d\Gamma({\tilde \nu}_\tau \ra \nu_\tau \chi^0_1)$ &
 brntx02 & $d\Gamma({\tilde \nu}_\tau \ra \nu_\tau \chi^0_2)$ \\
 brntxp1 & $d\Gamma({\tilde \nu}_\tau \ra \tau^\mp \chi^\pm_1)$ &
 brtaux01 & $d\Gamma({\tilde \tau}_1 \ra \tau \chi^0_1)$ \\
 brtaux02& $d\Gamma({\tilde \tau}_1 \ra \tau \chi^0_1)$ & 
 brtauxp1& $d\Gamma({\tilde \tau}_1 \ra \nu_\tau \chi^+_1)$ 
\end{tabular}
\end{center} 
\caption {R-parity conserving partial widths in GeV - left handed
  sfermions. Block ``COL8''.}\label{ntp9}
\end{table}

\begin{table}[htb]
\begin{center}    
\begin{tabular}{l|l||l|l}     
 Variable & Description& Variable & Description \\
 \hline
 brupx01r & $d\Gamma({\tilde u}_R \ra u \chi^0_1)$   &
 brupx02r& $d\Gamma({\tilde u}_R \ra u \chi^0_2)$ \\
 brupxp1r& $d\Gamma({\tilde u}_R \ra d \chi^+_1)$ &
 brdownx01r& $d\Gamma({\tilde d}_R \ra d \chi^0_1)$ \\
 brdownx02r& $d\Gamma({\tilde d}_R \ra d \chi^0_2)$ &
 brdownxp1r& $d\Gamma({\tilde d}_R \ra u \chi^-_1)$ \\
 brelx01r & $d\Gamma({\tilde e}_R \ra e \chi^0_1)$ &
 brelx02r & $d\Gamma({\tilde e}_R \ra e \chi^0_2)$ \\
 brelxp1r & $d\Gamma({\tilde e}_R \ra \nu_e \chi^+_1)$ &
 brstx01r & $d\Gamma({\tilde t}_2 \ra c (or t) \chi^0_1)$ \\
 brstx02r & $d\Gamma({\tilde t}_2 \ra c (or t) \chi^0_1)$ &
 brstxp1r & $d\Gamma({\tilde t}_2 \ra b \chi^+_1)$ \\
 brbotx01r & $d\Gamma({\tilde b}_2 \ra b \chi^0_1)$ &
 brbotx02r & $d\Gamma({\tilde b}_2 \ra b \chi^0_2)$ \\
 brbotxp1r & $d\Gamma({\tilde b}_2 \ra c (or t) \chi^-_2)$ &
 brtaux01r& $d\Gamma({\tilde \tau}_2 \ra \tau \chi^0_1)$ \\
 brtaux02r& $d\Gamma({\tilde \tau}_2 \ra \tau \chi^0_1)$ &
 brtauxp1r& $d\Gamma({\tilde \tau}_2 \ra \nu_\tau \chi^+_1)$
\end{tabular}
\end{center} 
\caption {R-parity conserving partial widths in GeV - right handed
  sfermions. Block ``COL8''.}\label{ntp10}
\end{table}

\begin{table}[htb]
\begin{center}    
\begin{tabular}{l|l||l|l}     
 Variable & Description& Variable & Description \\
 \hline
 brx2a & $d\Gamma({\chi^0_2 \ra q {\bar q} \chi^0_1})$  &
 brx2b& $d\Gamma({\chi^0_2 \ra e^+ e^- \chi^0_1})$ \\
 brx2c& $d\Gamma({\chi^0_2 \ra \mu^+ \mu^- \chi^0_1})$ &
 brx2d& $d\Gamma({\chi^0_2 \ra \tau^+ \tau^- \chi^0_1})$ \\
 brx2e& $d\Gamma({\chi^0_2 \ra \nu {\bar \nu} \chi^0_1})$ &
 brx2f& $d\Gamma({\chi^0_2 \ra \gamma \chi^0_1})$ \\
 brx2g& $d\Gamma({\chi^0_2 \ra h (or H) \chi^0_1})$ &
 brx2h& $d\Gamma({\chi^0_2 \ra H^\mp \chi^\pm_1})$ \\
 brxp1a& $d\Gamma({\chi^+_1 \ra q q' \chi^0_1})$ &
 brxp1b& $d\Gamma({\chi^+_1 \ra e \nu \chi^0_1})$ \\
 brxp1c& $d\Gamma({\chi^+_1 \ra \mu \nu \chi^0_1})$ &
 brxp1d& $d\Gamma({\chi^+_1 \ra \tau \nu \chi^0_1})$ \\
 brxp1e& $d\Gamma({\chi^+_1 \ra H^+ \chi^0_1})$ &
 brxp1f& -  \\
 brxp1g& - &
 brxp1h& - 
\end{tabular}
\end{center} 
\caption {R-parity conserving partial widths in GeV - Charginos and Neutralinos. Block ``COL9''.}\label{ntp11}
\end{table}

\begin{table}[htb] 
\begin{center}    
\begin{tabular}{l|l||l|l||l|l}     
 Variable & Description& Variable & Description& Variable & Description \\
 \hline
 ex1x1 & $\epsilon(\chi^0_1 \chi^0_1)$ &
 ex1x2 & $\epsilon(\chi^0_1 \chi^0_2)$ &
 ex2x2 & $\epsilon(\chi^0_2 \chi^0_2)$ \\
 ex1x3 & $\epsilon(\chi^0_1 \chi^0_3)$ &
 ex2x3 & $\epsilon(\chi^0_2 \chi^0_3)$ &
 ex3x3 & $\epsilon(\chi^0_3 \chi^0_3)$ \\
 ex1x4 & $\epsilon(\chi^0_1 \chi^0_4)$ &
 ex2x4 & $\epsilon(\chi^0_2 \chi^0_4)$ &
 ex3x4 & $\epsilon(\chi^0_3 \chi^0_4)$ \\
 ex4x4 & $\epsilon(\chi^0_4 \chi^0_4)$ &
 ex1x1p & $\epsilon(\chi^+_1 \chi^+_1)$ &
 ex1x2p & $\epsilon(\chi^+_1 \chi^+_2)$ \\
 ex2x2p & $\epsilon(\chi^+_2 \chi^+_2)$ &
 enue & $\epsilon({\tilde \nu}_e {\tilde \nu}_e)$  &
 enutot & $\epsilon({\tilde \nu}_e {\tilde \nu}_e + {\tilde  \nu}_\mu {\tilde
 \nu}_\mu$  \\ 
 & & & & & $+{\tilde \nu}_\tau {\tilde \nu}_\tau)$ \\
 esel & $\epsilon({\tilde e}_L {\tilde e}_L)$  &
 eser & $\epsilon({\tilde e}_R {\tilde e}_R)$  &
 eselser & $\epsilon({\tilde e}_L {\tilde e}_R)$  \\
 emul & $\epsilon({\tilde \mu}_L {\tilde \mu}_L)$  &
 emur & $\epsilon({\tilde \mu}_R {\tilde \mu}_R)$  &
 etau1 & $\epsilon({\tilde \tau}_1 {\tilde \tau}_1)$  \\
 etau2 & $\epsilon({\tilde \tau}_2 {\tilde \tau}_2)$  &
 ebt1 & $\epsilon({\tilde b}_1 {\tilde b}_1)$  &
 ebt2 & $\epsilon({\tilde b}_2 {\tilde b}_2)$  \\
 etp1 & $\epsilon({\tilde t}_1 {\tilde t}_1)$  &
 etp2 & $\epsilon({\tilde t}_2 {\tilde t}_2)$  &
 eupl & $\epsilon({\tilde u}_L {\tilde u}_L)$ \\
 edowl & $\epsilon({\tilde d}_L {\tilde d}_L)$ &
 echal& $\epsilon({\tilde c}_L {\tilde c}_L)$ &
 estl& $\epsilon({\tilde s}_L {\tilde s}_L)$ \\
 eupr& $\epsilon({\tilde u}_R {\tilde u}_R)$ &
 edowr& $\epsilon({\tilde d}_R {\tilde d}_R)$ &
 echar& $\epsilon({\tilde c}_R {\tilde c}_R)$ \\
 estr& $\epsilon({\tilde s}_R {\tilde s}_R)$ &
 egx1 & $\epsilon(\chi^0_1 {\tilde G})$ &
 egx2 & $\epsilon(\chi^0_2 {\tilde G})$ \\
 egx3 & $\epsilon(\chi^0_3 {\tilde G})$ &
 egx4 & $\epsilon(\chi^0_4 {\tilde G})$ &
 erpvx1 & $\epsilon(\chi^0_1 { \nu})$ \\
 erpvx2 & $\epsilon(\chi^0_2 { \nu})$ &
 erpvx3 & $\epsilon(\chi^0_3 { \nu})$ &
 erpvx4 & $\epsilon(\chi^0_4 { \nu})$ \\
 erpvx1p & $\epsilon(\chi^+_1 {l^-})$ &
 erpvx2p & $\epsilon(\chi^+_2 {l^-})$ &
 ezh1 & $\epsilon(hZ)$ \\
 ezh2 & $\epsilon(HZ)$ &
 eah1 & $\epsilon(hA)$ &
 eah2 & $\epsilon(HA)$ \\
 ehphp & $\epsilon(H^+ H^- )$  & & & 
\end{tabular}
\end{center} 
\caption {Efficiencies (calculated in routine USER) for the various signals -
 block ``COL10''.}\label{ntp12}
\end{table}

\clearpage

\section{Test Run Input}

{\footnotesize\begin{verbatim}
**** SUGRA mode
MODES 1
M 80.
mu -50.
m0 100.
tanb 1.41
****
ECM 183.
* initial state radiation ON
ISR 1
* list events
DEBUG 1
* write o/p to fortran file unit 12
GENER 1
LUWRIT TRUE
**** generate one chargino pair event
WINO TRUE
SUSEVENTS 1.
END
\end{verbatim}}

\section{Test Run Output}
\subsection{The susygen.dat file}
{\footnotesize\begin{verbatim}
 INPUTS:
 M       =    80.000   mu       =   -50.000
 m0      =   100.000   TANB     =     1.410
 At      =     0.000   Ab       =     0.000 Atau =     0.000
 Ecm     =   183.000  EVENTS    =     1.000 RAD CORR=  1

  Sparticle masses 

 SUPR      253.  SUPL       260.

 SDNR      253.  SDNL       263.

 SELR      110.  SELL       125.

 SNU       117.

 STPL      312.  STPR       307.

 SBTL      263.  SBTR       253.

 STAL      125.  STAR       110.


 M1 =     40.107 M2 =     80.000

 NEUTRALINO m, CP, ph/zi/ha/hb 1 =   45.0  1.  0.755 -0.196 -0.083 -0.620
 NEUTRALINO m, CP, ph/zi/ha/hb 2 =   49.7  1.  0.625  0.014 -0.118  0.772
 NEUTRALINO m, CP, ph/zi/ha/hb 3 =   98.9 -1.  0.054 -0.476  0.872  0.099
 NEUTRALINO m, CP, ph/zi/ha/hb 4 =  124.4  1.  0.192  0.857  0.467 -0.100

 CHARGINO MASSES    =    82.205   122.500
 CHARGINO ETA      =     1.000     1.000

  U matrix   WINO    HIGGSINO 
 W1SS+        -0.076     0.997
 W2SS+         0.997     0.076

  V matrix   WINO    HIGGSINO 
 W1SS-         0.722    -0.692
 W2SS-         0.692     0.722

  
  
  HIGGS masses 
  

 Light CP-even Higgs =    56.372
 Heavy CP-even Higgs =   313.893
       CP-odd  Higgs =   300.000
       Charged Higgs =   310.070
       sin(a-b)      =    -0.613
       cos(a-b)      =     0.790

 PARENT -->     DAUGHTERS              WIDTH (eV)        BRANCHING RATIO

 MASS of Z1SS         =    44.9614410400391      GeV 
  
 Z2SS   -->  Z1SS   UP     UB                    0.019      2.12
 Z2SS   -->  Z1SS   DN     DB                    0.024      2.73
 Z2SS   -->  Z1SS   NUE    ANUE                  0.011      1.21
 Z2SS   -->  Z1SS   CH     CB                    0.002      0.25
 Z2SS   -->  Z1SS   ST     SB                    0.022      2.48
 Z2SS   -->  Z1SS   NUM    ANUM                  0.011      1.21
 Z2SS   -->  Z1SS   NUT    ANUT                  0.011      1.21
 Z2SS   -->  Z1SS   GAMMA                        0.785     88.79
  
 MASS of Z2SS         =    49.7181472778320      GeV 
 Total WIDTH of Z2SS  =   8.838237113852746E-010 GeV 
 Lifetime             =   2.2289512E-05 cm  
  
 Z3SS   -->  Z1SS   UP     UB                 7815.816      4.86
 Z3SS   -->  Z1SS   DN     DB                10375.482      6.45
 Z3SS   -->  Z1SS   NUE    ANUE               6575.981      4.09
 Z3SS   -->  Z1SS   E-     E+                 3426.354      2.13
 Z3SS   -->  Z1SS   CH     CB                 7759.725      4.82
 Z3SS   -->  Z1SS   ST     SB                10372.120      6.44
 Z3SS   -->  Z1SS   NUM    ANUM               6575.981      4.09
 Z3SS   -->  Z1SS   MU-    MU+                3426.271      2.13
 Z3SS   -->  Z1SS   BT     BB                 9638.576      5.99
 Z3SS   -->  Z1SS   NUT    ANUT               6575.981      4.09
 Z3SS   -->  Z1SS   TAU-   TAU+               3393.080      2.11
 Z3SS   -->  Z1SS   GAMMA                       16.500      0.01
 Z3SS   -->  Z2SS   UP     UB                 9585.074      5.95
 Z3SS   -->  Z2SS   DN     DB                11596.827      7.20
 Z3SS   -->  Z2SS   NUE    ANUE               5487.067      3.41
 Z3SS   -->  Z2SS   E-     E+                 4263.929      2.65
 Z3SS   -->  Z2SS   CH     CB                 9491.563      5.90
 Z3SS   -->  Z2SS   ST     SB                11592.267      7.20
 Z3SS   -->  Z2SS   NUM    ANUM               5487.067      3.41
 Z3SS   -->  Z2SS   MU-    MU+                4263.800      2.65
 Z3SS   -->  Z2SS   BT     BB                10586.754      6.58
 Z3SS   -->  Z2SS   NUT    ANUT               5487.067      3.41
 Z3SS   -->  Z2SS   TAU-   TAU+               4210.220      2.62
 Z3SS   -->  Z2SS   GAMMA                      958.936      0.60
 Z3SS   -->  W1SS+  DN     UB                  549.331      0.34
 Z3SS   -->  W1SS+  E-     ANUE                318.880      0.20
 Z3SS   -->  W1SS+  ST     CB                  515.701      0.32
 Z3SS   -->  W1SS+  MU-    ANUM                318.860      0.20
 Z3SS   -->  W1SS+  TAU-   ANUT                298.372      0.19
  
 MASS of Z3SS         =    98.9230117797852      GeV 
 Total WIDTH of Z3SS  =   1.609635813320035E-004 GeV 
 Lifetime             =   1.2238793E-10 cm  
  
 Z4SS   -->  Z1SS   UP     UB                64341.051      0.20
 Z4SS   -->  Z1SS   DN     DB                71851.289      0.23
 Z4SS   -->  Z1SS   NUE    ANUE             799706.625      2.53
 Z4SS   -->  Z1SS   E-     E+              1257176.125      3.98
 Z4SS   -->  Z1SS   CH     CB                64144.719      0.20
 Z4SS   -->  Z1SS   ST     SB                71837.734      0.23
 Z4SS   -->  Z1SS   NUM    ANUM             799710.813      2.53
 Z4SS   -->  Z1SS   MU-    MU+             1257128.750      3.98
 Z4SS   -->  Z1SS   BT     BB                70013.922      0.22
 Z4SS   -->  Z1SS   NUT    ANUT             800902.125      2.54
 Z4SS   -->  Z1SS   TAU-   TAU+            1239106.500      3.93
 Z4SS   -->  Z1SS   H0L                   11142074.000     35.31
 Z4SS   -->  Z2SS   UP     UB                70878.781      0.22
 Z4SS   -->  Z2SS   DN     DB               113081.391      0.36
 Z4SS   -->  Z2SS   NUE    ANUE              47392.527      0.15
 Z4SS   -->  Z2SS   E-     E+               651443.563      2.06
 Z4SS   -->  Z2SS   CH     CB                70611.727      0.22
 Z4SS   -->  Z2SS   ST     SB               113055.344      0.36
 Z4SS   -->  Z2SS   NUM    ANUM              47392.547      0.15
 Z4SS   -->  Z2SS   MU-    MU+              651432.125      2.06
 Z4SS   -->  Z2SS   BT     BB               109480.492      0.35
 Z4SS   -->  Z2SS   NUT    ANUT              47398.039      0.15
 Z4SS   -->  Z2SS   TAU-   TAU+             647318.063      2.05
 Z4SS   -->  Z2SS   H0L                    1357764.000      4.30
 Z4SS   -->  Z3SS   NUE    ANUE             500639.094      1.59
 Z4SS   -->  Z3SS   E-     E+                 8530.695      0.03
 Z4SS   -->  Z3SS   NUM    ANUM             500641.719      1.59
 Z4SS   -->  Z3SS   MU-    MU+                8529.966      0.03
 Z4SS   -->  Z3SS   NUT    ANUT             501390.094      1.59
 Z4SS   -->  Z3SS   TAU-   TAU+               8321.920      0.03
 Z4SS   -->  W1SS+  DN     UB                26603.908      0.08
 Z4SS   -->  W1SS+  E-     ANUE            2805715.750      8.89
 Z4SS   -->  W1SS+  ST     CB                26414.109      0.08
 Z4SS   -->  W1SS+  MU-    ANUM            2805695.500      8.89
 Z4SS   -->  W1SS+  TAU-   ANUT            2799321.750      8.87
  
 MASS of Z4SS         =    124.350257873535      GeV 
 Total WIDTH of Z4SS  =   3.155728429046366E-002 GeV 
 Lifetime             =   6.2426160E-13 cm  
  
 W1SS+  -->  Z1SS   UP     DB                11165.424     20.37
 W1SS+  -->  Z1SS   NUE    E+                 3578.382      6.53
 W1SS+  -->  Z1SS   CH     SB                11047.252     20.15
 W1SS+  -->  Z1SS   NUM    MU+                3578.319      6.53
 W1SS+  -->  Z1SS   NUT    TAU+               3548.031      6.47
 W1SS+  -->  Z2SS   UP     DB                 7352.349     13.41
 W1SS+  -->  Z2SS   NUE    E+                 2441.424      4.45
 W1SS+  -->  Z2SS   CH     SB                 7255.331     13.23
 W1SS+  -->  Z2SS   NUM    MU+                2441.417      4.45
 W1SS+  -->  Z2SS   NUT    TAU+               2417.434      4.41
  
 MASS of W1SS+        =    82.2054044085650      GeV 
 Total WIDTH of W1SS+ =   5.482536228723776E-005 GeV 
 Lifetime             =   3.5932274E-10 cm  
  
 W2SS+  -->  Z1SS   UP     DB               389890.656      7.89
 W2SS+  -->  Z1SS   NUE    E+               376353.063      7.62
 W2SS+  -->  Z1SS   CH     SB               389994.125      7.89
 W2SS+  -->  Z1SS   NUM    MU+              376345.625      7.62
 W2SS+  -->  Z1SS   NUT    TAU+             365778.000      7.40
 W2SS+  -->  Z2SS   UP     DB               626844.563     12.69
 W2SS+  -->  Z2SS   NUE    E+               381795.750      7.73
 W2SS+  -->  Z2SS   CH     SB               625884.000     12.67
 W2SS+  -->  Z2SS   NUM    MU+              381794.438      7.73
 W2SS+  -->  Z2SS   NUT    TAU+             381132.250      7.71
 W2SS+  -->  Z3SS   UP     DB                 1306.987      0.03
 W2SS+  -->  Z3SS   NUE    E+               154496.859      3.13
 W2SS+  -->  Z3SS   CH     SB                 1267.973      0.03
 W2SS+  -->  Z3SS   NUM    MU+              154476.875      3.13
 W2SS+  -->  Z3SS   NUT    TAU+             141892.547      2.87
 W2SS+  -->  W1SS+  UP     UB                 2861.109      0.06
 W2SS+  -->  W1SS+  DN     DB                 5163.812      0.10
 W2SS+  -->  W1SS+  NUE    ANUE              41784.742      0.85
 W2SS+  -->  W1SS+  E-     E+                15585.073      0.32
 W2SS+  -->  W1SS+  CH     CB                 2810.073      0.06
 W2SS+  -->  W1SS+  ST     SB                 5158.582      0.10
 W2SS+  -->  W1SS+  NUM    ANUM              41784.945      0.85
 W2SS+  -->  W1SS+  MU-    MU+               15583.285      0.32
 W2SS+  -->  W1SS+  BT     BB                 4413.938      0.09
 W2SS+  -->  W1SS+  NUT    ANUT              41842.730      0.85
 W2SS+  -->  W1SS+  TAU-   TAU+              14907.641      0.30
  
 MASS of W2SS+        =    122.500414227970      GeV 
 Total WIDTH of W2SS+ =   4.941149627879395E-003 GeV 
 Lifetime             =   3.9869262E-12 cm  
  
 MASS of UPL          =    259.504386750604      GeV 
  
 MASS of DNL          =    263.370394294766      GeV 
  
 NUEL   -->  Z1SS   NUE                   17929614.000     11.16
 NUEL   -->  Z2SS   NUE                      88574.359      0.06
 NUEL   -->  Z3SS   NUE                   11599536.000      7.22
 NUEL   -->  W1SS+  E-                   131070248.000     81.57
  
 MASS of NUEL         =    116.793041222755      GeV 
 Total WIDTH of NUEL  =   0.160687973351000      GeV 
 Lifetime             =   1.2259785E-13 cm  
  
 EL-    -->  Z1SS   E-                   147906416.000     49.85
 EL-    -->  Z2SS   E-                   140011936.000     47.19
 EL-    -->  Z3SS   E-                     4365978.000      1.47
 EL-    -->  Z4SS   E-                       87047.195      0.03
 EL-    -->  W1SS-  NUE                    1985265.250      0.67
 EL-    -->  W2SS-  NUE                    2332605.750      0.79
  
 MASS of EL-          =    125.486749674947      GeV 
 Total WIDTH of EL-   =   0.296689237161868      GeV 
 Lifetime             =   6.6399441E-14 cm  
  
 MASS of CHL          =    259.504386750604      GeV 
  
 MASS of STL          =    263.370394294766      GeV 
  
 NUML   -->  Z1SS   NUM                   17929614.000     11.16
 NUML   -->  Z2SS   NUM                      88574.359      0.06
 NUML   -->  Z3SS   NUM                   11599536.000      7.22
 NUML   -->  W1SS+  MU-                  131069416.000     81.57
  
 MASS of NUML         =    116.793041222755      GeV 
 Total WIDTH of NUML  =   0.160687141212113      GeV 
 Lifetime             =   1.2259848E-13 cm  
  
 MUL-   -->  Z1SS   MU-                  147906352.000     49.85
 MUL-   -->  Z2SS   MU-                  140012720.000     47.19
 MUL-   -->  Z3SS   MU-                    4365900.500      1.47
 MUL-   -->  Z4SS   MU-                      86681.125      0.03
 MUL-   -->  W1SS-  NUM                    1985265.250      0.67
 MUL-   -->  W2SS-  NUM                    2332605.750      0.79
  
 MASS of MUL-         =    125.486749674947      GeV 
 Total WIDTH of MUL-  =   0.296689524561566      GeV 
 Lifetime             =   6.6399380E-14 cm  
  
 MASS of BT1          =    263.417851693807      GeV 
  
 MASS of TP1          =    312.439636958576      GeV 
  
 NUTL   -->  Z1SS   NUT                   17929614.000     11.17
 NUTL   -->  Z2SS   NUT                      88574.359      0.06
 NUTL   -->  Z3SS   NUT                   11599536.000      7.23
 NUTL   -->  W1SS+  TAU-                 130831880.000     81.54
  
 MASS of NUTL         =    116.793041222755      GeV 
 Total WIDTH of NUTL  =   0.160449605025549      GeV 
 Lifetime             =   1.2277999E-13 cm  
  
 TAU1-  -->  Z1SS   TAU-                 147912640.000     49.83
 TAU1-  -->  Z2SS   TAU-                 140262848.000     47.25
 TAU1-  -->  Z3SS   TAU-                   4347101.000      1.46
 TAU1-  -->  W1SS-  NUT                    1986060.125      0.67
 TAU1-  -->  W2SS-  NUT                    2351768.000      0.79
  
 MASS of TAU1-        =    125.499288741710      GeV 
 Total WIDTH of TAU1- =   0.296860418547054      GeV 
 Lifetime             =   6.6361155E-14 cm  
  
 MASS of UPR          =    253.087070455429      GeV 
  
 MASS of DNR          =    253.391977765737      GeV 
  
 ER-    -->  Z1SS   E-                   221465200.000     67.88
 ER-    -->  Z2SS   E-                   103275600.000     31.65
 ER-    -->  Z3SS   E-                     1522374.125      0.47
  
 MASS of ER-          =    109.855352062128      GeV 
 Total WIDTH of ER-   =   0.326263181300843      GeV 
 Lifetime             =   6.0380703E-14 cm  
  
 MASS of CHR          =    253.087070455429      GeV 
  
 MASS of STR          =    253.391977765737      GeV 
  
 MUR-   -->  Z1SS   MU-                  221465760.000     67.88
 MUR-   -->  Z2SS   MU-                  103275296.000     31.65
 MUR-   -->  Z3SS   MU-                    1522204.250      0.47
  
 MASS of MUR-         =    109.855352062128      GeV 
 Total WIDTH of MUR-  =   0.326263714477010      GeV 
 Lifetime             =   6.0380601E-14 cm  
  
 MASS of TP2          =    307.130371718121      GeV 
  
 MASS of BT2          =    253.439037459913      GeV 
  
 TAU2-  -->  Z1SS   TAU-                 221675248.000     67.89
 TAU2-  -->  Z2SS   TAU-                 103215376.000     31.61
 TAU2-  -->  Z3SS   TAU-                   1477561.500      0.45
 TAU2-  -->  W1SS-  NUT                     131231.641      0.04
  
 MASS of TAU2-        =    109.869716964501      GeV 
 Total WIDTH of TAU2- =   0.326499416297990      GeV 
 Lifetime             =   6.0337016E-14 cm  
  
 MASS of GLSS         =    148.387594191548      GeV 
  
 MASS of H0L          =    56.3720788464574      GeV 
  
 MASS of H0H          =    313.892691024644      GeV 
  
 MASS of A0           =    300.000000000000      GeV 
  
 MASS of H+           =    310.070287501751      GeV 
  
 MASS of TP           =    174.000000000000      GeV 
  
  
  Cross section of W1SS+  W1SS-   =     0.89911E+00 pb 
    
\end{verbatim}}

\subsection{The susygen.log file}
{\footnotesize\begin{verbatim}
                            Event listing (summary)

    I  particle/jet KS     KF orig    p_x      p_y      p_z       E        m

    1  !E--!        21     11    0    0.000    0.000   91.500   91.500    0.000
    2  !E-+!        21    -11    0    0.000    0.000  -91.500   91.500    0.000
    3  GAMMA         1     22    0    0.000    0.000   -0.964    0.964    0.000
    4  (SusyProd0)  11     79    0    0.000    0.000    0.964  182.036  182.033
    5  (W1SS++)     11     75    4  -17.668  -31.556   15.259   91.096   82.205
    6  (W1SS-)      11     77    4   17.668   31.556  -14.295   90.940   82.205
    7  NUT           1     16    5    5.178   -0.308   -7.449    9.077    0.000
    8  (TAU-+)      11    -15    5    1.541    4.919   14.326   15.328    1.777
    9  (Z2SS0)      11     72    5  -24.386  -36.167    8.383   66.691   49.718
   10  NUT~          1    -16    6   16.947   16.397    1.743   23.645    0.000
   11  (TAU--)      11     15    6   10.948   -2.815   -2.115   11.637    1.777
   12  (Z2SS0)      11     72    6  -10.227   17.974  -13.923   55.657   49.718
   13  GAMMA         1     22    9   -3.387   -8.227    3.401    9.525    0.000
   14  Z1SS0         1     71    9  -20.999  -27.940    4.981   57.166   44.961
   15  GAMMA         1     22   12   -4.034   -1.489   -2.487    4.967    0.000
   16  Z1SS0         1     71   12   -6.193   19.463  -11.435   50.690   44.961
   17  NUT~          1    -16    8   -0.015    1.559    3.381    3.723    0.000
   18  pi+           1    211    8    1.160    1.789    5.627    6.019    0.140
   19  (eta)        11    221    8    0.396    1.571    5.318    5.587    0.548
   20  NUT           1     16   11    3.007   -0.958    0.086    3.157    0.000
   21  (rho-)       11   -213   11    7.941   -1.857   -2.202    8.480    0.741
   22  GAMMA         1     22   19    0.076    1.008    3.876    4.006    0.000
   23  GAMMA         1     22   19    0.319    0.563    1.442    1.581    0.000
   24  pi-           1   -211   21    0.341   -0.086   -0.036    0.380    0.140
   25  (pi0)        11    111   21    7.600   -1.771   -2.165    8.099    0.135
   26  GAMMA         1     22   25    0.863   -0.245   -0.241    0.929    0.000
   27  GAMMA         1     22   25    6.736   -1.526   -1.924    7.170    0.000
                   sum:  0.00         0.000    0.000    0.000  183.000  183.000

  Susygen has now finished ... Bye
\end{verbatim}}

\end{document}